\documentclass[11pt,letterpaper]{article}
 \pdfoutput=1

\usepackage{jcappub}

\usepackage{caption}
\usepackage{floatrow}
\usepackage{journals}

\usepackage{feynmp}

\usepackage{subfig}

\usepackage{bm}

\def\be{\begin{eqnarray}}
\def\ee{\end{eqnarray}}

\def\E{\bigstar}

\def\k{\bm{k}}

\def\x{\bm{x}}
\def\q{\bm{q}}
\def\s{\bm{s}}
\def\sd{\dot{\bm{s}}}
\def\H{\mathcal{H}}
\def\xs{{\bm{x}}_s}
\def\d{\delta_D}

\def\la{\left (}
\def\ra{\right )}
\def\v{\bm{v}}

\newsavebox{\tempbox}

\title{Estimating CDM Particle Trajectories in the Mildly Non-Linear Regime of Structure Formation.\\Implications for the Density Field in Real and Redshift Space}
\author{Svetlin Tassev$^{a}$ and Matias Zaldarriaga$^{b}$}
\affiliation{ \sl $^{a}$ Center for Astrophysics, Harvard University, Cambridge, MA 02138, USA\\
 \sl $^{b}$ School of Natural Sciences, Institute for Advanced Study, Olden Lane, Princeton, \\NJ 08540, USA
}

\abstract{ We obtain approximations for the CDM particle trajectories starting from Lagrangian Perturbation Theory. These estimates for the CDM trajectories result in  approximations for the density in real and redshift space, as well as for the momentum density that are better than what standard Eulerian and Lagrangian perturbation theory give. For the real space density, we find that our  proposed approximation gives a good cross-correlation ($>95\%$) with the non-linear density down to scales almost twice smaller than the non-linear scale, and six times smaller than the corresponding scale obtained using linear theory. This allows for a speed-up of an order of magnitude or more in the scanning of the cosmological parameter space with N-body simulations for the scales relevant for the baryon acoustic oscillations.  Possible future applications of our method include baryon acoustic peak reconstruction, building mock galaxy catalogs, momentum field reconstruction.
} 

\notoc
\usepackage{graphicx}
\begin{document}
\maketitle



\section{Introduction}\label{intro}

Understanding the evolution of the large-scale structure (LSS) of the universe is a crucial ingredient of present-day cosmology. Because of the larger accessible volume, the importance of the LSS will only grow in the future as it has the potential to become an even more accurate cosmological probe than the Cosmic Microwave Background (CMB). Thus observations of the LSS alone or combined with CMB observations could provide even stronger constraints on the equation of state of dark energy, modifications to gravity and primordial non-Gaussianities.


On the theoretical side, the largest scales in the universe ($\gtrsim 100\,$Mpc at redshift $z=0$) can be treated perturbatively (e.g. \cite{jain}). However, the perturbative treatment of the LSS breaks down when the overdensity (the standard parameter controlling the perturbative expansions) becomes of order 1. This corresponds to the mildly non-linear (MNL) regime: scales $\lesssim 50\,$Mpc at $z=0$. The dynamics at these scales is important as it introduces percent-level shifts to the acoustic peak caused by the Baryon Acoustic Oscillations (BAO) in the matter correlation function \cite{2005ApJ...633..560E}.  On the observational side, such an accuracy in the determination of the position of the acoustic peak is required by ongoing and future surveys of LSS, such as BOSS\footnote{http://www.sdss3.org/surveys/boss.php} and WFIRST\footnote{http://wfirst.gsfc.nasa.gov/}. Such surveys target biased tracers of the underlying density field in redshift space. Thus, to make contact with observations, what is required is a good understanding not only of the MNL regime, but of the non-linear (NL) regime as well, since finger-of-god (FoG) effects from NL structures, such as galaxy clusters, leak into the MNL scales.

Most theoretical studies of structure formation in the MNL and NL regimes are currently done through numerical simulations. However, computational time is still a very important constraint on simulations. Sampling variance requires one to perform numerous simulations of different realizations of the same cosmology; or alternatively simulate large volumes. This prevents an efficient sampling of the cosmological parameter space (e.g. \cite{Angulo:2007fw}), and thus, it is still difficult to get a complete handle on the uncertainties that would arise from measurements at the MNL scales relevant to the BAO, and the NL scales relevant to e.g. weak lensing and galaxy clustering. 

Having an accurate model for the (M)NL regimes of structure formation is therefore important for the utilization of observations of galaxy clustering.
One important application of such a model will be for improving the reconstruction techniques \cite{2009PhRvD..79f3523P,2010ApJ...720.1650S} of the baryon acoustic peak from observations. One can argue that reconstruction pulls the information stored in the higher-$n$-point functions and puts it back into the power spectrum, and thus has been proven to reduce the errors in the determination of the position of the acoustic peak \cite{2009PhRvD..79f3523P,2010ApJ...720.1650S} -- a necessary improvement for utilizing current and future BAO observations.

In a recent paper \cite{tassev} (TZ from now on), we offered a model for the matter distribution in the MNL regime. We showed that through an expansion around the Zel'dovich approximation (ZA) \cite{zeldovich} (or higher-order Lagrangian Perturbation Theory (LPT) \cite{catelan}) one can construct robust approximations to the MNL density field in real space. Therefore, one can write the true overdensity $\delta$ in terms of the cheap-to-compute density field in the ZA, $\delta_z$, as (in Fourier space):
\be\label{first}
\delta=R_\delta\delta_z+\delta_{MC}\ ,
\ee
where $R_\delta$ is a non-random scale- and time-dependent transfer function; and $\delta_{MC}$ ($MC$ standing for mode coupling) is a small random residual at MNL scales. We showed that the resulting matter power spectrum estimator reduces sample variance by more than an order of magnitude in the MNL regime. We found that $\delta$ and $\delta_z$ are strongly correlated in that regime, which implies a small $\delta_{MC}$ and an $R_\delta$ which has small sample variance: a result which is key to the success of our treatment in TZ.

In TZ we also showed that other methods (e.g. \cite{rpt,mlpt}) based on expansions similar to eq.~(\ref{first}) but done around the linear density field, $\delta_L$, do not offer similar improvements to cosmic variance. We found that this is caused by the fact that $\delta_L$ and $\delta$ become decorrelated for scales smaller than the rms particle displacements (corresponding to a wavevector $k\approx0.12h/$Mpc). We refer the reader to TZ for a detailed discussion of this result.

In general, in this paper we refer to quantities $Q_\E$ which are highly-correlated to the corresponding non-linear quantities $Q$, as \textit{approximations} of $Q$. This is because, once we know the transfer function $R_Q$ relating $Q$ and $Q_\E$, we can construct approximations to $Q$ by writing $R_Q Q_\E$. Obviously, to build unbiased \textit{estimators}, one needs to include the effect of the MC terms: $Q=R_QQ_\E+Q_{MC}$, as done in eq.~(\ref{first}) and in TZ. However, since by construction $Q_\E$ and $Q$ are strongly correlated, we have that $Q_{MC}$ is small and $R_Q$ has small sample variance and can be obtained by using only few modes. In this terminology, $\delta_z$ is a good approximation of $\delta$, while $\delta_L$ is not.

One problem with the TZ treatment is that it is not suitable for including the effects of Redshift Space Distortions (RSD). It is not particle-based, i.e. it constructs an approximation of the nonlinear density by treating it as a field, and not as generated by discrete particles with known velocities. Furthermore, this prevents one from extending the TZ method for constructing mock galaxy catalogs based on estimated particle positions and velocities. Another way of saying this is that working only with the density field misses the complicated phase-space structure of Cold Dark Matter (CDM), which can be recovered in the particle picture. 


In this paper, we address this issue by concentrating on the CDM particle trajectories in LPT, and relating those to the trajectories obtained from N-body simulations.  The reason we choose to work around the LPT results is again related to the discussion in TZ of the effects of the bulk flows, which cancel explicitly at each order in LPT. We  find that using \textit{approximations} of the particle velocities and positions based on LPT  results in good \textit{approximations} for the density in real and redshift space, which  allows us to go well into the MNL regime down to scales comparable and even smaller than the non-linear scale ($k_{NL}\approx 0.25h/$Mpc for redshift of $z=0$). 

\begin{figure}[t!]
  \centering
\includegraphics[width=0.6\textwidth]{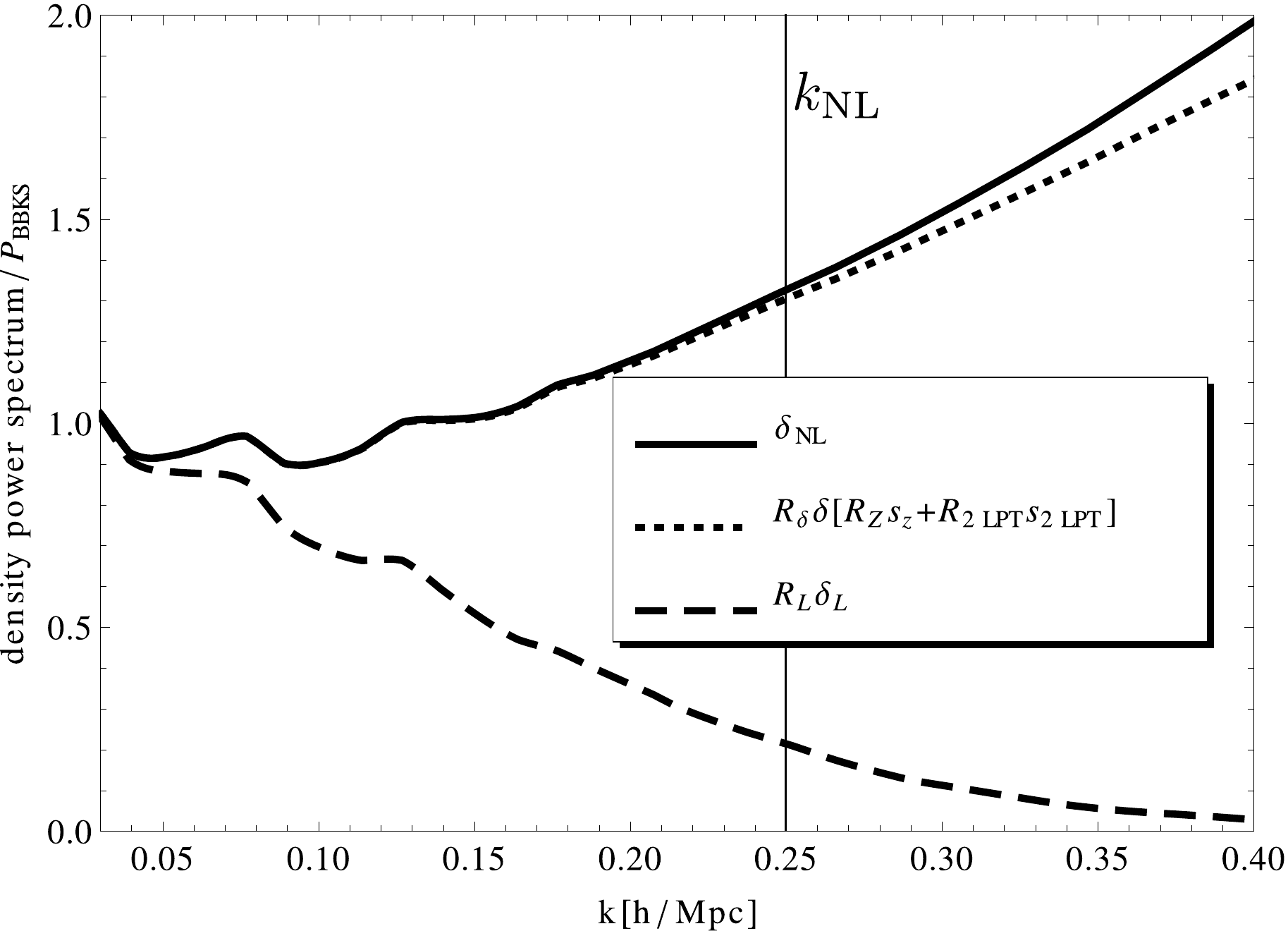}   
  \caption{Shown are various matter power spectra at $z=0$ for $\Lambda$CDM, similar to Fig. 1 of \cite{tassev} (errorbars are suppressed for clarity). The power spectra are divided by a smooth BBKS \cite{BBKS} power spectrum with shape parameter $\Gamma=0.15$ in order to highlight the wiggles due to the BAO. The solid curve shows the non-linear power spectrum (i.e. the ``exact'' power spectrum obtained from N-body simulations); the part of the NL power spectrum due to the ``memory of the initial conditions'' \cite{rpt} is shown with the dashed line; the power due to the projection of the non-linear density field on our best approximation (cf. second line of eq.~(\ref{QE2})) is given by the dotted curve. The residual mode-coupling power is 13.5\% of the NL power at $k=0.5h/$Mpc for $z=0$. Here $R_\delta$ corresponds to the transfer function relating the NL density and the density given by our best approximation. A vertical line denotes the non-linear scale. The current Hubble expansion rate in units of 100\,km/s/Mpc is given by  $h$. }
  \label{fig:power}
\end{figure}

In a nutshell, the method used in this paper for building an approximation of a quantity, $Q[\x,\bm{v}]$, which is a functional of the particle positions ($\bm{x}$) and velocities ($\bm{v}$) is as follows.\footnote{Here $Q$ can stand for the matter density in real and redshift space, as well as the momentum and kinetic energy densities.} It consists of first calculating the displacement (related to the positions $\bm{x}_{LPT}$) and velocity ($\bm{v}_{LPT}$) fields in LPT. Those are then transformed using transfer functions obtained after a calibration with simulations. These transfer functions, $R$ and $R^v$, relate the LPT displacement and velocity fields to their respective fully non-linear counterparts. The effect of these transfer functions is twofold: 1) To capture higher-order LPT effects, which are not included under a truncation of the LPT expansion; 2) To capture some of the effects of stream crossing, which are missed by LPT (see below). Therefore, schematically we can write ${\bm{x}}=\bm{x}_\E+\x_{MC}\equiv R*\x_{LPT}+\x_{MC}$ and ${\bm{v}}=\bm{v}_\E+\bm{v}_{MC}\equiv R^v*\bm{v}_{LPT}+\bm{v}_{MC}$, where $*$ denotes a convolution in Lagrangian space, and quantities with subscript $MC$ are random residuals. One can then construct $Q_\E\equiv Q[\x_\E,\bm{v}_\E]$. The effect of the $MC$ terms can then be captured by relating $Q_\E$ to $Q$ by using yet another transfer function ($R_Q$) and a mode coupling term ($Q_{MC}$): $Q=R_Q Q_\E+Q_{MC}$ (in Fourier space). The main benefit of first transforming the LPT displacement and velocity fields is that, as we will show, the phase-space structure of the CDM is recovered better. This means that $Q_\E$ as calculated above is better correlated to $Q$, than if $Q_\E$ was built from the uncorrected LPT quantities (such that $Q_\E=Q_{LPT}\equiv Q[\x_{LPT},\bm{v}_{LPT}]$) as we did in TZ. 

One may be worried that the proliferation of transfer functions compared to TZ is an overkill. However, those can be obtained from calibration with modest N-body runs as the transfer functions have rather small sample variance. Another point we would like to stress is that in this paper we concentrate on the CDM dynamics only, neglecting the effects of baryon physics. Those, however, should be possible to capture by calibrating the displacement and velocity transfer functions with N-body simulations, which include baryon physics.

To demonstrate the power of the method described above, in Fig.~\ref{fig:power} we show the NL power spectrum obtained from simulations, along with the power spectrum of $R_\delta\delta_\E$ (without the MC power) obtained from our method (cf. Fig.~1 in TZ). For comparison, we include the power spectrum of the piece of the overdensity ($R_L\delta_L$) corresponding to the memory of the initial conditions \cite{rpt}. Thus, the residual mode-coupling power for the method described above is indeed rather small well into the NL regime -- it is 13.5\% of the NL power at $k=0.5h/$Mpc for $z=0$. As we will show later on, the estimators for the NL power spectrum based on the method above (see also TZ) have a negligible variance in the linear regime (the variance is suppressed by a factor of $\sim10^5$ for $k\sim0.02h/$Mpc at $z=0$). The cross-correlation coefficient between our best $\delta_\E$ and the NL $\delta$ is $>0.95$ up to $k\approx 0.45h/$Mpc at $z=0$. One should compare this number with the corresponding scale for the pure 2LPT-based approximation (the best approximation used in TZ): $k\approx 0.37h/$Mpc.

\begin{figure}[t!]
  \centering
\includegraphics[width=0.6\textwidth]{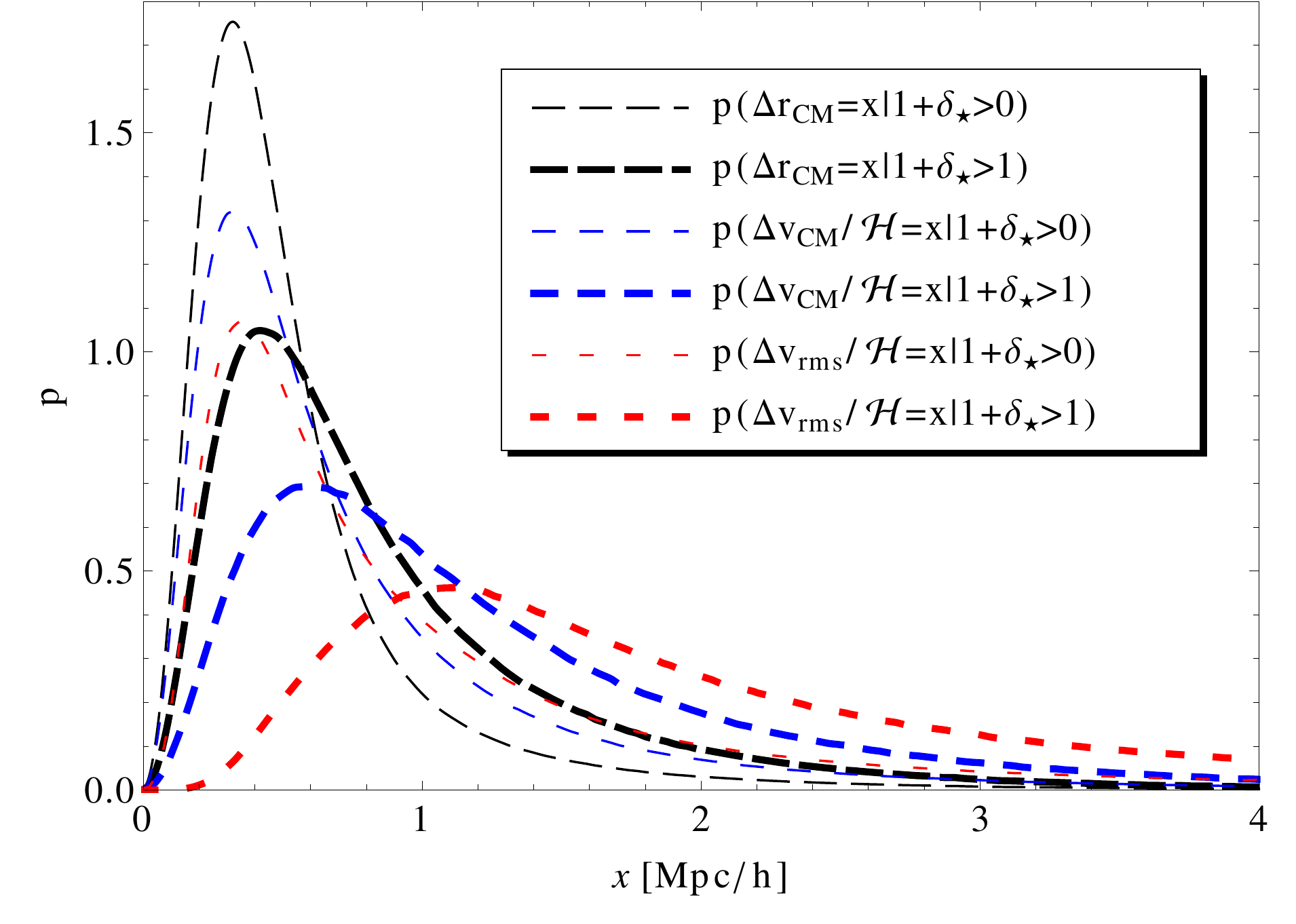}   
  \caption{Probability density, $p$, for the distribution of errors in determining the center-of-mass (CM) position/velocity of a given volume element; as well as the error in determining the rms velocity ($z=0$). We show the distributions after imposing the denoted cutoffs in the approximated density field, $\delta_\E$. The approximate velocities, positions and density are calculated according to our best model (cf. second line of eq.~(\ref{QE2})). CM velocities and positions for high-density regions are determined with worse precision than for voids, but still the error is within a couple of Mpc. The rms velocity responsible for the FoG effect is determined with worse precision but again within several Mpc. The rms linear displacement corresponding to $z=0$ is $\approx15\,$Mpc/h, or about an order of magnitude larger than the errors depicted in the plot. The probabilities are obtained after Gaussian smoothing on a scale of $1.3\,$Mpc$/h$. }
  \label{fig:hist}
\end{figure}

However, one is ultimately interested in tracers of the underlying density field. Therefore, in Fig.~\ref{fig:hist}, we show the probability density for the distribution of errors in approximating the center-of-mass (CM) positions and velocities of Eulerian volume elements; as well as the distribution of errors in estimating the root-mean-square (rms) velocity, responsible for the FoG effect. We show the probabilities both for all space ($1+\delta_\E>0$) and for overdense regions ($1+\delta_\E>1$) only. We obtained Fig.~\ref{fig:hist} after Gaussian smoothing on a scale of $1.3\,$Mpc$/h$ and therefore $1+\delta_\E>1$  corresponds to regions containing approximately a galaxy mass or more. From the figure one can argue that our method should be robust to within a couple of Mpc in predicting tracer positions both in real and in redshift space (neglecting the FoG). The FoG effects can lead to somewhat larger errors. However, the BAO scale is determined by averaging the cross-correlation between numerous tracers, which implies that the precision in measuring the BAO scale should be better than the precision in the position of individual tracers. Comparing the curves for $1+\delta_\E>0$ and $1+\delta_\E>1$ one can also argue that our method works much better in underdense regions -- a result which we confirm later on.

In Section~\ref{prelim} we write down our model for the particle trajectories, and their relation to the density in real and redshift space. In Section~\ref{sec:num} we show the displacement and velocity transfer functions obtained from simulations, and show their relation (at lowest order in LPT) by building a simple model for their time dependence. Then, we plot the resulting density transfer and cross-correlation functions in real and redshift space (as well as the ones for the momentum and kinetic energy densities), and compare our results with the results in TZ. We summarize in Section~\ref{sec:summary}.

\section{Expanding Perturbatively the CDM Equation of Motion}\label{prelim}
\subsection{Density in real and redshift space}\label{sec:density}
In this section we will relate the CDM particle trajectories (those can be perturbative or obtained from N-body simulations) to the density in real and redshift space. These relations will later allow us to discuss the resulting approximations for the density, starting from the approximations for the particle trajectories, which we will construct later on.
 
 We start off by writing the relation between the Lagrangian coordinates, $\q$, of a particle and its Eulerian comoving coordinates, $\x$:
\be\label{EL}
\x(\q,\eta)=\q+\s(\q,\eta)\ ,
\ee
where $\eta$ is conformal time; and $\s$ is a stochastic displacement vector field. As an illustration, in the Zel'dovich approximation \cite{zeldovich}, $\s$ is given by $\s_z(\q,\eta)=D(\eta)\s_z(\q,\eta_0)$, where $D$ is the growth factor with $D(\eta_0)=1$. Here $\partial_{\q}\cdot\s_z=-\delta_L$, where $\delta_L$ is the fractional overdensity given by linear theory.

The fully non-linear fractional CDM overdensity in real space is then given by
\be\label{delta}
\delta(\x,\eta)=\int d^3 q \d(\x-\q-\s)-1\ ,
\ee
where $\delta_D$ is the Dirac delta function. The integral above is simply a sum over all CDM streams of the number of particles falling in an Eulerian volume element around $\x$. 

Similarly, we can write the redshift space coordinates, $\xs(\q,\eta)$, of a particle as $\xs=\x+\sd_{||}/\H=\q+\s+\sd_{||}/\H$, where $\sd_{||}$ is the line-of-sight (LoS) piece of the peculiar velocity, $\sd$. A dot represents a derivative with respect to conformal time, and $\H\equiv\dot a/a= a H$, where $a$ is the cosmological scale factor, and $H$ is the Hubble parameter. Therefore, the fractional overdensity in redshift space, $\delta_s$, is given by
\be\label{deltaRSD}
\delta_s(\xs,\eta)&=&\int d^3 q \d\left(\xs-\q-\s-\frac{\sd_{||}}{\H}\right)-1\\\nonumber
&=&\int d^3x d^3 q \d\left(\xs-\x-\frac{\sd_{||}}{\H}\right)\d(\x-\q-\s)-1\ .
\ee
We can Fourier transform with respect to $\xs$ and expand with respect to the velocity to get ($k\neq 0$):
\be\label{redshiftDelta}
\delta_s(\k,\eta)=\sum\limits_{n=0}^\infty\frac{1}{n!}\left(\frac{-ik_{||}}{\H}\right)^n T_{||}^{(n)}(\k,\eta)\ ,
\ee
where $T_{||}^{(n)}(\k, \eta)$ is the Fourier transform with respect to $\x$ (not $\xs$) of the quantity:
\be
T_{||}^{(n)}(\x,\eta)=\int d^3 q \d\bigg(\x-\q-\s(\q,\eta)\bigg){\dot{s}}_{||}^n(\q,\eta)\ .
\ee
The result, eq.~(\ref{redshiftDelta}), matches that of \cite{SeljakRSD} (up to a Fourier convention).

Therefore, as in \cite{SeljakRSD} we can see that at zero order in the velocity, $\delta_s$ is given by the real-space density $\delta$. At first order it is related to the LoS momentum density; while at second order -- to the LoS kinetic energy density.

\subsection{The CDM equation of motion}\label{sec:EoM}
To see how the displacement field can be expanded perturbatively, we have to start with the equation of motion for CDM particles:
\be\label{eomX}
\ddot \x+\H \dot \x=-\nabla\phi\ ,
\ee
where $\phi$ is the gravitational potential sourced by $\delta$ according to the Poisson equation. To write eq.~(\ref{eomX}), we restrict ourselves to modes well inside the horizon, and neglect the effect of baryons. Using eq.~(\ref{EL},\ref{delta}), the above equation can be rewritten as
\be\label{S}
\s''(\q,\eta(\tau))&=&-\frac{D''}{D}\nabla \nabla^{-2}\delta\\
&=&-\frac{D''}{D}\int \frac{d^3\tilde qd^3k}{(2\pi)^3}(-i)\frac{\k}{k^2}e^{i\k\cdot(\q-\tilde\q)}\exp\bigg( i\k\cdot\la\s(\q,\eta)-\s(\tilde \q,\eta)\ra\bigg)\nonumber \ ,
\ee
using Jeans swindle in the form of $\d(\k)\k/k^2=0$ (i.e. a uniform density field generates zero acceleration).
Here primes denote a derivative with respect to $\tau$ defined as $d\tau\equiv d\eta/a$. Unlike LPT, the equation above is exact, containing all information about stream crossing.\footnote{\label{foot:LPT}To compare eq.~(\ref{S}) with LPT, we need to take the divergence (in Eulerian space) of that equation. We obtain:
\be\label{divS}
\mathrm{Tr}\left[\frac{1}{\bm{1}+\bm{M}}\bm{M}''(\q)\right]&=&-\frac{D''}{D}\delta=\frac{D''}{D}-\frac{D''}{D}\left.\sum_{\tilde q}\frac{1}{\mathrm{det}\left[\bm{1}+\bm{M}(\tilde \q)\right]}\right|_{\tilde \q+\s(\tilde q,\eta)=\q+\s(\q,\eta)}\\
&=&\frac{D''}{D}-\frac{D''}{D}\int \frac{d^3q'd^3k}{(2\pi)^3}e^{i\k\cdot(\q-\tilde\q)}\exp\bigg( i\k\cdot\la\s(\q,\eta)-\s(\tilde\q,\eta)\ra\bigg)\nonumber\ ,
\ee
where $(\bm{M})_{ij}=\partial_{q_i}s_j$ is the deformation tensor. The sum is restricted over all streams originating from $\tilde \q$ which arrive at $\x=\q+\s(\q,\eta)$ at time $\eta$. LPT (e.g. \cite{catelan}) is obtained by writing the density entering above in the single-stream approximation. That implies restricting the sum on the first line of eq.~(\ref{divS}) to $\tilde \q=\q$, and replacing $s_i(\q)-s_i(\tilde\q)$ on the second line with $(q-\tilde q)_j M_{ji}(\q)$.}
Since we want to go beyond LPT, another route for solving eq.~(\ref{S}) is needed. 

The above equation can in principle be solved using the Born approximation starting from the Zel'dovich solution $\s_z$. This automatically recovers the particle interpretation of the Helmholtz hierarchy developed in \cite{HHpaper} (see their eq.~(8.11) with $m=n-1$). However, solving this hierarchy still requires intensive N-body calculations (see \cite{HHpaper}).

Another way of thinking about eq.~(\ref{S}) is as an equation of motion for the vector field $\s$ in a non-linear theory, whose first order (``free theory'') solution is the Zel'dovich approximation in which $\s$ is a Gaussian random field. However, as can be seen from eq.~(\ref{S}), the interaction vertices for $\s$ are infinitely many and of infinite order. Therefore, treating $\s$ analytically in the interaction picture and \textit{not} using the single-stream approximation for the density is still an immensely complicated task. 

Therefore, in the rest of this paper, we will try to solve eq.~(\ref{S}) by guessing a solution for $\bm{s}$, which order by order is given by the LPT result, but convolved with a transfer function to be obtained by a calibration with simulations. The effect of the transfer functions will be twofold: 1) To capture higher-order LPT effects, which are not included under a truncation of the LPT expansion; 2) To capture some of the effects of stream crossing, which are missed by LPT (see footnote \ref{foot:LPT}).

\subsection{Expanding the displacement field}

In this section we will write down the approximations for the CDM trajectories following the prescription at the end of Section~\ref{sec:EoM}. We build a linear model for the fully non-linear displacements and velocities starting from the LPT results (in analogy with the linear model for the density we built in TZ). For simplicity, we restrict ourselves to second order LPT. Thus, we can write (in Fourier space with respect to $\q$) :
\be\label{splitD}
\s(\k,\eta)&=&\s_{\E}(\k,\eta)+\s_{MC}(\k,\eta)\\ \nonumber
&=&R_z(k,\eta)\s_z(\k,\eta)+R_{2LPT}(k,\eta)\s_{2LPT}(\k,\eta)+\s_{MC}(\k,\eta)\\
\v(\k,\eta)&=&\dot\x(\k,\eta)=\dot\s(\k,\eta)=\v_{\E}(\k,\eta)+\v_{MC}(\k,\eta)\label{splitVel}\\ \nonumber &=&R^v_z(k,\eta)\dot\s_z(\k,\eta)+R^v_{2LPT}(k,\eta)\dot\s_{2LPT}(\k,\eta)+\v_{MC}(\k,\eta)\ ,
\ee
where  $\s_\E\equiv \s-\s_{MC}$ and  $\v_\E\equiv \v-\v_{MC}$ are the displacement and velocity approximations\footnote{Note that we do not need extra transfer functions $R_Q$ (see eq.~(\ref{Q})) when $Q$ stands for $\s$ or $\v$, because of the linearity of our model (\ref{splitD}, \ref{splitVel}), which means that any $R_Q$ can be absorbed in the $R$'s and $R^v$'s.}, respectively; a dot denotes a derivative with respect to $\eta$; and $\k$ above denotes wavevectors corresponding to $\q$ (not $\x$); $R_{z/2LPT}(k,\eta)$ is the transfer function acting on the displacement, $\s_{z/2LPT}(\k,\eta)$, obtained at the corresponding order in LPT.\footnote{The lowest order in LPT is the the ZA (subscripts $z$), and the next is 2LPT (subscripts $2LPT$).} We denote with $\s_{MC}$ the residual (``mode coupling'') displacement, defined such that $\langle \s_{MC}(\k,\eta) \s_{z/2LPT}^*(\k,\eta)\rangle=0$, where angular brackets denote ensemble averages. The displacement transfer functions are given by a least squares fit in Lagrangian space:\footnote{Note that the displacement and velocity transfer functions should in principle be tensors. Thus, for example, the term $R_zs_{z,i}$ should really be $R_{z,ij}s_{z,j}$. However, due to isotropy $R_{z,ij}$ is proportional to the identity matrix, which allows us to replace it with the scalar transfer function, $R_z$. One can make the same argument for the rest of the transfer functions.}
\be\label{Rdispl}
R_z(k,\eta)&=&\frac{\langle s_i(\k,\eta)s_{z,i}^*(\k,\eta)\rangle}{\langle s_{z,i}(\k,\eta)s_{z,i}^*(\k,\eta)\rangle} \ \ \hbox{(no summation)}\nonumber\\
R_{2LPT}(k,\eta)&=&\frac{\langle s_i(\k,\eta)s_{2LPT,i}^*(\k,\eta)\rangle}{\langle s_{2LPT,i}(\k,\eta)s_{2LPT,i}^*(\k,\eta)\rangle} \ \ \hbox{(no summation)}\ ,
\ee
where  we used that $\langle\s_z\s_{2LPT}\rangle=0$ in the absence of primordial non-Gaussianities, since $\s_z$ is first order in $\delta_L$, while $\s_{2LPT}$ is second order. Obviously, the transfer functions above depend on the order at which we truncate the LPT expansion. However, since $\langle\s_z\s_{2LPT}\rangle=0$, $R_z$ and $R_z^v$ are the same whether we truncate at first or second order in LPT. Following TZ, in order to make the above transfer functions unbiased, when numerically calculating the ensemble averages one has to average over all available modes (i.e. both over all simulation boxes and over all $\hat {k}$).
 As long as $\s$  and $\s_{z/2LPT}$ are well correlated, $\s_{MC}$ is small, and the $R$'s have small sample variance, and can be obtained from relatively few modes in the same way we obtained the density transfer function in TZ. The corresponding quantities entering in eq.~(\ref{splitVel}) for the velocity are defined analogously.

Below, we will investigate the consequences of the linear model, eq.~(\ref{splitD},\ref{splitVel}), on the resulting approximations for the density (in both real and redshift space), as well as for the momentum and kinetic energy densities. All of those quantities are functionals of $\x(\q,\eta)$ and $\v(\q,\eta)$, and therefore we denote them as $Q[\x,\v]$ as we did in the Introduction. We already wrote down $\delta[\x,\v]$ and $\delta_s[\x,\v]$ in Section~\ref{sec:density}. There, we can also read off the LoS momentum density (divided by the CDM particle mass) given by $T^{(1)}_{||}$, and the LoS kinetic energy density (divided by the CDM particle mass) given by $T^{(2)}_{||}/2$.

To build an approximation for $Q$, one has to calculate the displacement fields $\s_z$ and $\s_{2LPT}$ and their time derivatives for a given realization, and scale them with the growth factor functions to the redshift of interest. This can be done with any 2LPT code for calculating the initial conditions for N-body simulations. One then multiplies (in Lagrangian Fourier space) the Lagrangian displacement/velocity fields with the displacement/velocity transfer functions to obtain $\s_\E$ and $\v_\E$.  The transfer functions themselves can be obtained from a small set of realizations from eq.~(\ref{Rdispl}) and the analogous expression for the $R^v$'s.

The next step is to include the effects of the mode coupling terms. One possible way of dealing with that problem is to make a model for $\s_{MC}$ and $\v_{MC}$, e.g. by assuming they are Gaussian for simplicity. Another route (which works better as we will show later) for including their effects is to first calculate the approximation $Q_\E(\k,\eta)\equiv Q[\q+\s_\E,\v_\E](\k,\eta)$ (where $\k$ can be a  wavevector corresponding to $\x$, $\x_s$ or $\q$ depending on the quantity $Q$) and then write the following linear model for $Q$:
\be\label{Q}
Q(\k,\eta)=R_Q(k,\eta)Q_\E(\k,\eta)+Q_{MC}(\k,\eta)\ , \ \ \hbox{with}\\ \nonumber
R_Q(k,\eta)=\frac{\langle Q(\k,\eta)Q_\E^*(\k,\eta)\rangle}{\langle Q_\E(\k,\eta)Q_\E^*(\k,\eta)\rangle} \ ,
\ee
where again the transfer function $R_Q$ is defined in the least-squares sense, and $Q_{MC}$ is given by $\langle Q_\E(\k,\eta)Q_{MC}^*(\k,\eta)\rangle=0$. 
In this paper, we follow this second route as it requires no assumptions on the statistics of $\s_{MC}$ and $\v_{MC}$. Moreover, as we will show, $Q_\E$ and $Q$ are very well correlated, implying that $Q_\E$ is indeed a good approximation for $Q$.

To summarize, the model we pursue in this paper is given by the equations of this section: eq.~(\ref{splitD}, \ref{splitVel}, \ref{Q}). Therefore, estimators (denoted by a hat) of fully non-linear quantities, $Q$, can be built for each realization according to the following scheme:
\be
(\s_{LPT},\dot\s_{LPT}) \to (\s_\E,\v_\E) \to Q_\E \to \hat Q\ ,
\ee
once one has calculated the corresponding transfer functions. The same applies if one wants to construct unbiased estimators of statistics of $Q$ such as its power spectrum. In that case, the mode coupling term, $Q_{MC}$, can be dealt with in the same way as done in TZ. Note that since eq.~(\ref{Q}) is a linear model for $Q$, it should be possible to invert the arrows in the scheme above, thus obtaining a self consistent reconstruction technique for the BAO peak. However, we postpone such a study to future work.

\section{Numerical results}\label{sec:num}

In this section we present numerical results based on the  model given by eq.~(\ref{splitD}, \ref{splitVel}, \ref{Q}). 
The results that follow are extracted from 26 N-body simulations, made using GADGET-2 \cite{gadget}, with $256^3$ particles each, in a box with side $L_{\mathrm{box}}=500\,$Mpc$/h$. The cosmological parameters  we used are (in the standard notation) as follows
\be\label{cosmo}
(\Omega_b,\Omega_{\mathrm{matter}},\Omega_\Lambda,h,n_s,\sigma_8)=(0.046,0.28,0.72,0.70,0.96,0.82)\ .
\ee
The initial conditions were set at a redshift of $49$, using the 2LPT code provided by \cite{2006MNRAS.373..369C}. 

 Following TZ, we quantify the quality of the model given by eq.~(\ref{splitD}, \ref{splitVel}, \ref{Q}) by plotting the cross-correlation coefficient, $\rho_{Q}$, between a fully non-linear quantity $Q$ and its approximation, $Q_\E$:
\be\label{rhoQ}
\rho_{Q}(k,\eta)\equiv \frac{\langle Q(\k,\eta)Q_\E^*(\k,\eta)\rangle}{\sqrt{\langle |Q_\E(\k,\eta)|^2\rangle\langle |Q(\k,\eta)|^2\rangle} }\ ,
\ee
where in this work $Q$ stands for any of the following: $\s$, $\v$, $\delta$, $\delta_s$, 
$k_{||}T^{(1)}_{||}$, $k_{||}^2T^{(2)}_{||}$; and the wavevector $\k$ above corresponds to $\q$, $\q$, $\x$, $\x_s$, $\x$, $\x$, respectively.

The plots that follow are done for two redshifts: $z=0$ and $z=1$. Unless stated otherwise, for each quantity $Q[\x,\v]$ we show $R_Q$ and $\rho_Q$ (eq's.~(\ref{Q}, \ref{rhoQ})) for four different approximations, $Q_\E$. Two of them correspond to the ZA and 2LPT results:
\be\label{QETZ}
Q_\E=Q_z\equiv Q\left[\q+\s_z,\v_z\right] \ \ \hbox{and} \ \ Q_\E=Q_{2LPT}\equiv Q\left[\q+\s_z+\s_{2LPT},\v_z+\v_{2LPT}\right]\ ,
\ee
 which also correspond to the model in TZ, where the particle positions and velocities were not corrected with transfer functions. The next two $Q_\E$'s are given by\footnote{A natural extension to our models is calculating the 2LPT terms using an optimally smoothed $\delta_L$. However, after applying such a smoothing, we did not find an appreciable improvement to the resulting non-linear $\delta$, which is why we stick to the simpler models above.}: 
 \be\label{QE2}
Q_\E&=& Q\left[\q+R_z*\s_z,R_z^v*\v_z\right]\ \ \hbox{and} \ \\\nonumber  Q_\E&=&Q\left[\q+R_z*\s_z+R_{2LPT}*\s_{2LPT},R_z^v*\v_z+R_{2LPT}^v*\v_{2LPT}\right]
\ ,
 \ee
 which correspond to the first and second order of the model of this paper, eq.~(\ref{splitD}, \ref{splitVel}, \ref{Q}).
The quality of any of the $Q_\E$'s at a given scale is measured by how close $\rho_Q(k,\eta)$ is to one.

It is important to stress that by analyzing all the approximations above, we do not aim to search for the best of them. It is by construction that the second approximation in eq.~(\ref{QE2}) must be the best. If any of the other approximations happen to be better for some of the quantities we analyze, that must be entirely for the wrong reasons. We will see examples of this as we proceed. Therefore, the comparison presented in the rest of this section should ultimately be treated in two ways: 1) As providing a better physical intuition of what effects are captured by which order in the LPT expansion (with or without transfer functions); 2) As giving us some idea of what improvements one may expect if one goes to higher order LPT.

\subsection{Displacement and velocity transfer functions}\label{sec:dv}
\begin{figure}[t!]
\centering
  $z=0$\hspace{0.5\textwidth} $z=1$
  \\
  \subfloat{\includegraphics[width=0.5\textwidth]{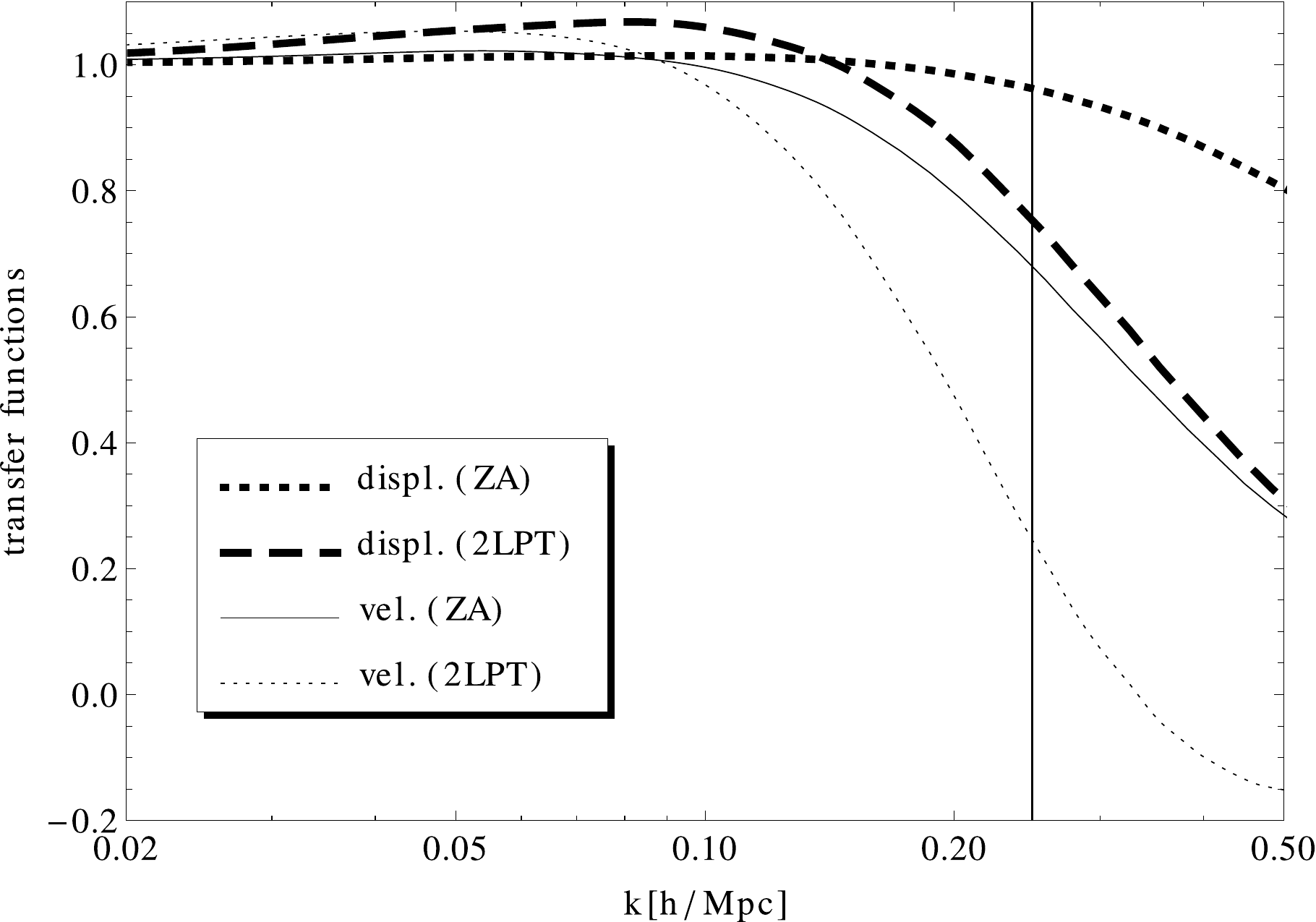}}
  \subfloat{\includegraphics[width=0.523\textwidth]{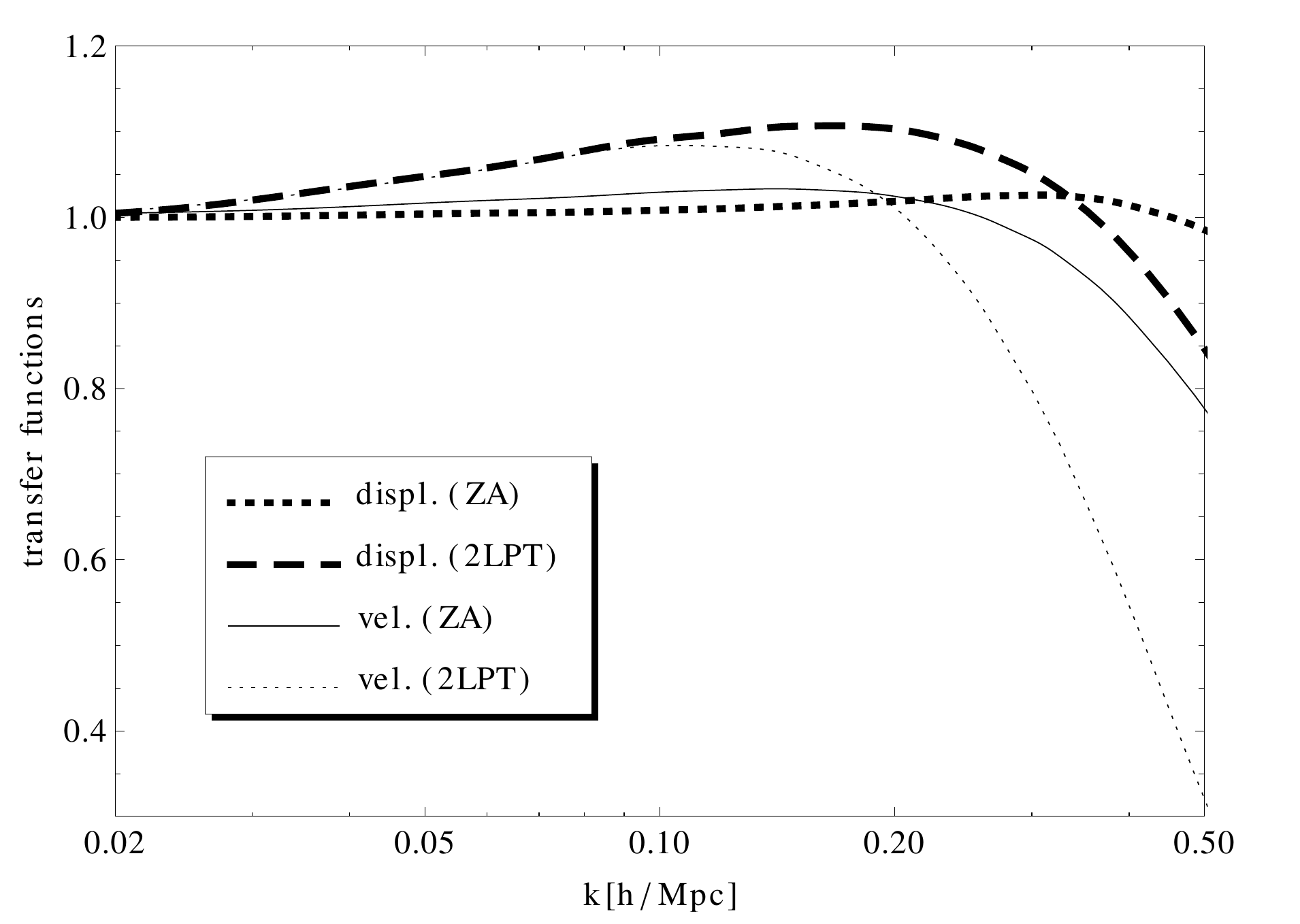}
\hfill
  }
  \\
    \subfloat{\includegraphics[width=0.5\textwidth]{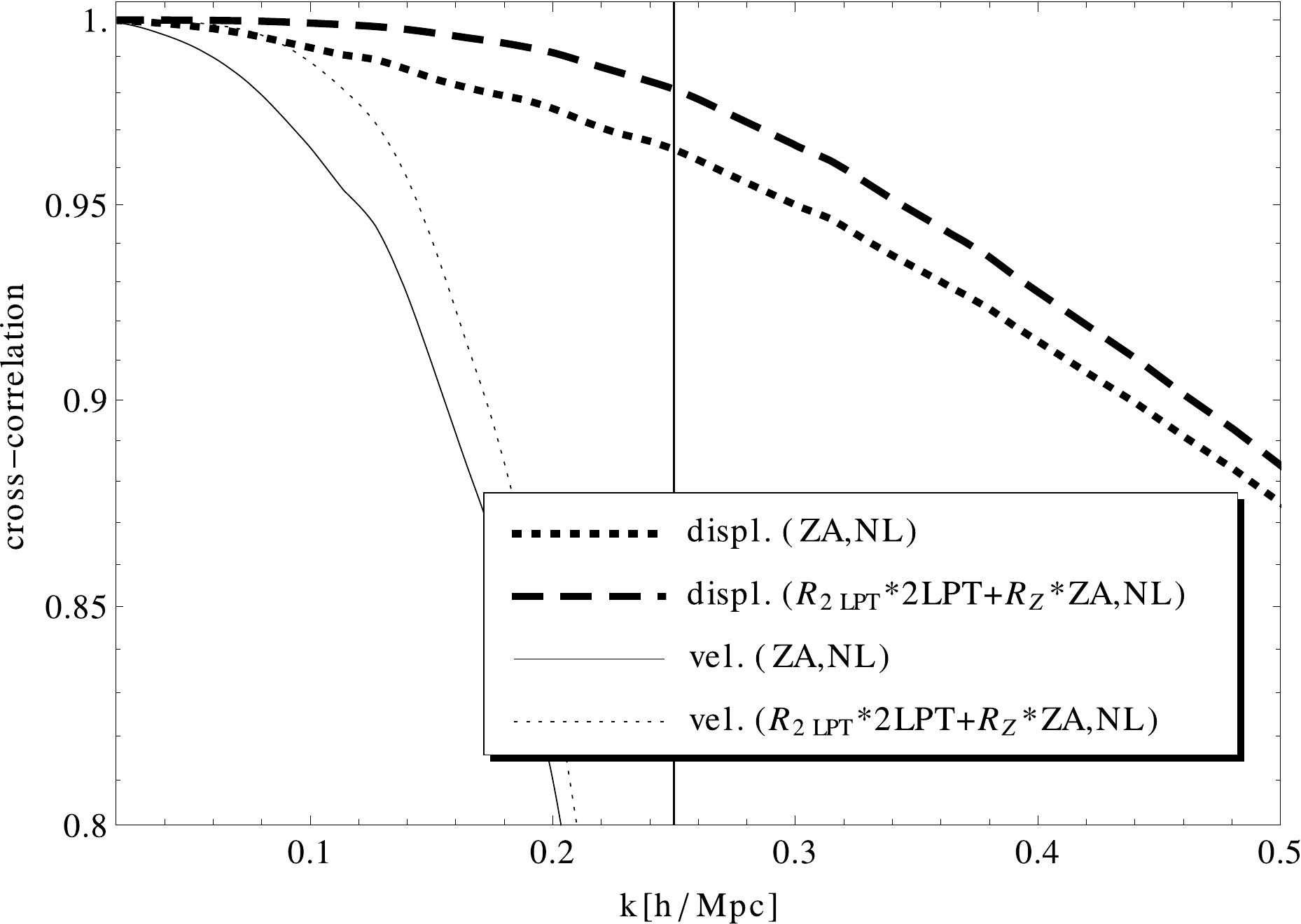}}
  \subfloat{\includegraphics[width=0.51\textwidth]{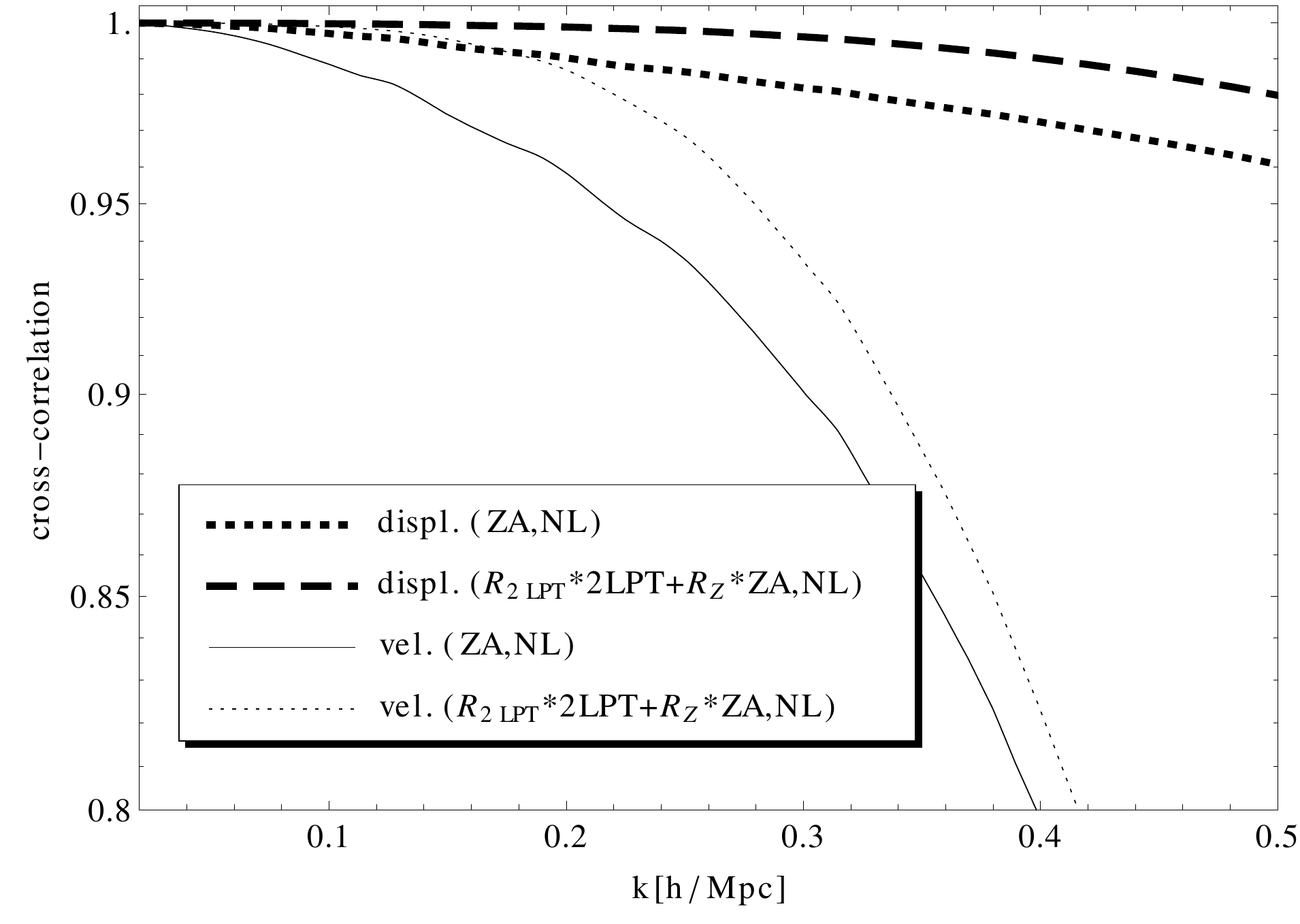}}
\hfill
  \\
 \subfloat{\includegraphics[width=0.5\textwidth]{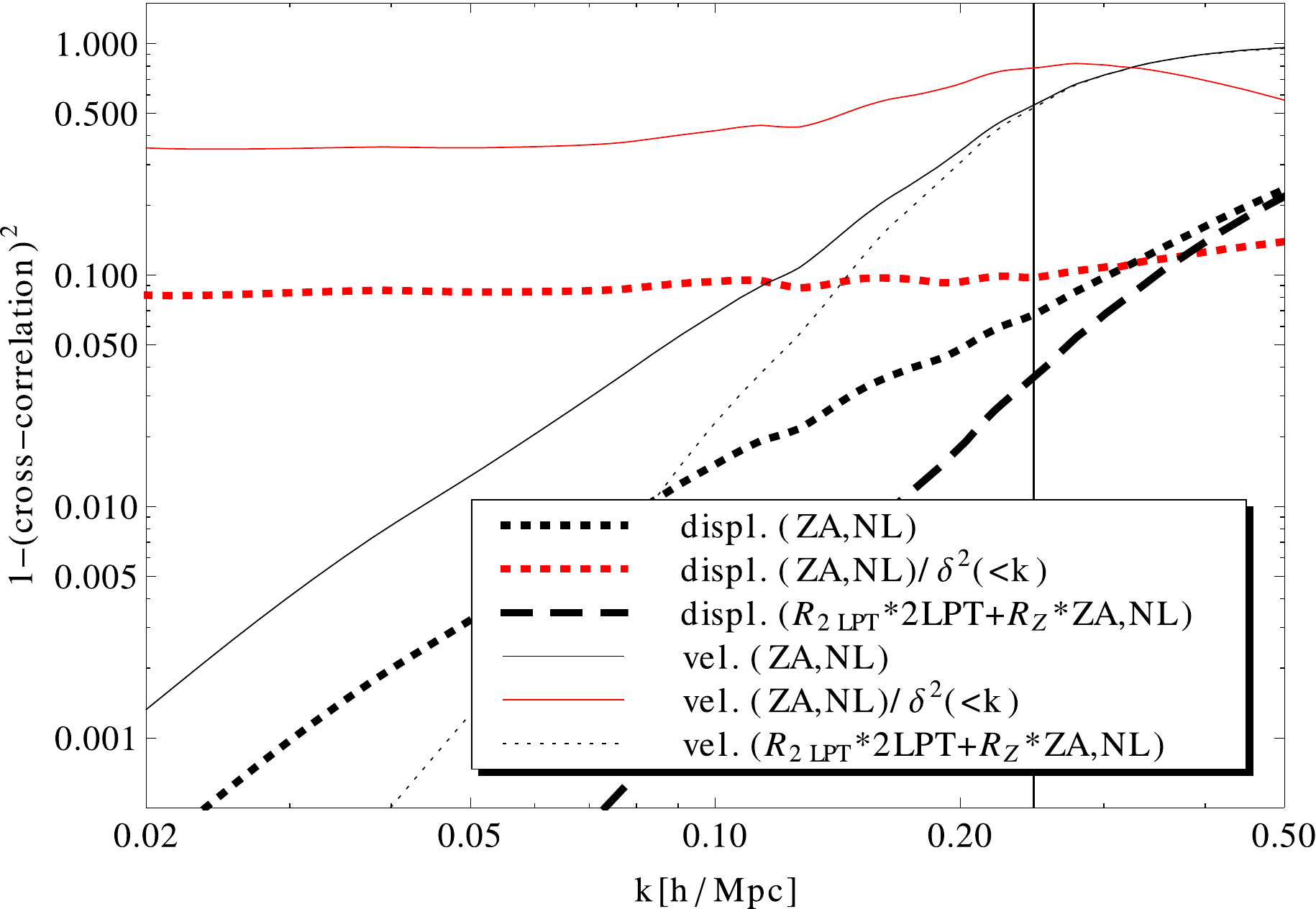}}
  \subfloat{\includegraphics[width=0.52\textwidth]{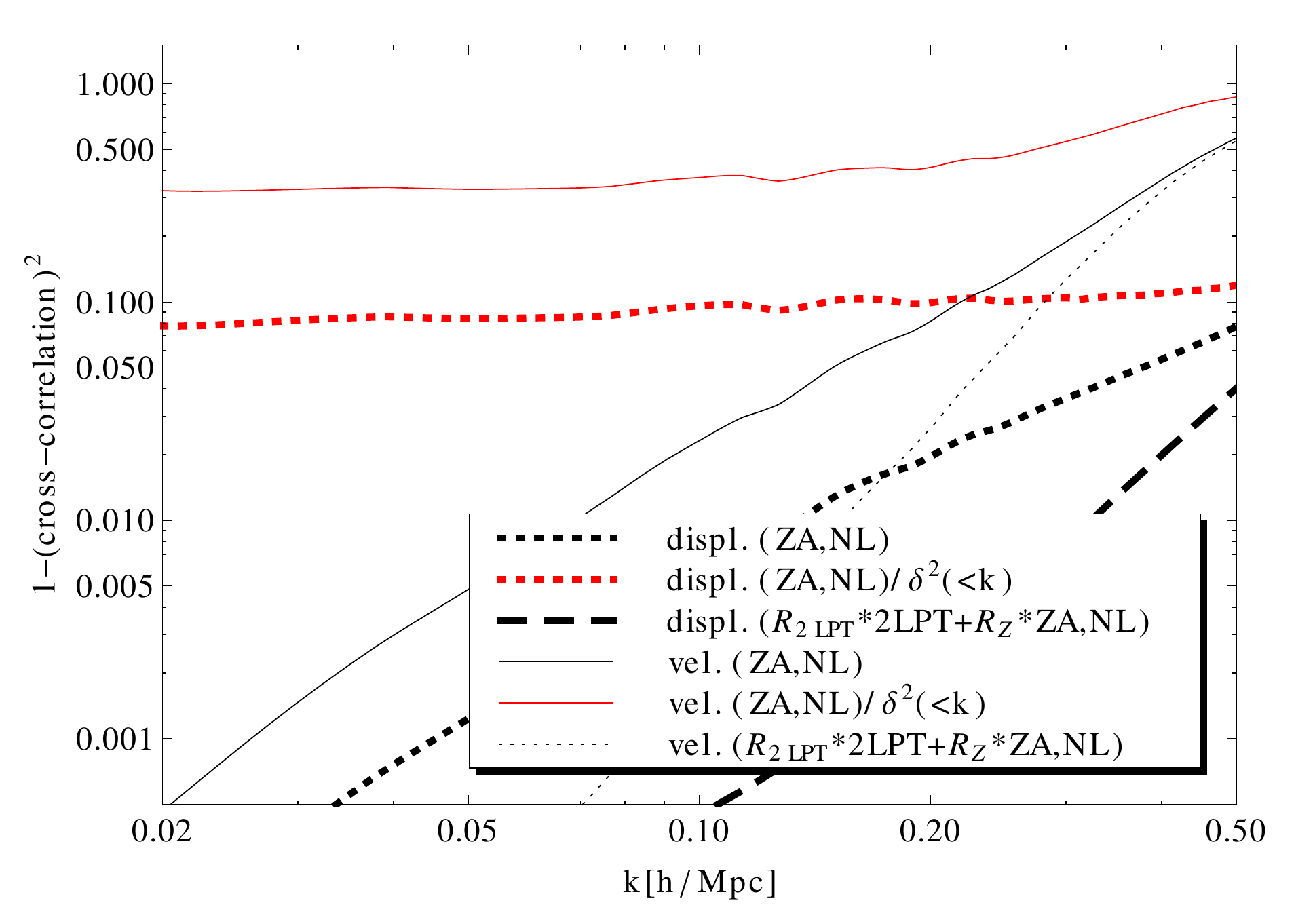}
  \hfill
  }
  \caption{Cross-correlation and transfer functions for the CDM displacements and velocities. In each plot we show the denoted quantities for the four different approximations given in eq.~(\ref{QETZ}, \ref{QE2}). The red curves show the corresponding cross-correlation function divided by $\delta^2(<k,\eta)$, to highlight the low-$k$ behavior. The non-linear scale (defined here as the linear power per logarithmic $k$-bin to be 1) is at $k_{NL}=0.25h/$Mpc for $z=0$ (denoted with vertical line) and $k_{NL}=0.74h/$Mpc for $z=1$.}
  \label{fig:displ}
\end{figure}

In the top row of Figure~\ref{fig:displ} we show the velocity and displacement transfer functions\footnote{Here we provide analytical fits (for the cosmology given in (\ref{cosmo})) for the two displacement transfer functions and the density transfer function for our best density approximation (second line of eq.~(\ref{QE2})):
\be
R_z(k,\eta)&=& \exp\left(-0.085 \left(\frac{k}{k_{NL}(\eta)}\right)^2\right)\nonumber\\
R_{2LPT}(k,\eta)&=& \exp\left( 0.64 \left(\frac{k}{k_{NL}(\eta)}\right) - 1.7 \left(\frac{k}{k_{NL}(\eta)}\right)^2  + 0.623 \left(\frac{k}{k_{NL}(\eta)}\right)^3  - 
  0.078 \left(\frac{k}{k_{NL}(\eta)}\right)^4\right)\nonumber
  \ee
 \be\label{fits}
R_\delta (k,\eta) &=& \exp\left(0.58 d\right)\ \ ,\ \ \mathrm{with}\ \ 
d\equiv  \delta^2\left(<\frac{k/2}{ k_{NL}(\eta)}k_{NL}(\eta_0),\eta_0\right)\ ,
\ee
with $\eta_0$ evaluated at $z=0$, and $k_{NL}$ and $\delta^2$ can be read off from eq.~(\ref{kNLeq}). 
These equations were checked for $z=0$ and $z=1$ and give a respective accuracy of within (2\%,4\%,1\%) at $z=0$ and (4\%,4\%,1\%) at $z=1$ in the range $0.02h$/Mpc$<k<0.5h/$Mpc.   The velocity transfer functions and the transfer function for the density in redshift space do not have a simple time dependence as above, and so we do not include fits for them.
} ($R_z$,  $R_z^v$, $R_{2LPT}$, $R_{2LPT}^v$); while the bottom two rows show the cross-correlation functions between $\s$ and its approximation $\s_\E$, as well as between $\v$ and $\v_\E$, for the four different approximations listed in eq.~(\ref{QETZ}, \ref{QE2}), in which one should substitute $Q$ with either $\s$ or $\v$.

Since at large scales the single-stream assumption behind LPT is not broken and the higher orders become negligible, all transfer functions and cross-correlations should go to 1 for $k/k_{NL}\to 0$, which they indeed do (Fig.~\ref{fig:displ}). As can be seen from the red curves in the plots at the bottom of Fig.~\ref{fig:displ}, at low $k$ the cross-correlation coefficients between the ZA and the NL displacements and velocities approach 1 as $\delta^2(<k)$ (defined below), which is the density variance integrated up to $k$. This is not surprising since the ZA in Lagrangian space misses second-order terms of order the large-scale density variance (see TZ). 

Including the 2LPT contributions clearly makes the low-$k$ power-law dependence much steeper, since 2LPT accounts for the terms in the transfer functions of order $\delta^2(<k)$, and misses only higher-order contributions. This result will appear again and again considering other quantities, such as the CDM density, as well as the momentum density (the kinetic energy density is a special case, which we will consider in detail below).

\begin{figure}[t!]
  \centering
 \subfloat{\includegraphics[width=0.5\textwidth]{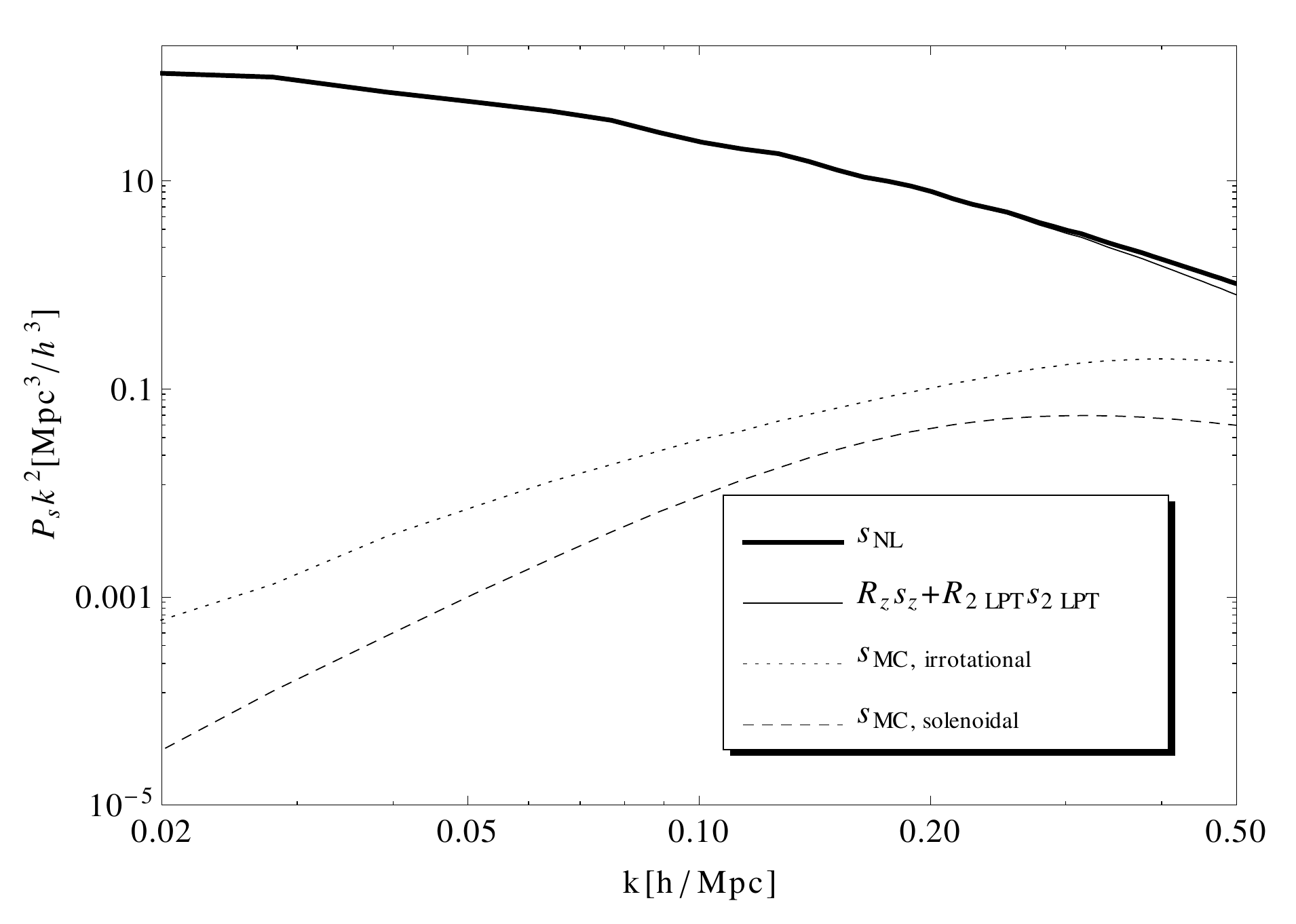}}                
  \subfloat{\includegraphics[width=0.5\textwidth]{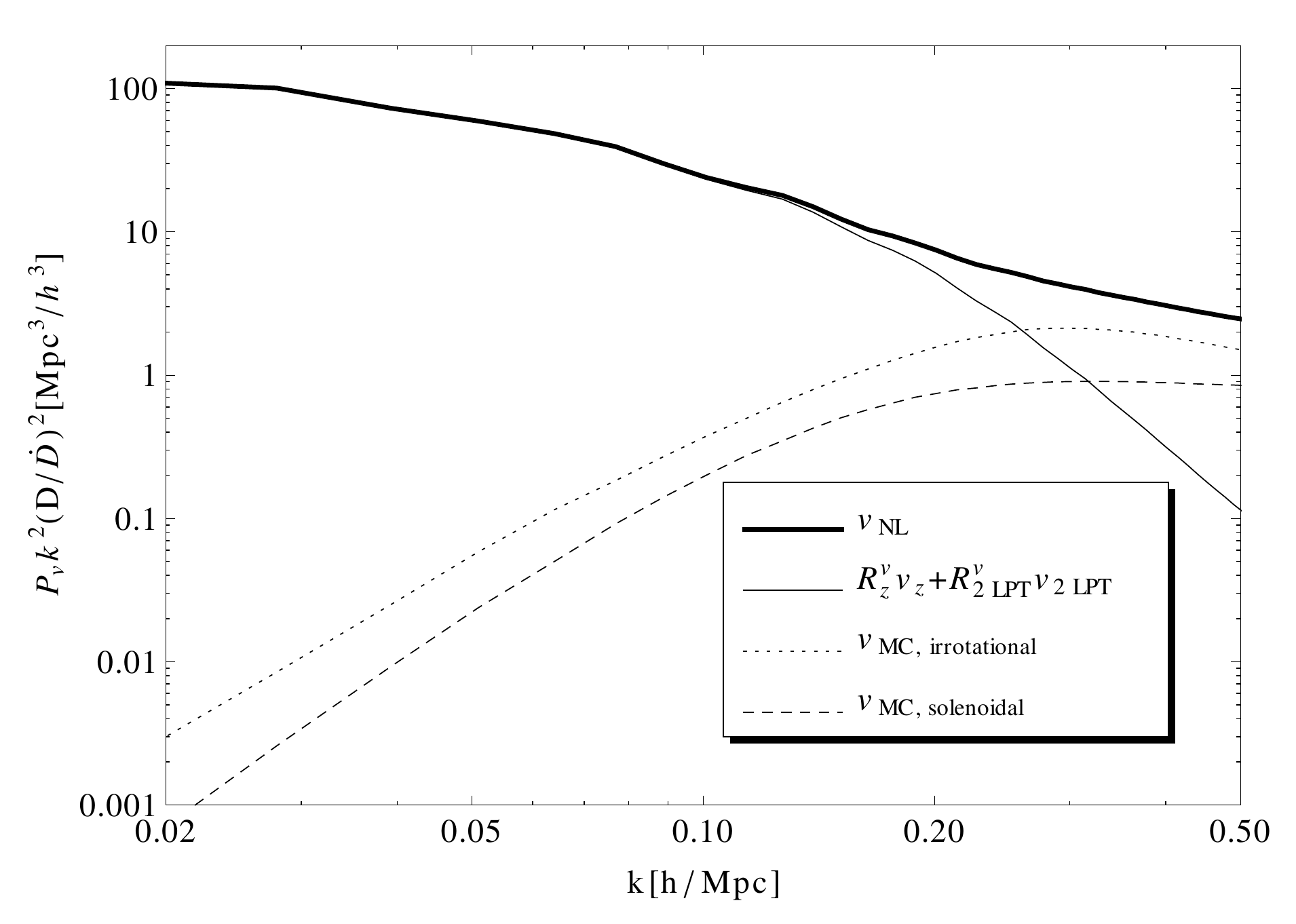}}    
  \caption{In this figure we show the different contributions to the power spectrum of the displacements (left panel) and velocities (right panel). The total quantities are denoted with $NL$, and the mode-coupling vector fields are split (in Lagrangian space) into irrotational and solenoidal parts. The plot is for $z=0$; and here $k$ is the wave-vector corresponding to Lagrangian space.}
  \label{fig:MCs}
\end{figure}

The different contributions to the displacement and velocity fields for our best model (second line of eq.~\ref{QE2}) are shown in Fig.~\ref{fig:MCs}. The MC vector fields are split (in Lagrangian space) into irrotational and solenoidal parts. The solenoidal power is smaller than the irrotational power both for the velocity and the displacements, since virialization is not complete at these scales. The mode-coupling power for the velocity becomes comparable to the approximated velocity, $v_\E$, at the non-linear scale ($k_{NL}=0.25h/$Mpc), while this is not the case for the displacement vector field. This can easily be explained by looking at Fig.~\ref{fig:displ} (see also Section~\ref{sssec:why} below): At high $k>k_{NL}$ the LPT and NL quantities decorrelate since LPT breaks down after shell-crossing. The velocity cross-correlation and transfer functions decay faster than the corresponding quantities for the displacements, resulting in correspondingly larger velocity MC power. This result will be addressed next by constructing a simple model for the time dependence of $R_z^v$.

\subsubsection{Why are the displacements better reconstructed than the velocities?}\label{sssec:why}

In this section we offer a toy model relating the velocity and displacement transfer functions in the ZA. It should help us understand better the decay of the velocity transfer function, relative to the higher-$k$ decay of the displacement transfer function, as observed in Fig.~\ref{fig:displ}. 

\begin{figure}[h!]
  \centering
  \includegraphics[width=0.7\textwidth]{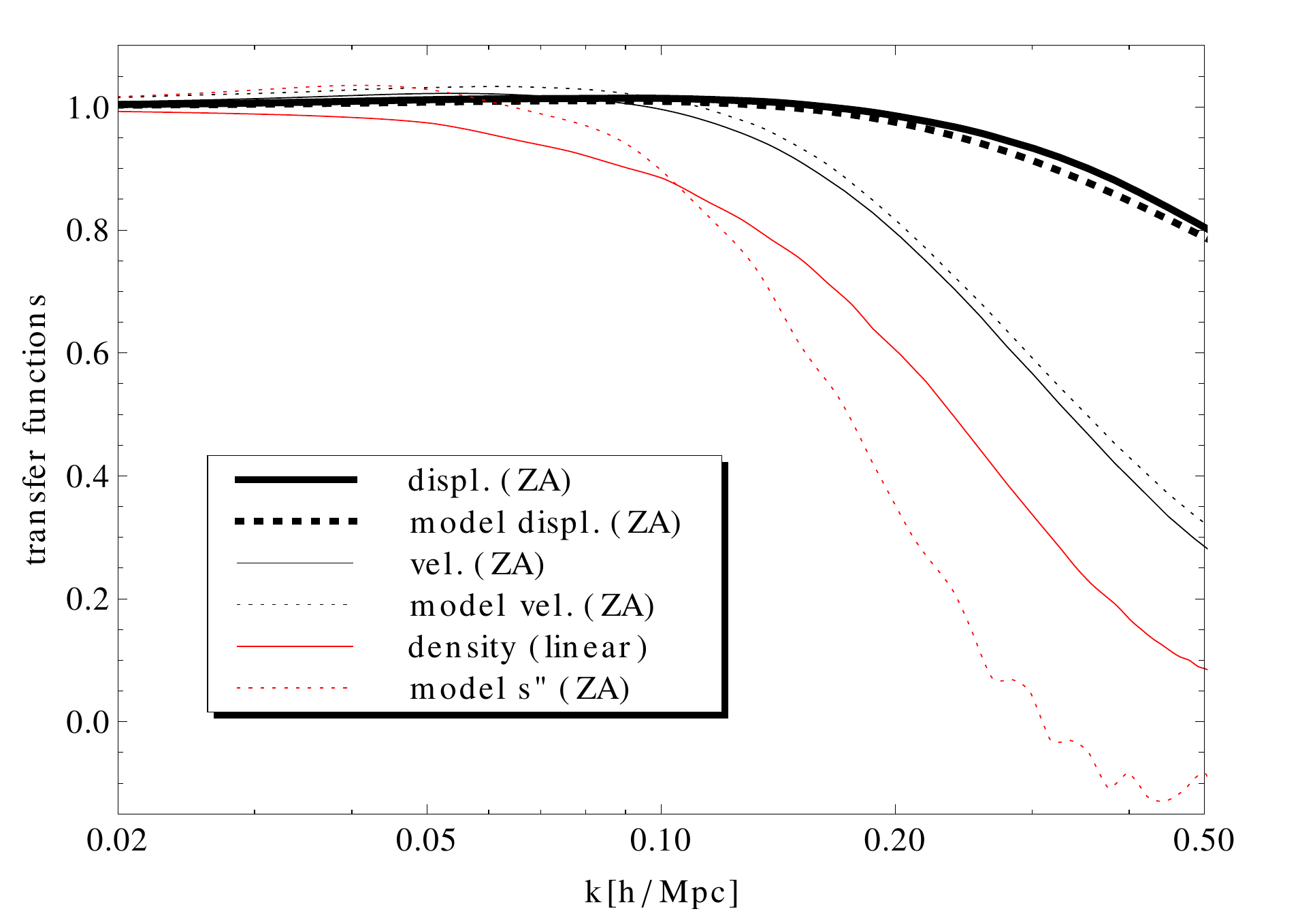}
  \caption{Toy model for the ZA displacement and velocity transfer functions evaluated at $z=0$.}
  \label{fig:toy}
\end{figure}

Starting from 
\be
\dot R_z=\partial_\eta\frac{\langle \s_z(\eta) \cdot \s(\eta)\rangle}{\langle| s_z(\eta)|^2\rangle}=\partial_\eta\left[\frac{1}{D(\eta)}\frac{\langle \s_z(\eta_0) \cdot \s(\eta)\rangle}{\langle| s_z(\eta_0)|^2\rangle}\right]\ ,
\ee
and using $\dot \s_z(\eta)=\dot D\s_z(\eta_0)$, we can write:
\be\label{rz}
\dot R_z=\frac{\dot D}{D}\left(R_z^v-R_z\right), \ \hbox{with}\ \ 
R_z(k,\eta= 0)=1\ .
\ee
Since $k_{NL}$ is the relevant scale of the problem, we can write the time dependence for $R_z^v(k,\eta)$ as follows (see eq.~(\ref{fits}) as well):
\be\label{rzv}
R_z^v(k,\eta)\approx R_z^v\left(k\frac{k_{NL}(\eta_0)}{k_{NL}(\eta)},\eta_0\right)\ ,
\ee 
where $k_{NL}(\eta)$ is the solution to\footnote{Elsewhere we quote $k_{NL}$ as the scale where the linear power per logarithmic $k$-bin becomes 1. However, we find that (\ref{kNLeq}) gives better agreement in eq.~(\ref{rzv}).}
\be\label{kNLeq}
\delta^2(<k_{NL},\eta)\equiv 4\pi\int_0^{k_{NL}(\eta)} dk k^2 P_L(k,\eta_0)D^2(\eta)=1\ ,
\ee
and $P_L$ is the linear power spectrum, defined as $\langle \delta_L(\k,\eta)
\delta_L(\tilde\k,\eta)\rangle\equiv \d(\k+\tilde\k)P_L(\k,\eta)$.

As a check of the proposed time dependence of $R_z^v$, in Figure~\ref{fig:toy} we plot $R_z^v$  for $z=0$ as obtained in two ways: directly from the N-body simulations (line denoted ``vel. (ZA)''); as well as from (\ref{rzv})  evaluated with $\eta_0$ corresponding to $z=1$, and $\eta$ to $z=0$  (line denoted ``model vel. (ZA)''). We can see that there is a very good agreement between the two curves, which indicates that we capture most of the time- and scale-dependence of $R^v_z$ with (\ref{rzv}).  

Using (\ref{rzv}), we can then integrate (\ref{rz}) to $z=0$ starting from the initial conditions at $\eta=0$, eq.~(\ref{rz}). Thus, we obtain an $R_z$ at $z=0$, which we plot in Figure~\ref{fig:toy} (line denoted ``model displ. (ZA)''). This should be compared with the line denoted ``displ. (ZA)'', showing the true $R_z$. Again we see a very good match between the two. So, we find that the time integral relating the velocity and the displacement moves the decay in the transfer function to higher $k$. This can be expected, since displacements are more insensitive to late time non-linearities than the velocity. To see this, one simply has to consider a virialized halo, where the velocities (and accelerations) can grow arbitrarily large, while the NL contribution to the displacement is bound to the size of the halo and is typically much smaller than the large-scale (linear) displacements.\footnote{
The reader may be interested in the transfer function for the accelerations. So, let us look into it in some detail. Above we obtained the time dependence of $R_z$ to a good approximation by integrating (\ref{rz}). One can then easily calculate the transfer function, $R_{s_z''}$, relating the accelerations $\s_z''$ and $\s''$. Using the properties of the Zel'dovich displacement as we did above, we can write $R_{s_z''}=(DR_z)''/D''$. The result is the curve denoted ``model $s''$ (ZA)'' in Figure~\ref{fig:toy}. 
One then may be tempted to use the equation of motion (\ref{S}) and write $R_{s_z''}\approx R_{L}$,  where $R_L$ is the transfer function relating the linear and NL overdensities. This relation comes about because naively the proportionality ($k$- and time-dependent) constant between $\s''$ and $\delta$ cancels out when calculating those transfer functions. 

However, this guess is not correct, since the $k$ in $R_{s_z''}(k,\eta)$ corresponds to Lagrangian Fourier space, while the $k$ in $R_{L}(k,\eta)$ is in Eulerian Fourier space. Thus, the large-scale dependence of $R_{s_z''}(k,\eta)$ can be only on the overdensity. That can be seen from eq.~(\ref{S}), where adding an overall constant vector to $\s$ leaves the equation invariant. Therefore, large-scale bulk motions are irrelevant for $R_{s_z''}$. However, the large-scale dependence of $R_{L}(k,\eta)$ (shown with the curve ``density (linear)'' in Fig.~\ref{fig:toy}) depends on the large-scale flows as well, and thus on the large-scale velocity dispersion (see TZ). Therefore, it is not surprising that $R_{s_z''}$ and $R_{L}$ are qualitatively different -- the former exhibiting a steeper decay than the latter.
}

\begin{figure}
  \centering
  \subfloat[ZA]{\includegraphics[width=0.5\textwidth]{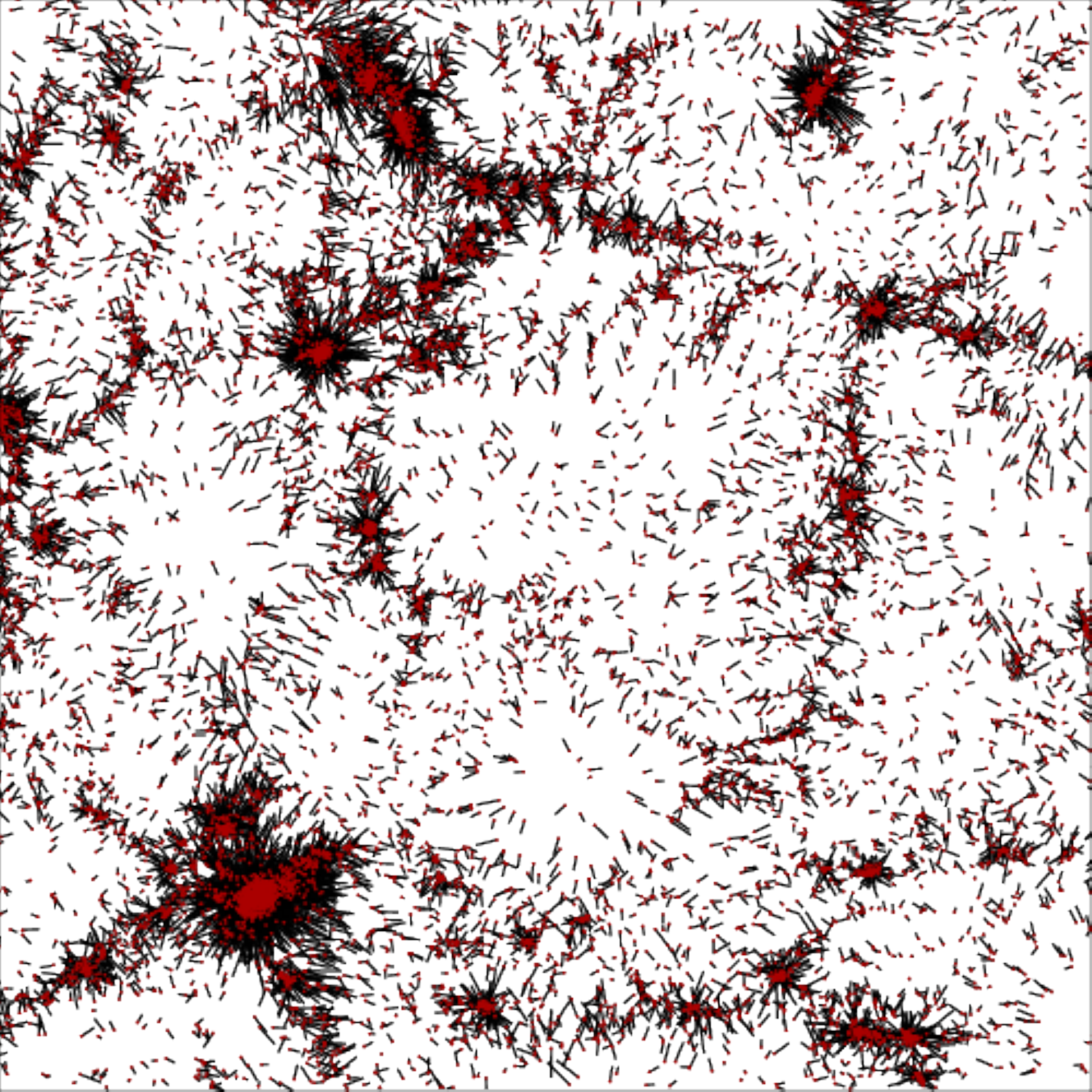}}                
  \subfloat[2LPT]{\includegraphics[width=0.5\textwidth]{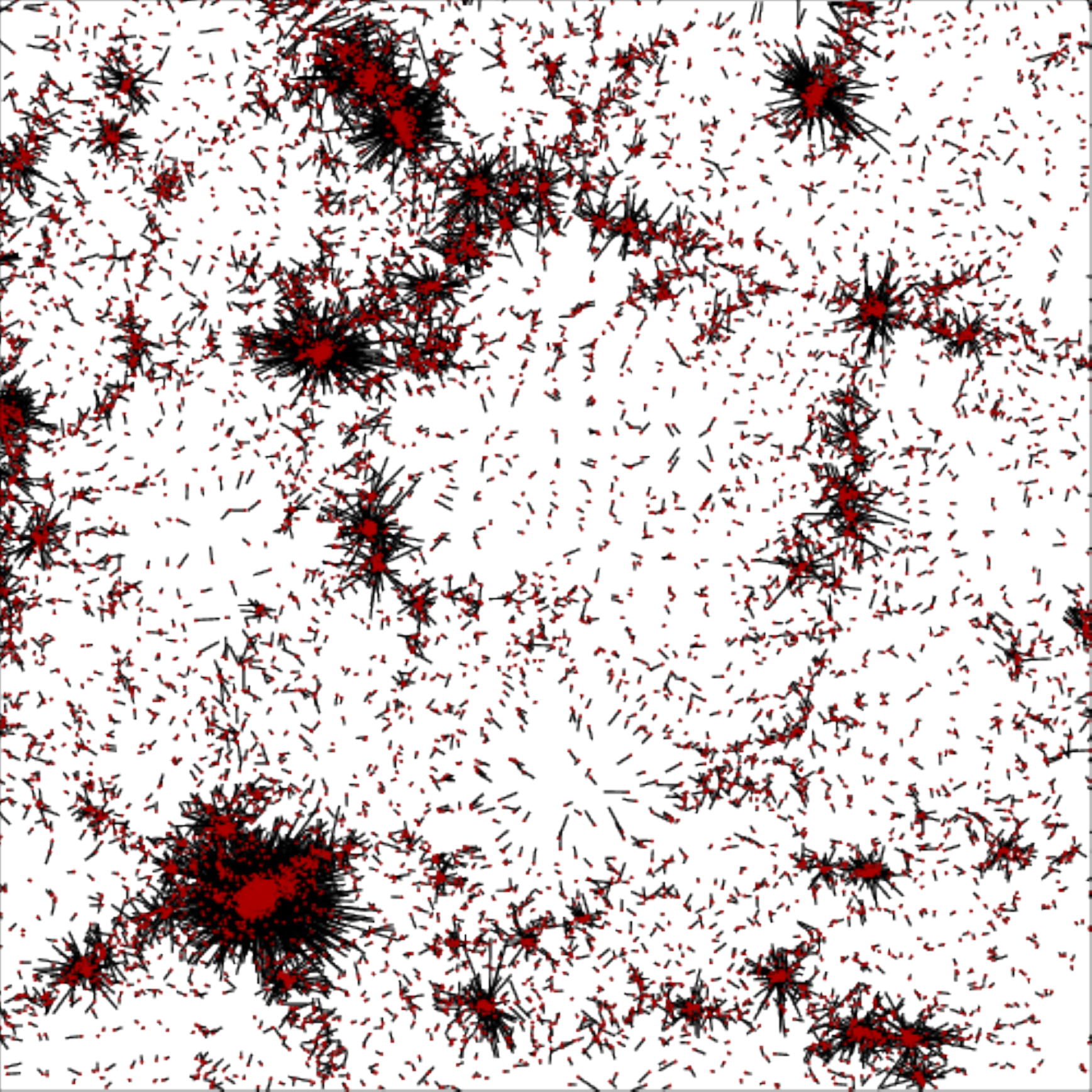}}
  \\
    \subfloat[ZA with transfer function]{\includegraphics[width=0.5\textwidth]{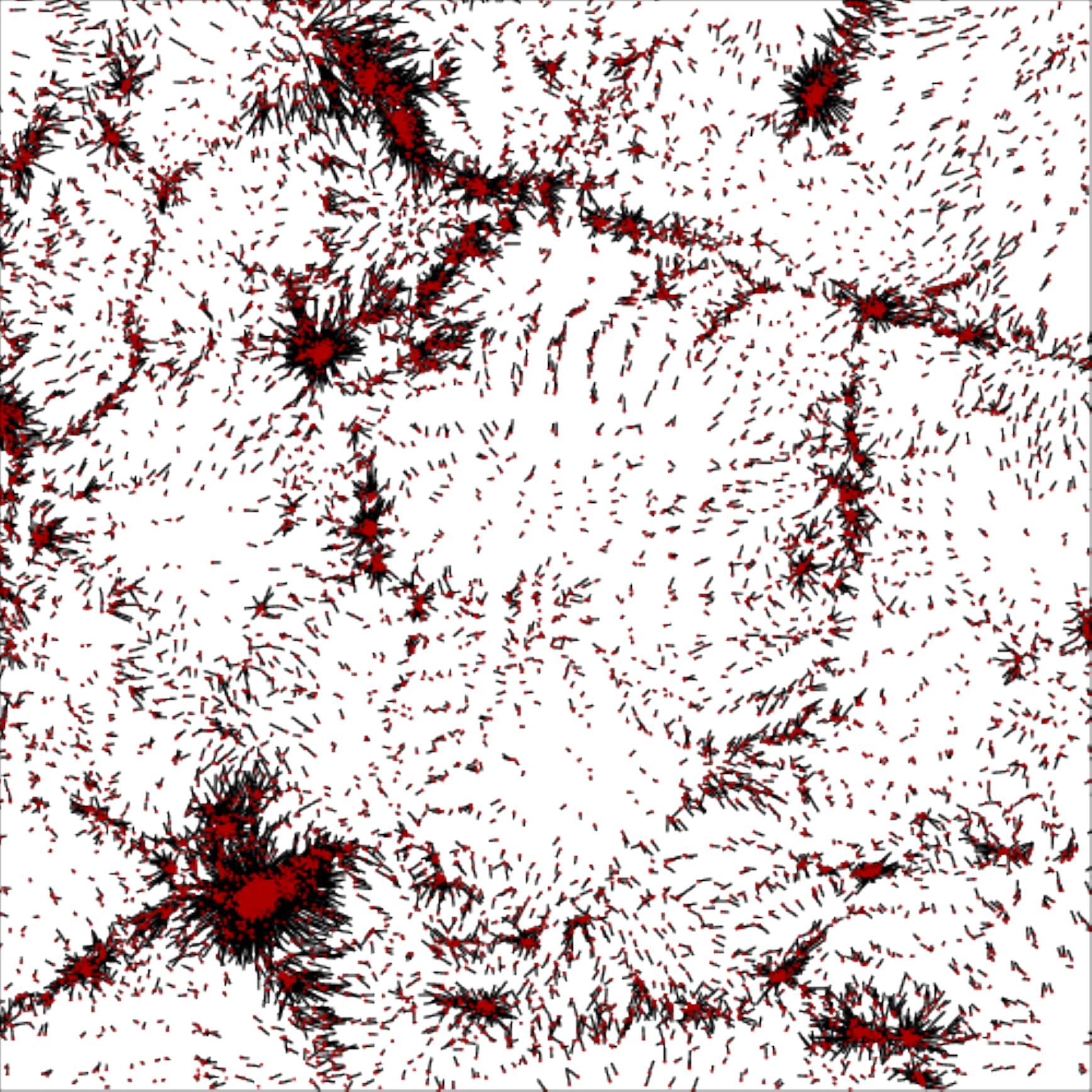}}                
        \subfloat[2LPT with transfer functions]{\includegraphics[width=0.5\textwidth]{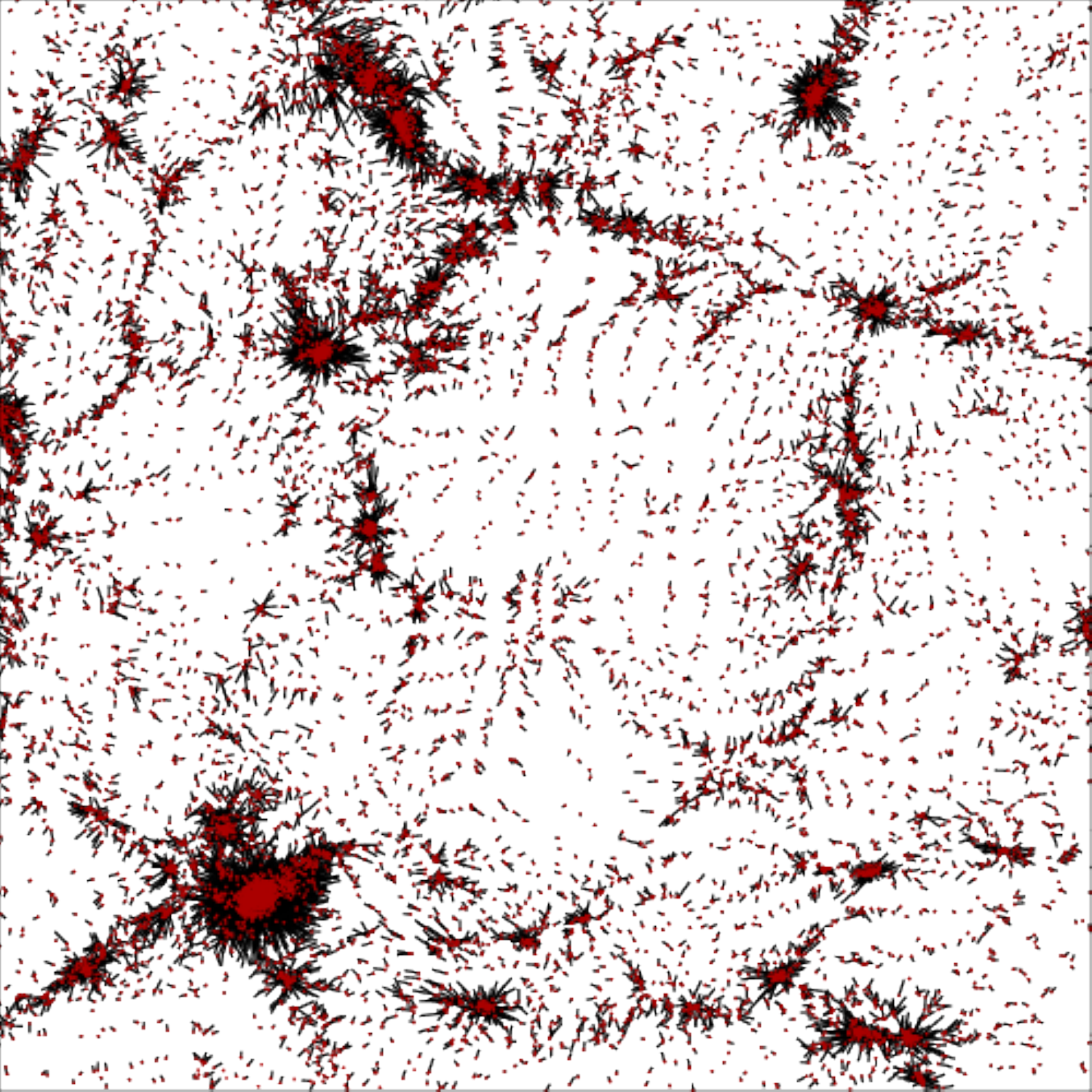}}                
  \caption{Slices in real space, 100$\,$Mpc$/h$ on the side ($z=0$).}
  \label{fig:slice}
\end{figure}

\subsubsection{Accuracy of the reconstructed particle positions.}

Let us now analyze the accuracy of the predicted positions and velocities, relative to the ones obtained by N-body simulations. To illustrate the effects of including the displacement and velocity transfer functions, in Figure \ref{fig:slice} we plot slices, $100\,$Mpc$/h$ on the side, $7.8\,$Mpc$/h$ thick, through one of our N-body simulations in real space. The red dots denote the ``true'' particle positions as calculated from the N-body code. Each red dot is connected with a black segment to the approximated particle position ($\x=\q+\s_\E$), obtained according to one of the four approximations given in eq.~(\ref{QETZ} and \ref{QE2}), corresponding to the four different panels in each figure.

One can confirm that even the ZA, the simplest approximation we use, reproduces the qualitative features of structure formation. We can also see that 2LPT over-corrects the particle positions in the voids (i.e. most black segments flip from one side to the other of the red dots when going from the upper left to the upper right panel of each figure), while in the overdense regions we see that halos become more diffuse in 2LPT. After the displacement  transfer functions are included (lower two panels), we can see that the voids are modeled almost perfectly at the resolution of the plots (as anticipated in the Introduction); and halos and filaments are modeled to a better accuracy than using pure LPT.

\begin{figure}
  \centering
  \subfloat{\includegraphics[width=0.32\textwidth]{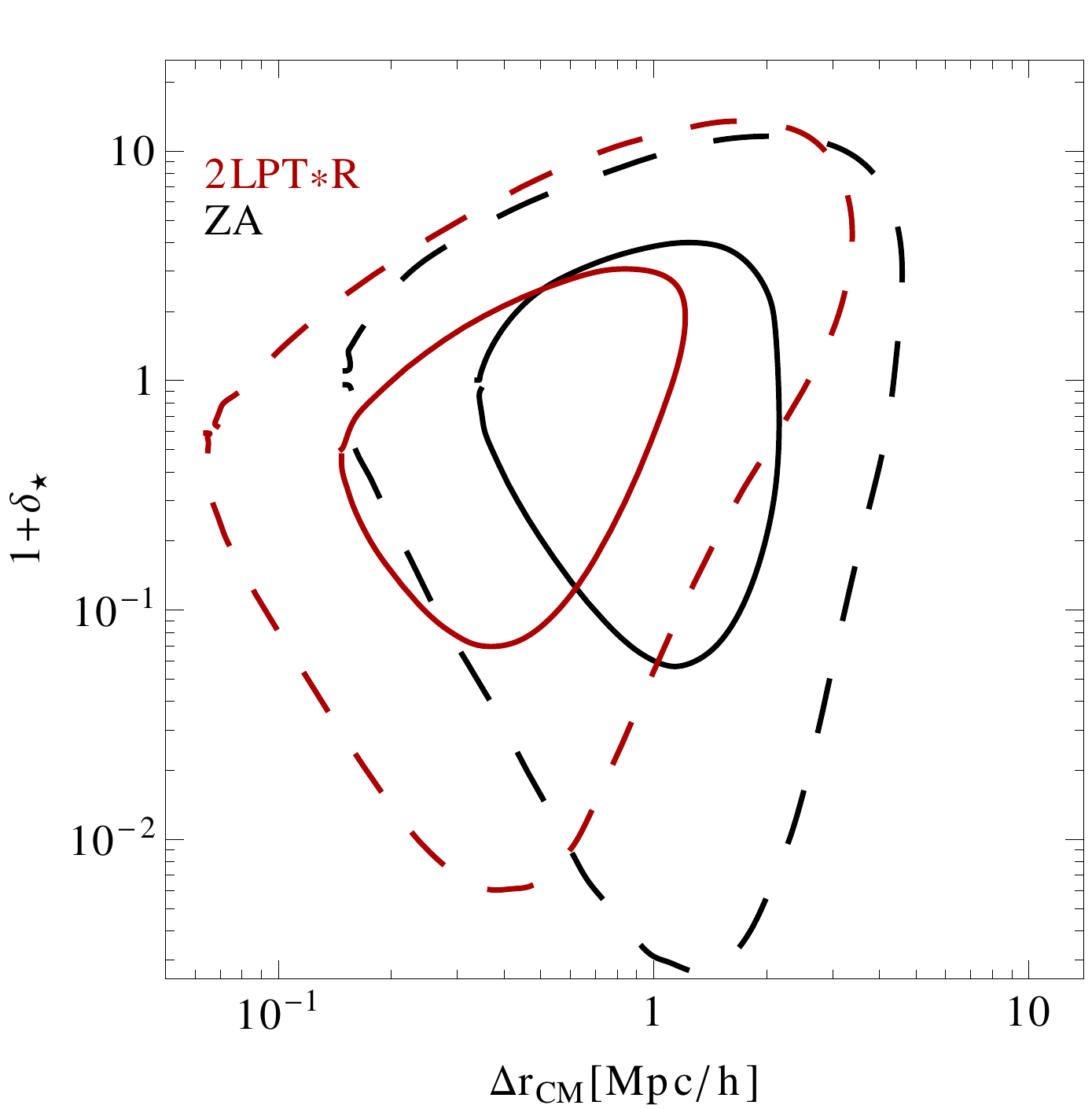}}  \hspace{0.1cm}  
  \subfloat{\includegraphics[width=0.32\textwidth]{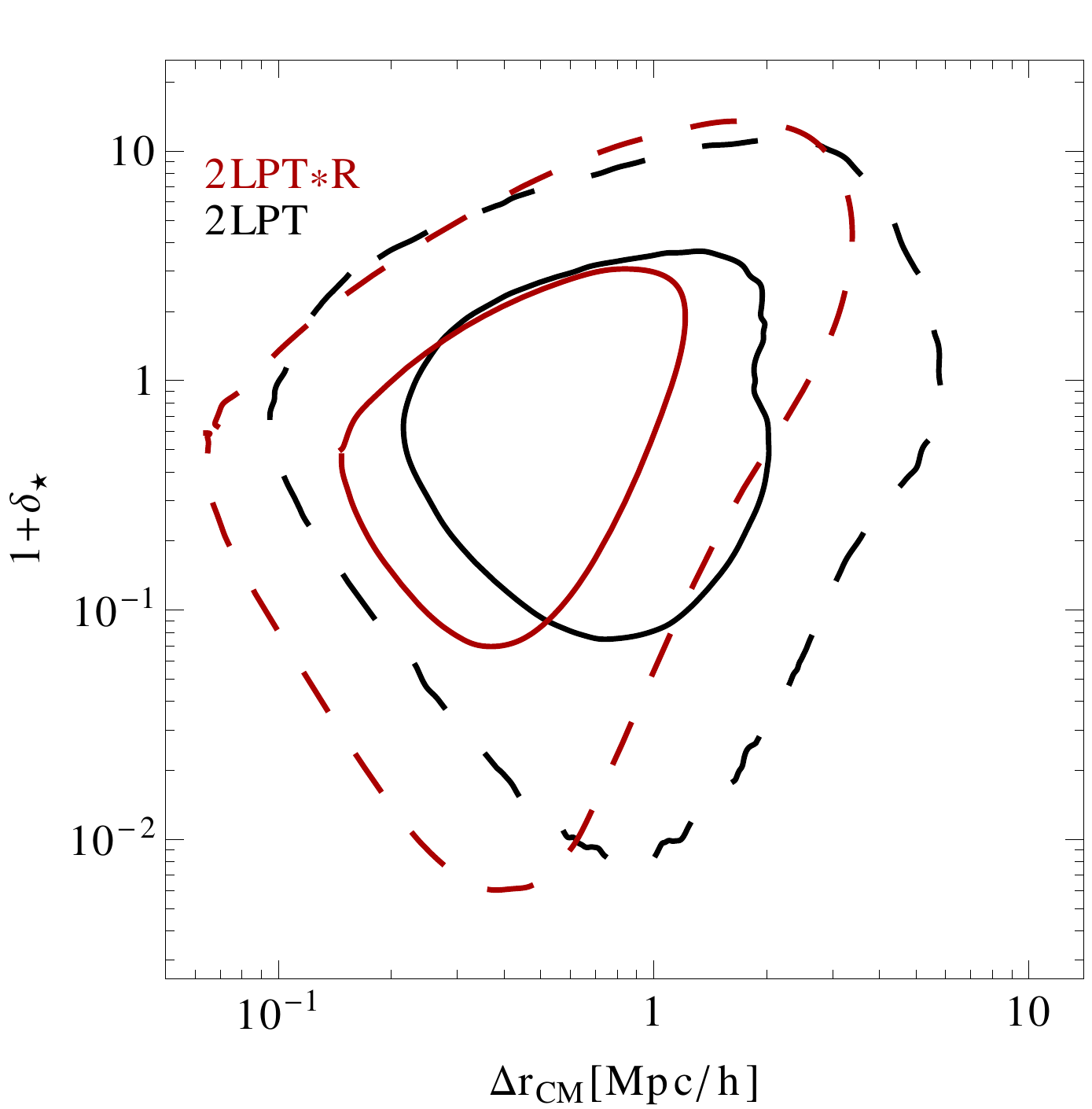}}\hspace{0.1cm}
    \subfloat{\includegraphics[width=0.32\textwidth]{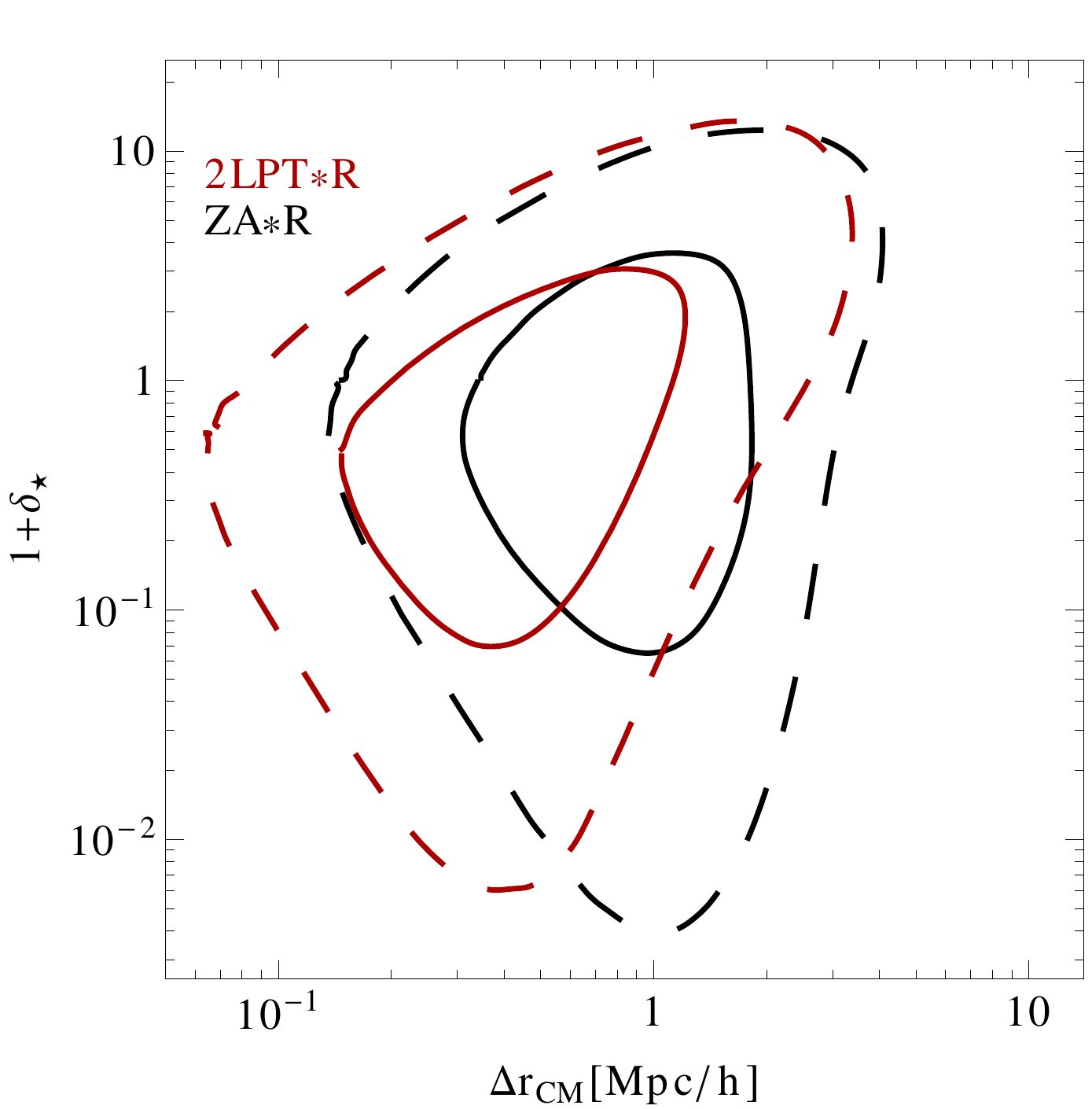}} \\
      \subfloat{\includegraphics[width=0.32\textwidth]{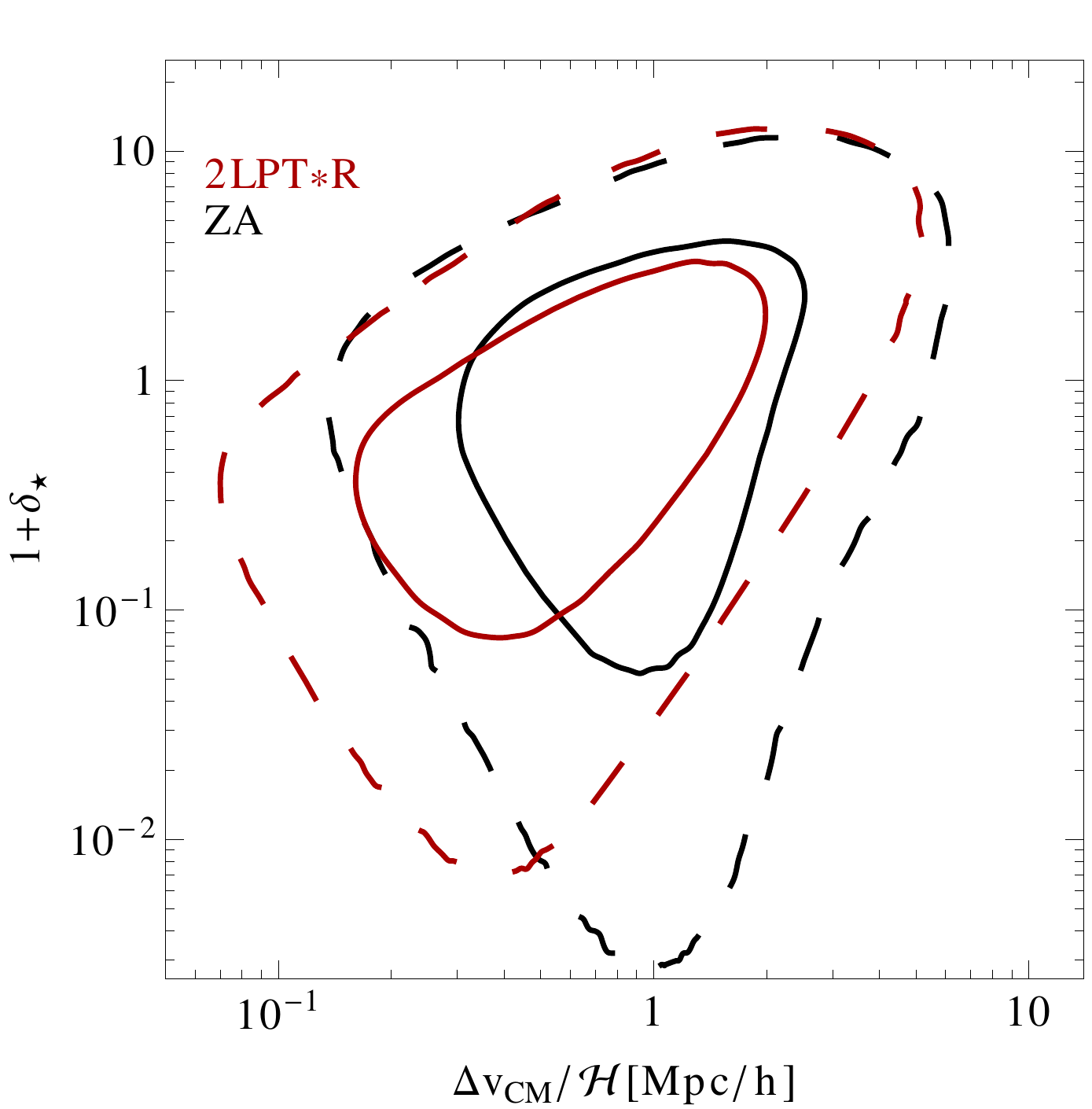}} \hspace{0.1cm}  
  \subfloat{\includegraphics[width=0.32\textwidth]{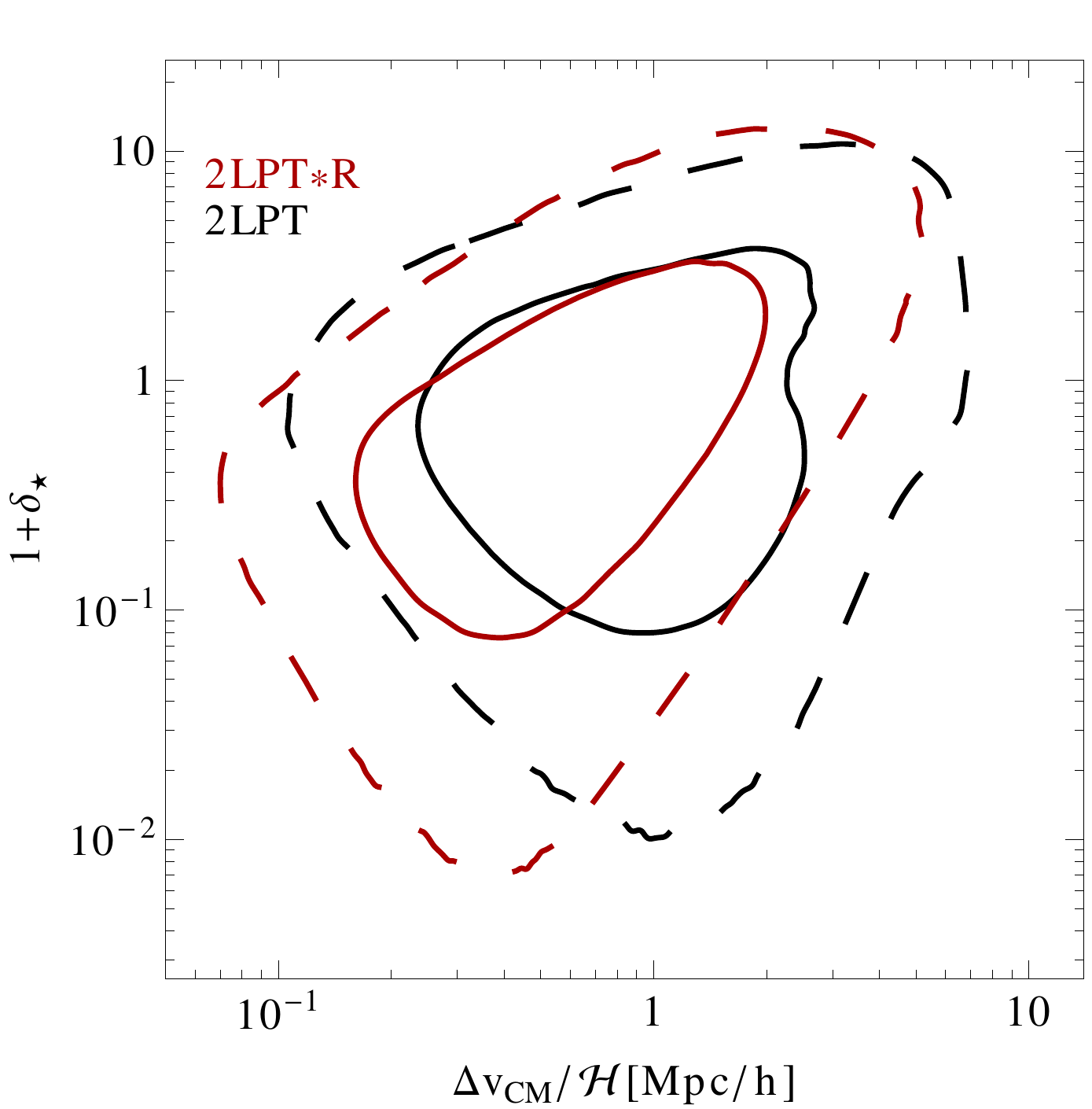}}\hspace{0.1cm}
    \subfloat{\includegraphics[width=0.32\textwidth]{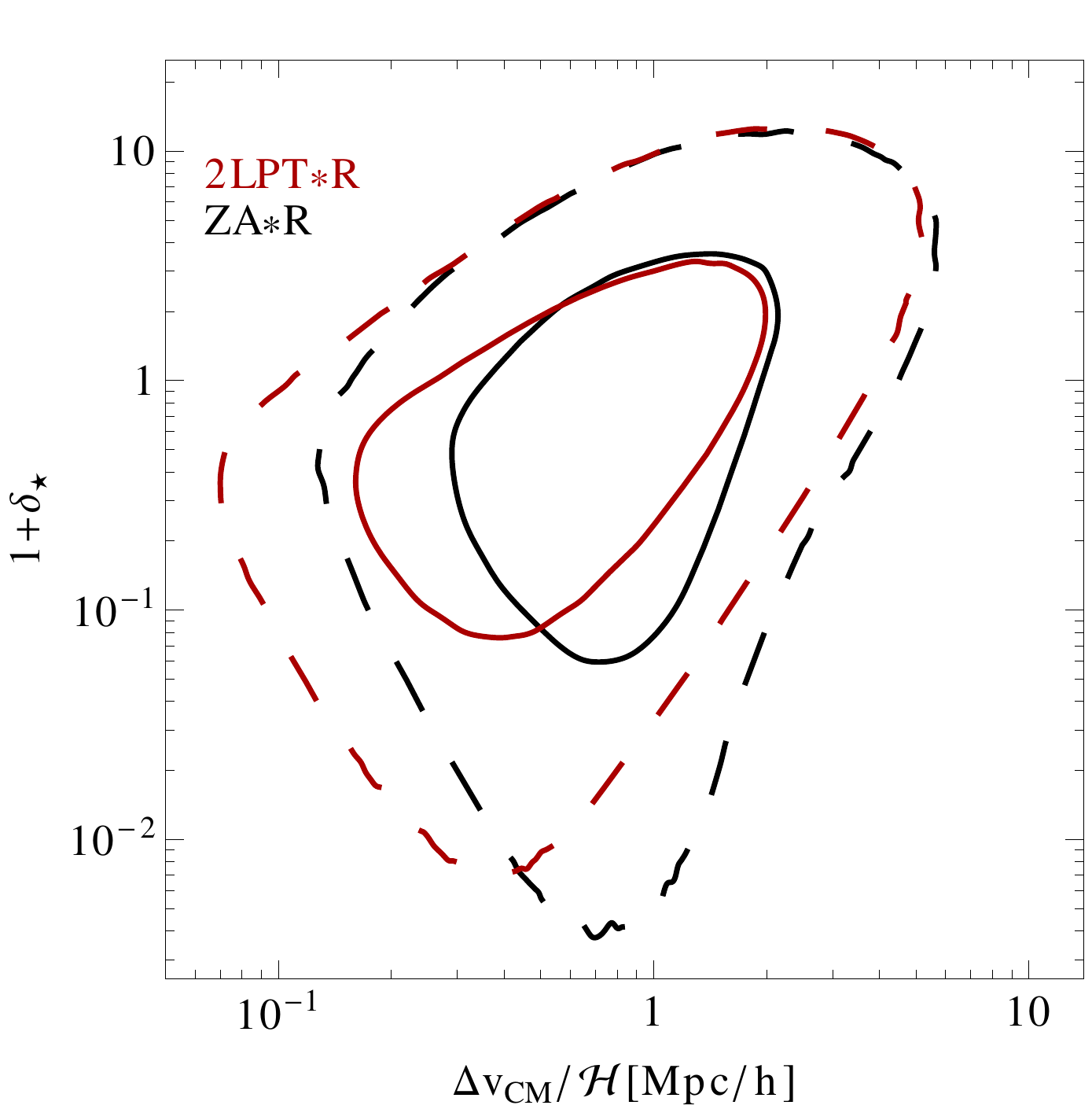}}                
     \\
      \subfloat{\includegraphics[width=0.32\textwidth]{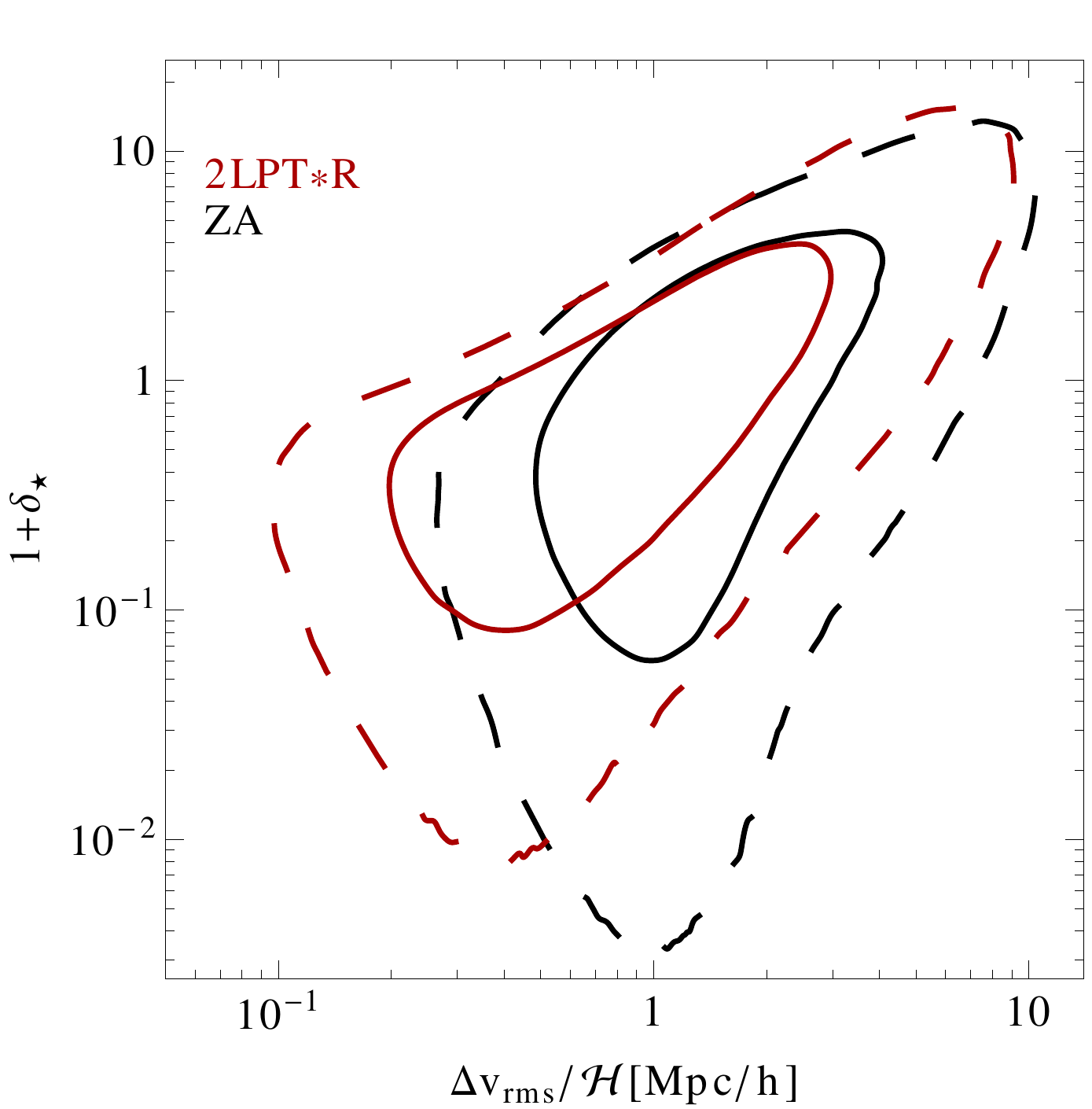}} \hspace{0.1cm}   
  \subfloat{\includegraphics[width=0.32\textwidth]{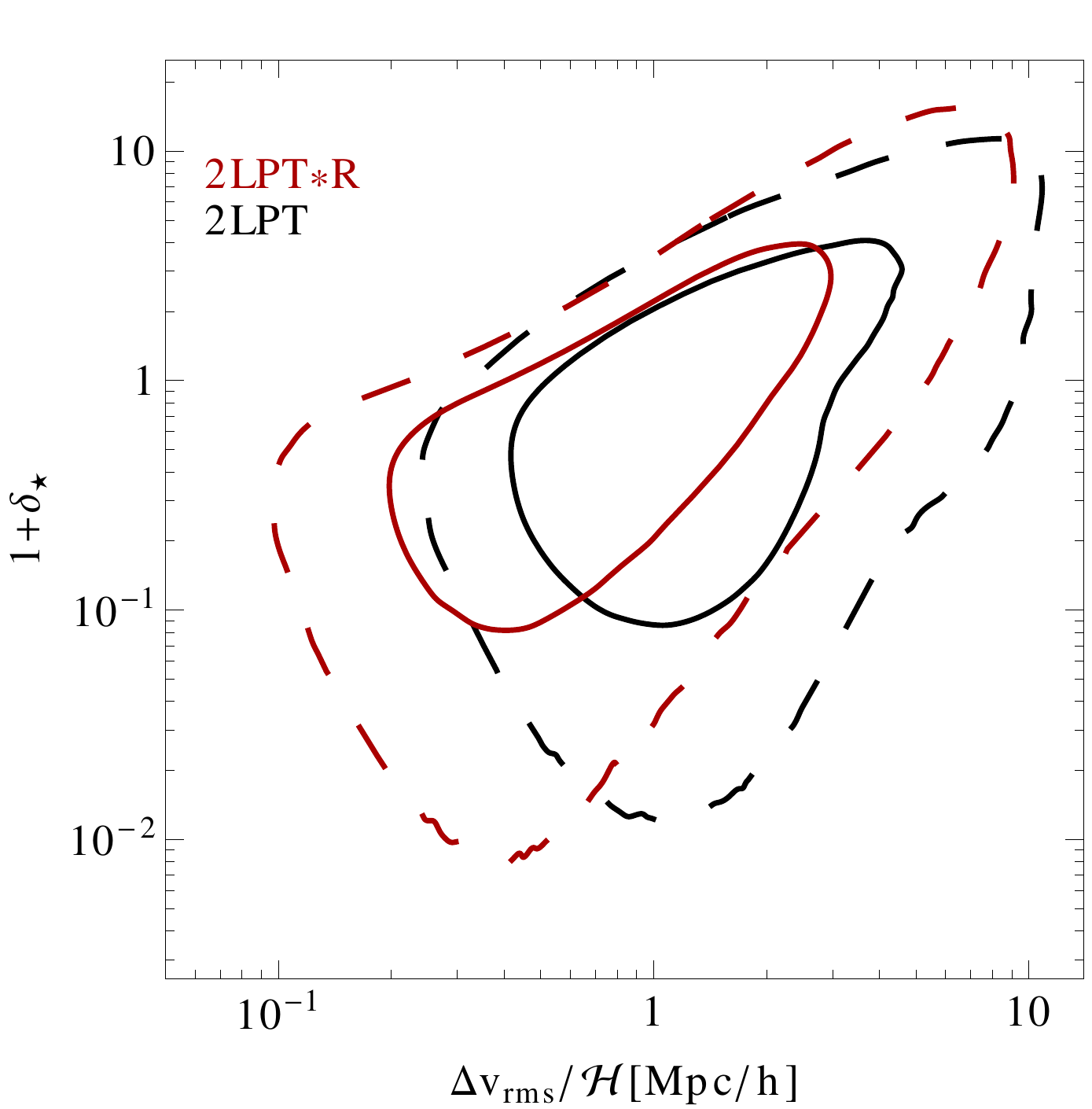}}\hspace{0.1cm}
    \subfloat{\includegraphics[width=0.32\textwidth]{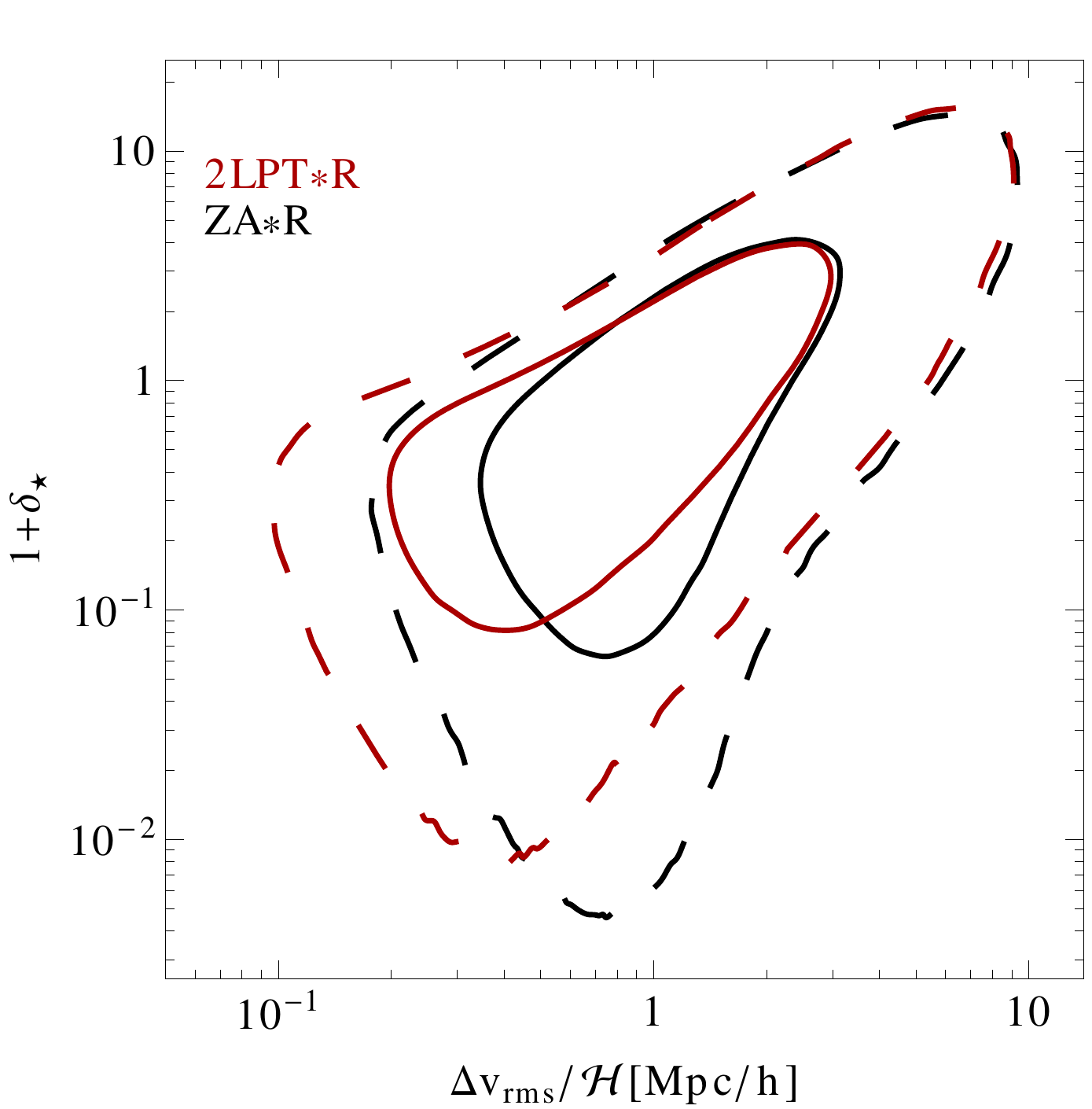}}                
  \caption{Probability density in the ($\log(1+\delta_\E)$,$\log(\Delta r_{\mathrm{CM}})$) plane (upper row), in the ($\log(1+\delta_\E)$,$\log(\Delta v_{\mathrm{CM}}/\mathcal{H})$) plane (middle row), and in the ($\log(1+\delta_\E)$,$\log(\Delta v_{\mathrm{rms}}/\mathcal{H})$) plane (bottom row). Here, $\delta_\E$ is the density in real space (obtained after Gaussian smoothing on a scale of $1.3\,$Mpc$/h$) as predicted by the approximation denoted in the plot. The residual displacement (not captured by our approximations) of the center of mass (CM) of a given cell in Eulerian space is given by $\Delta r_{\mathrm{CM}}$. The residual CM and rms velocities of that cell are given by $\Delta v_{\mathrm{CM}}$ and $\Delta v_{\mathrm{rms}}$, respectively. For each approximation we show the 1- and 2-sigma contours with solid and dashed lines, respectively. The plot is for $z=0$. For that redshift the rms linear displacement is $\approx15\,$Mpc/h, or about an order of magnitude larger than the residuals depicted in the plot. }
  \label{fig:p}
\end{figure}

To further quantify how well we reproduce the particle positions in real space, in the upper panels of Fig.~\ref{fig:p} we plot the probability density for finding a volume element in Eulerian space quantified by $(\log(1+\delta_\E)$,$\log(\Delta r_{\mathrm{CM}})$) at  $z=0$, where both quantities are calculated after Gaussian smoothing on a scale of $1.3\,$Mpc$/h$. Here $\Delta r_{\mathrm{CM}}$ is the error we make in predicting the center of mass (CM) position of that volume element by using the displacement approximations (given by eq.~(\ref{QETZ}, \ref{QE2})). In the middle row of Fig.~\ref{fig:p}, we show the same but for the probability density in the $(\log(1+\delta_\E)$,$\log(\Delta v_{\mathrm{CM}}/\mathcal{H})$) plane; while in the bottom row we show the probability density in the $(\log(1+\delta_\E)$,$\log(\Delta v_{\mathrm{rms}}/\mathcal{H})$) plane.
This figure then is the 2D analog of Fig.~\ref{fig:hist}, showing explicitly the density dependence.

One can see that the 2LPT approximation including the displacement transfer function is the best at predicting the CM position both in underdense and overdense regions, the former being markedly better reproduced (especially at the 95\% level) in agreement with our qualitative assessment from Figure~\ref{fig:slice}. The high-density region of this plot can be used as a guide for the accuracy of our displacement approximations to reproduce halo positions both in real and in redshift space. Thus, we find that for our choice of smoothing, the position of most (looking at the 2-sigma, or 95\% contour) overdense regions are found to better than $\sim$3Mpc$/h$. Concentrating on the 1-sigma contours in the upper middle panel, where we compare the 2LPT results with and without the transfer functions, we find an improvement in the approximated CM positions which is about a factor of 2 better (corresponding $\lesssim 1\,$Mpc$/h$) in real space. The FoG effects may introduce problems in redshift space as their effects are not captured as well: Looking at the bottom row, we can see that the errors in the rms velocity can be as large as 10$\,$Mpc/h (when mapped to real space)  in high density regions, which is comparable to the Doppler displacement from the velocity dispersion in galaxy clusters. However, the 1-sigma contour is again within a couple of Mpc, or a factor of a few smaller. 

\subsection{Density in real space}

\begin{figure}[t!]
  \centering
  $z=0$\hspace{0.5\textwidth} $z=1$
  \\
  \subfloat{\includegraphics[width=0.48\textwidth]{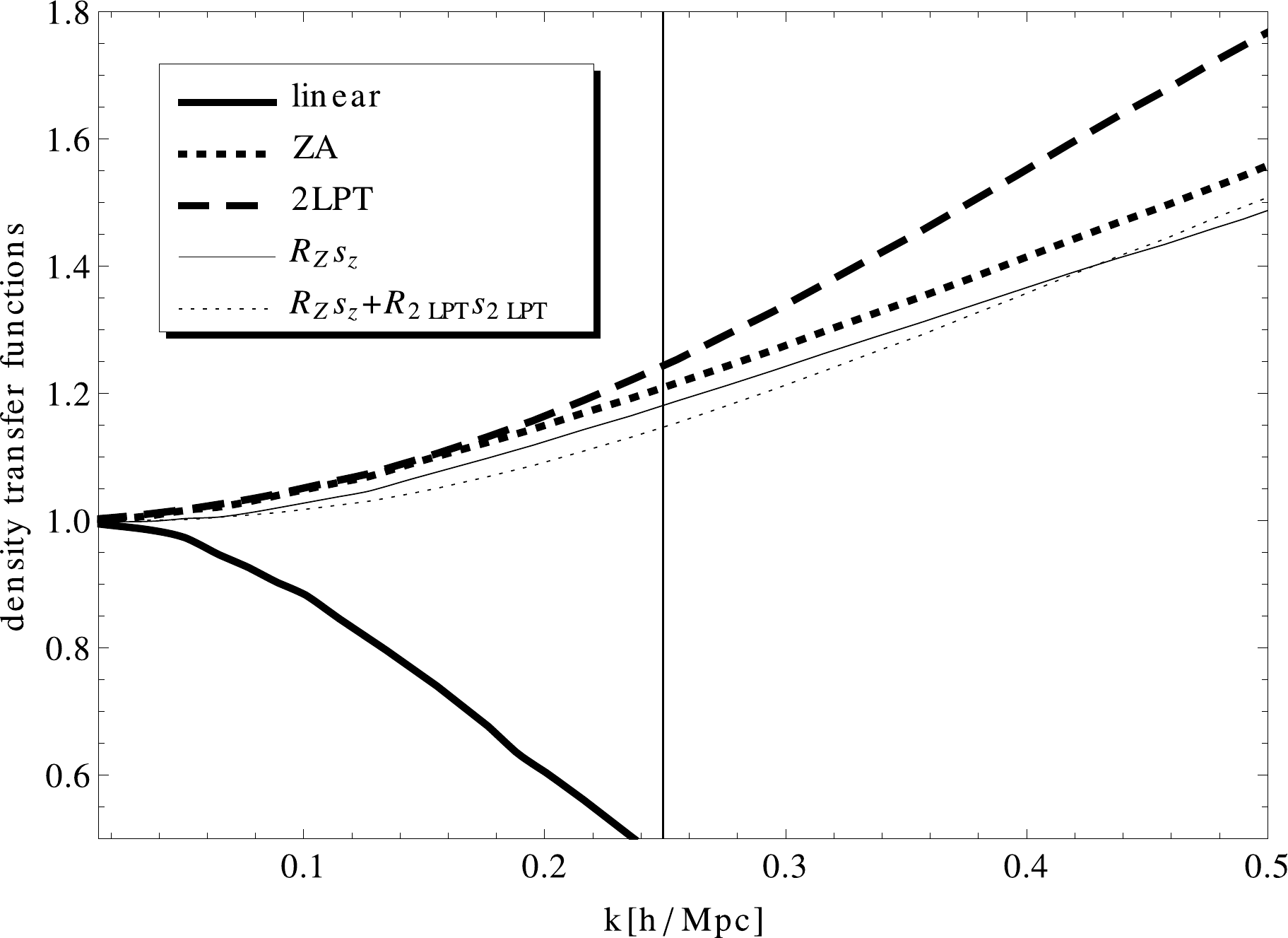}}
  \subfloat{\includegraphics[width=0.51\textwidth]{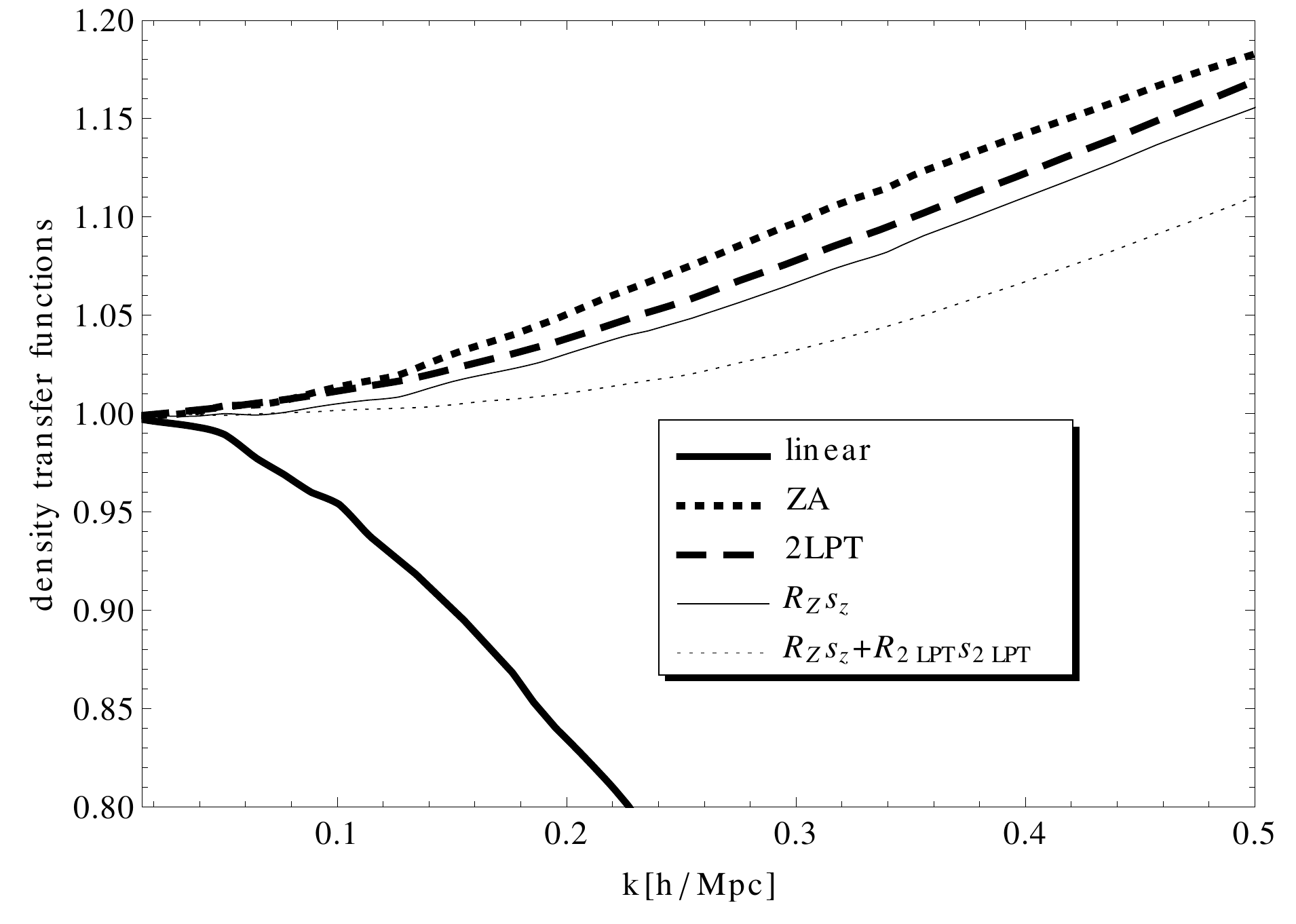}
  }
  \\
    \subfloat{\label{fig:gull}
    \includegraphics[width=0.49\textwidth]{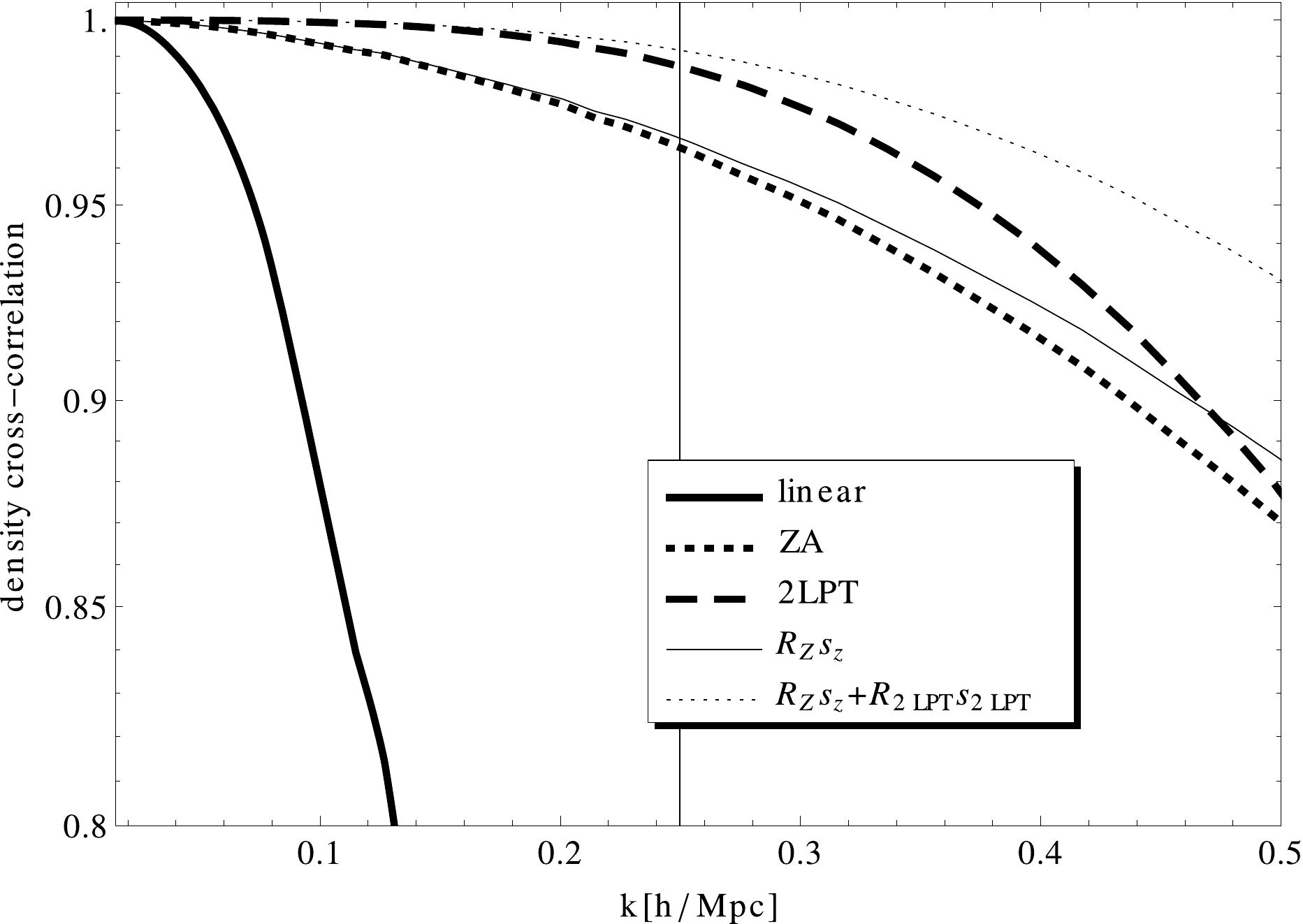}}
  \subfloat{\includegraphics[width=0.5\textwidth]{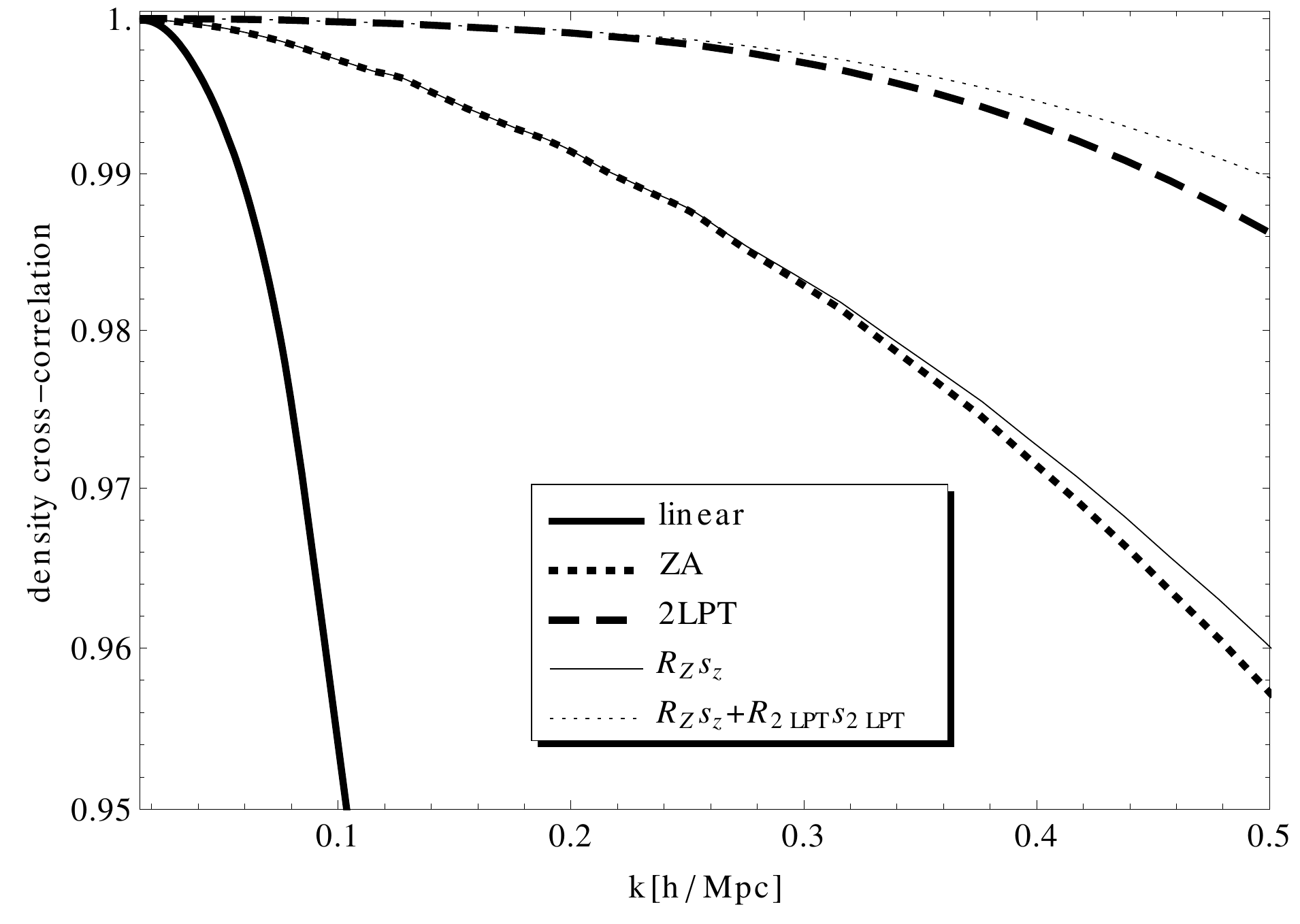}}
  \\
 \subfloat{\includegraphics[width=0.5\textwidth]{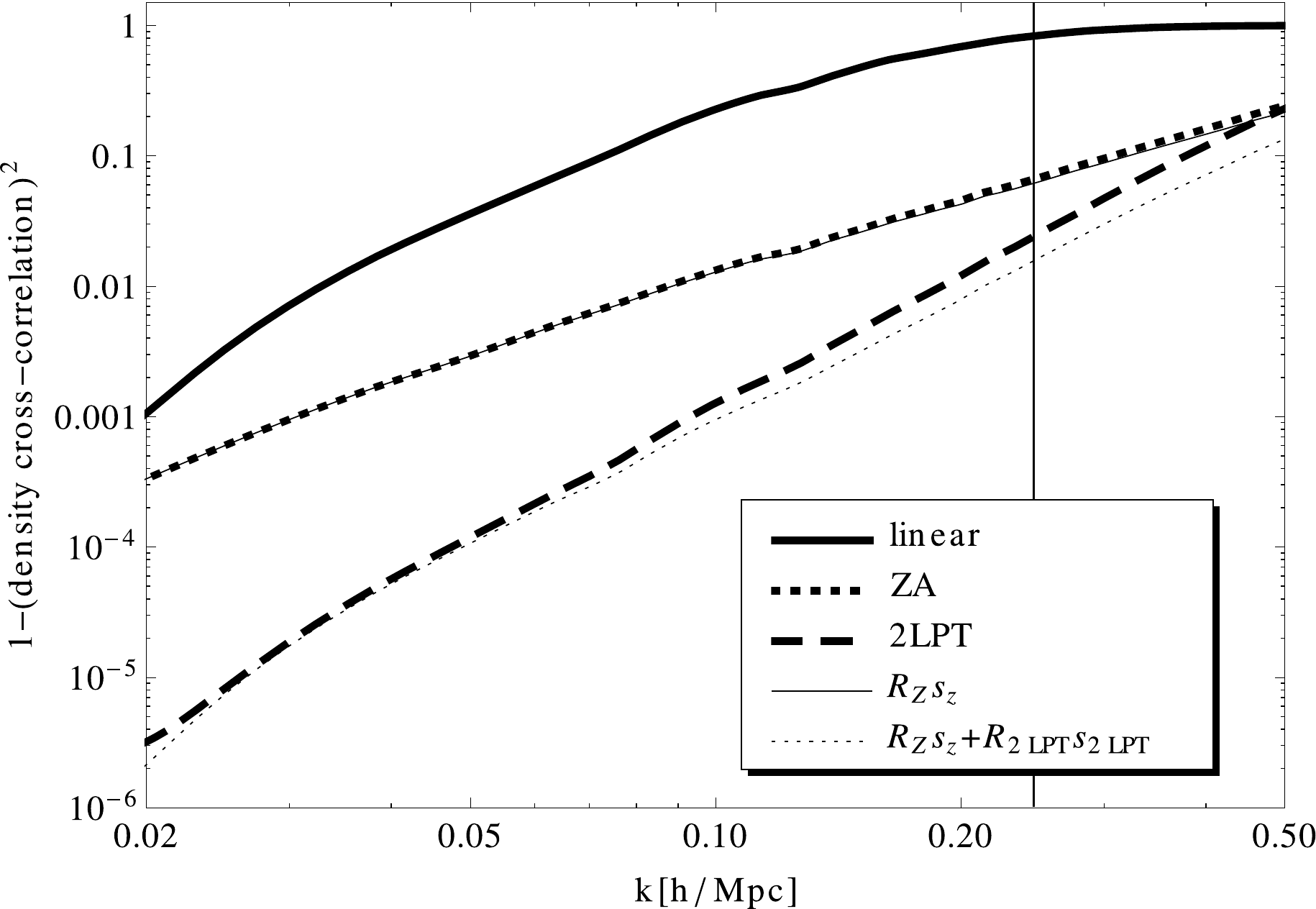}}
  \subfloat{\includegraphics[width=0.52\textwidth]{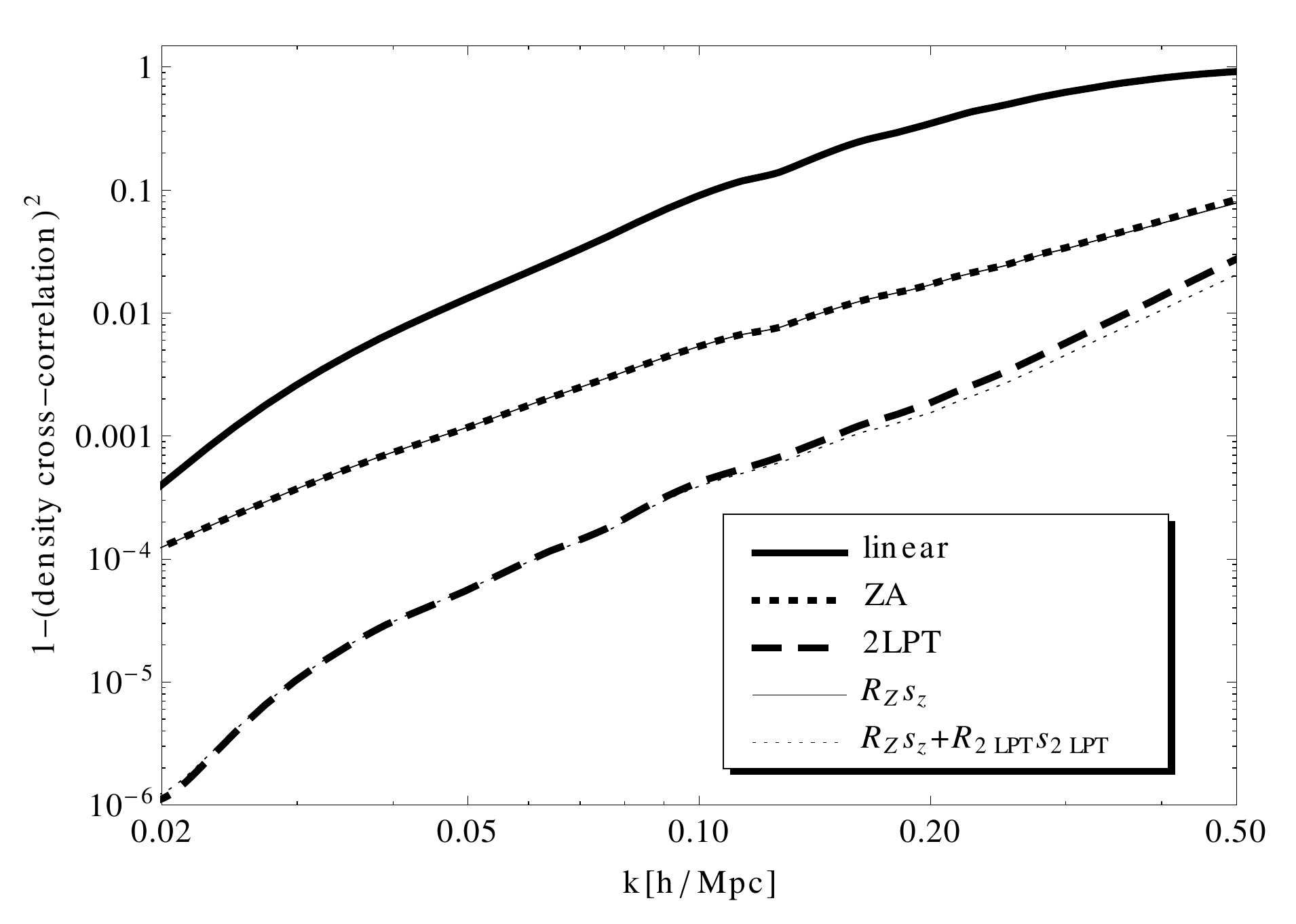}}
  \caption{The same as in Fig.~\ref{fig:displ} but for the density in real space.}
  \label{fig:dens}
\end{figure}

Next, in Fig.~\ref{fig:dens} we plot the cross-correlation and transfer functions for the density in real space for the four approximations of $\delta$ given in  eq.~(\ref{QETZ}, \ref{QE2}) (with Q replaced by $\delta$, which is given by eq.~(\ref{delta})). For comparison, we also include the transfer function $R_{L}$ relating $\delta_{L}$ with $\delta$, as well as their cross-correlation, $\rho_{L}$. As argued in TZ, $\delta_L$ is proportional to the initial overdensity, and therefore $R_{L}$ and $\rho_{L}$ involve statistics of quantities at unequal times. Thus, for scales smaller than the rms particle displacements ($k\sim 0.12h/$Mpc), the cross-correlation $\rho_{L}$ is destroyed (see TZ for detailed discussion). This is not the case for the $\delta_\E$'s based on LPT, where the only relevant scale is $k_{NL}$ (see TZ).

Concentrating on the rest of the transfer and cross-correlation functions, we can notice several interesting results. We find that including the displacement transfer functions (the velocity is irrelevant for $\delta$) as per eq.~(\ref{QE2}) has no effect on the large scales. That result is expected, because before shell-crossing LPT is exact and the displacement transfer functions go to 1 as discussed before. Another thing to note is that including both $R_z$ and $R_{2LPT}$ makes the low-$k$ $R_\delta$ (top row of Fig.~\ref{fig:dens}) flatter, lying closer to 1.

At small scales, we improve significantly the cross-correlation between $\delta_\E$ and $\delta$. Thus, a level of $\rho_\delta=0.95$, is reached at $k\approx 0.37h/$Mpc for the 2LPT approximation (the best approximation used in TZ), while including the displacement transfer functions to second order pushes that scale to $k\approx 0.45h/$Mpc. Note that without the displacement transfer functions, the ZA and 2LPT cross-correlation curves intersect in the NL regime, while this is no longer the case after including the displacement transfer functions. This result can be understood by the fact that $R_z$ and $R_{2LPT}$ capture some of the higher-order LPT contributions (missed by our second order truncation), as well as some of the stream-crossing effects missed by LPT altogether. 

In TZ we showed how one can build an approximation for $\delta$ using LPT to reduce the sample variance of the matter power spectrum. The amount of the sample variance reduction is approximately given by the ratio of $R_\delta^2\langle|\delta_\E|^2\rangle$ to the mode coupling power, $\langle|\delta_{MC}|^2\rangle$. This boils down to $\rho_\delta^2/(1-\rho_\delta^2)$ which for $\rho_\delta\sim 1$ is simply $\sim1/(1-\rho_\delta^2)$. This sample variance reduction corresponds to a speed-up  of the scanning of the cosmological parameter space one can gain.  Thus, one should pay close attention to the bottom row of Fig.~\ref{fig:dens}, showing $1-\rho^2$. We can see that
by using our best $\delta_\E$ to build an estimator for the matter power spectrum as done in TZ, 
we can gain a speed-up of a factor of $10^5$
 at scales ($k\sim 0.025h/$Mpc) ten times larger than the non-linear scale; while at the non-linear scale the expected speed-up is about a factor of ten to a hundred. Therefore, we see that this speed-up is significant well into the NL regime, especially if one uses the transfer-function-corrected displacements.

\begin{figure}
  \centering
  $z=0$\hspace{0.5\textwidth} $z=1$
  \\
  \subfloat{\includegraphics[width=0.5\textwidth]{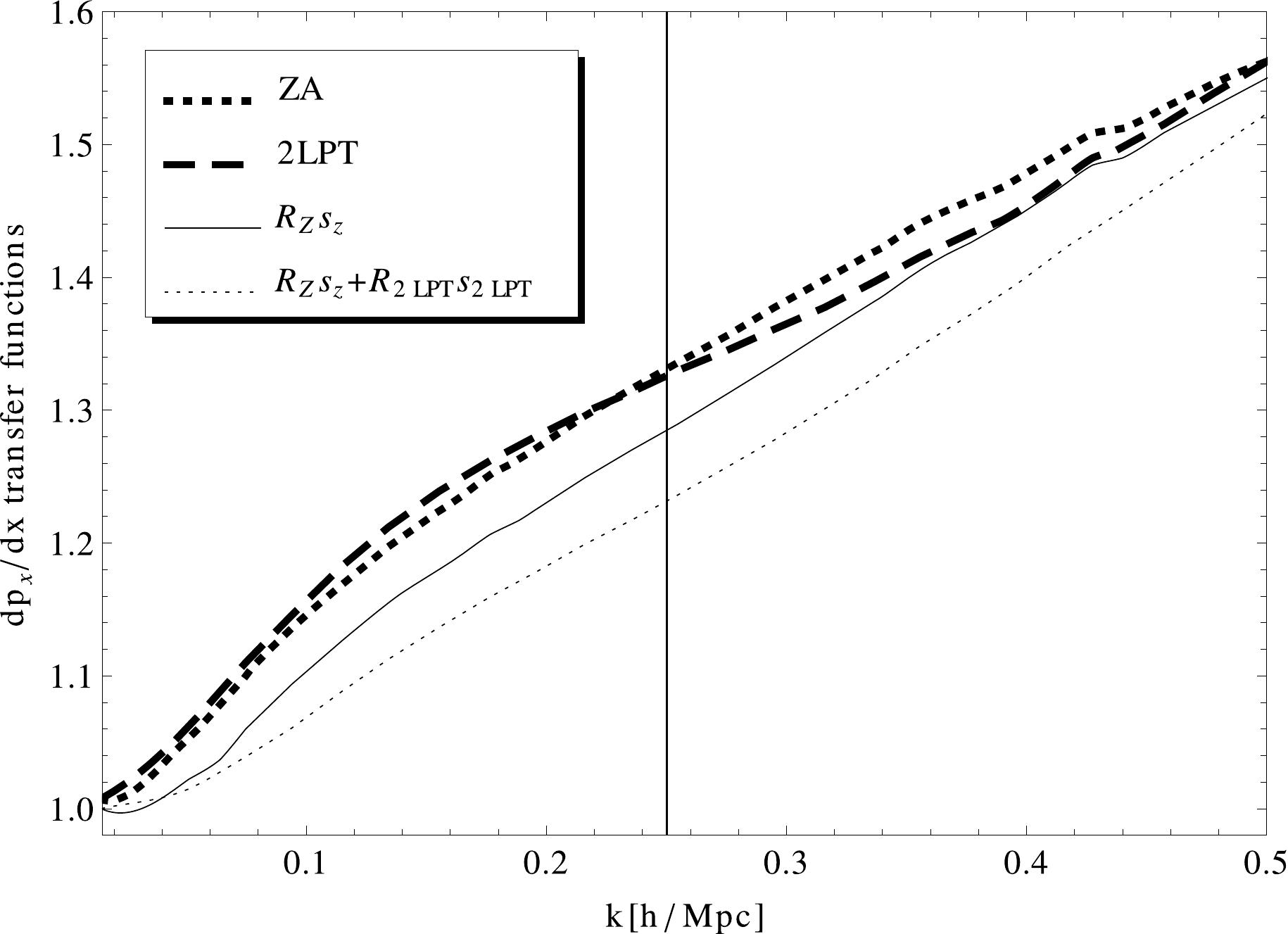}}
  \subfloat{\includegraphics[width=0.5\textwidth]{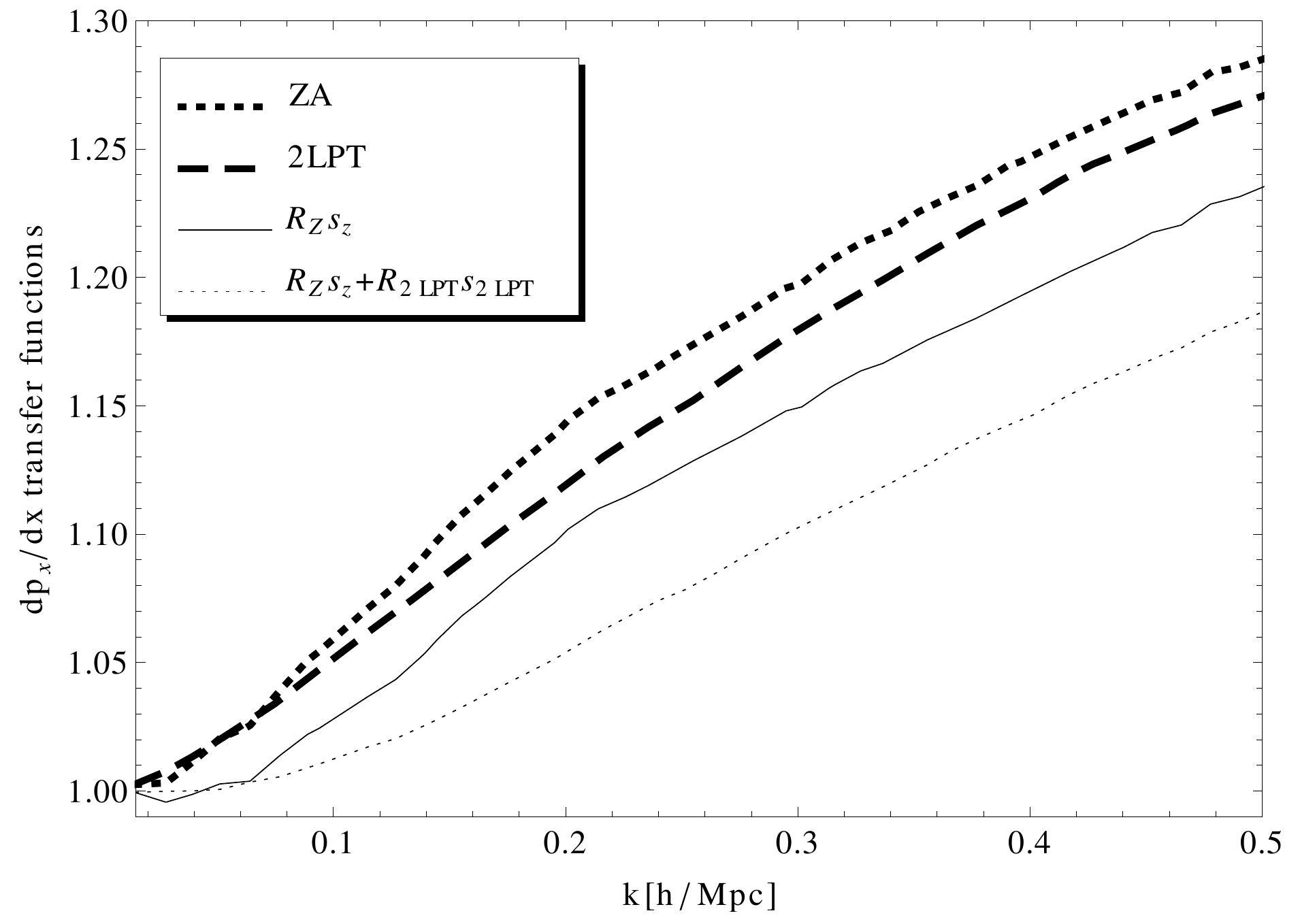}}
  \\
    \subfloat{\includegraphics[width=0.5\textwidth]{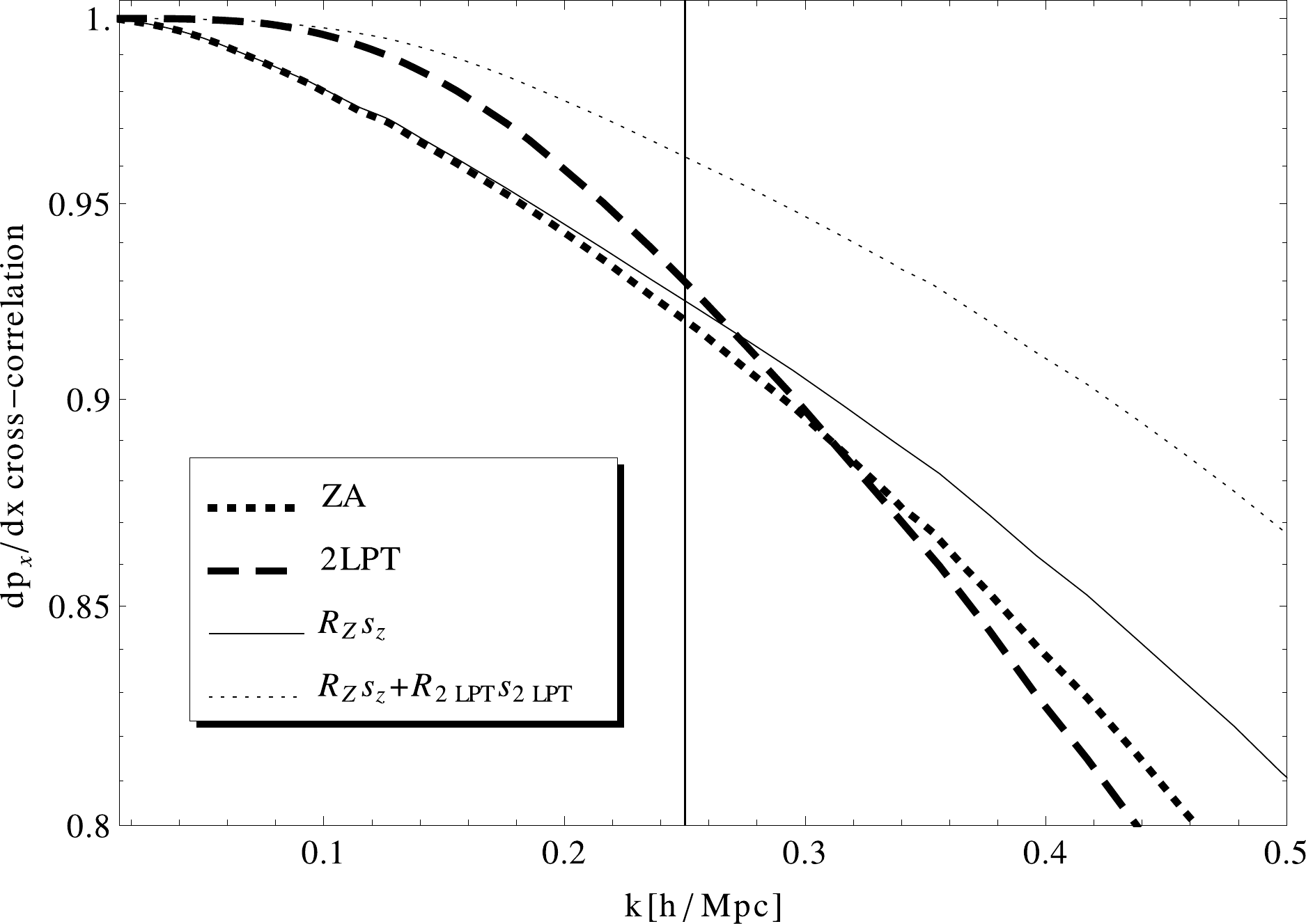}}
  \subfloat{\includegraphics[width=0.5\textwidth]{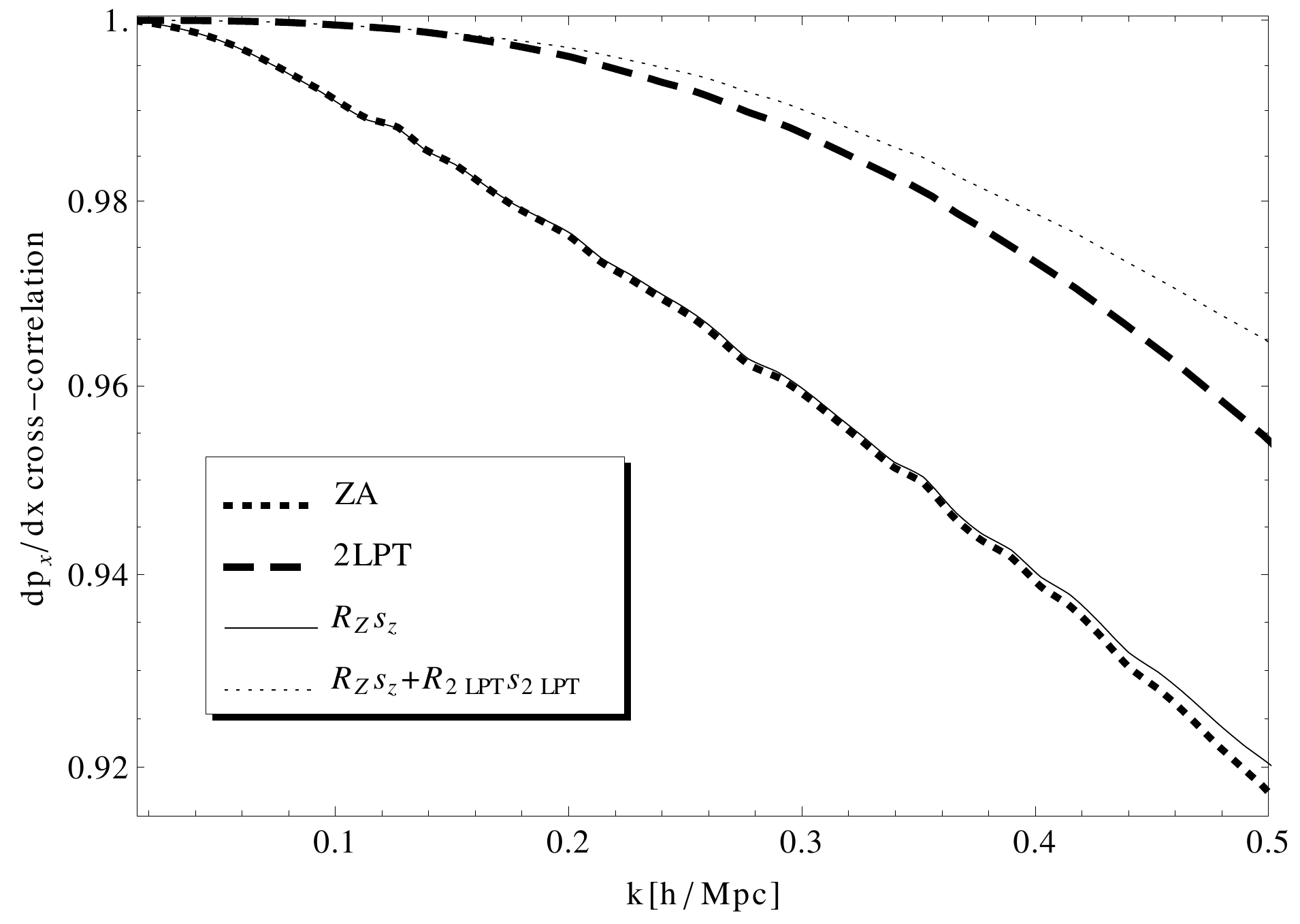}}
  \\
 \subfloat{\includegraphics[width=0.5\textwidth]{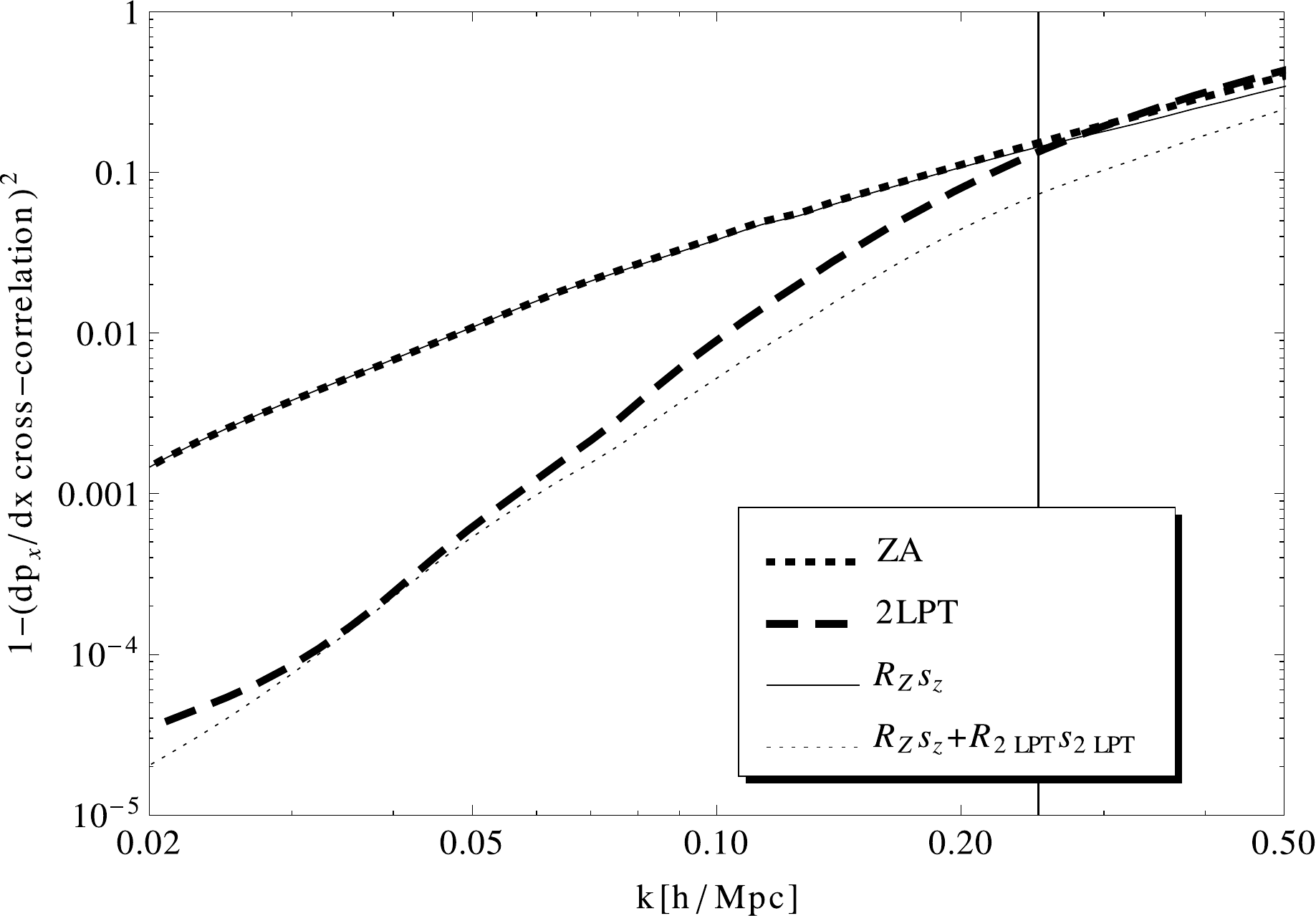}}
  \subfloat{\includegraphics[width=0.51\textwidth]{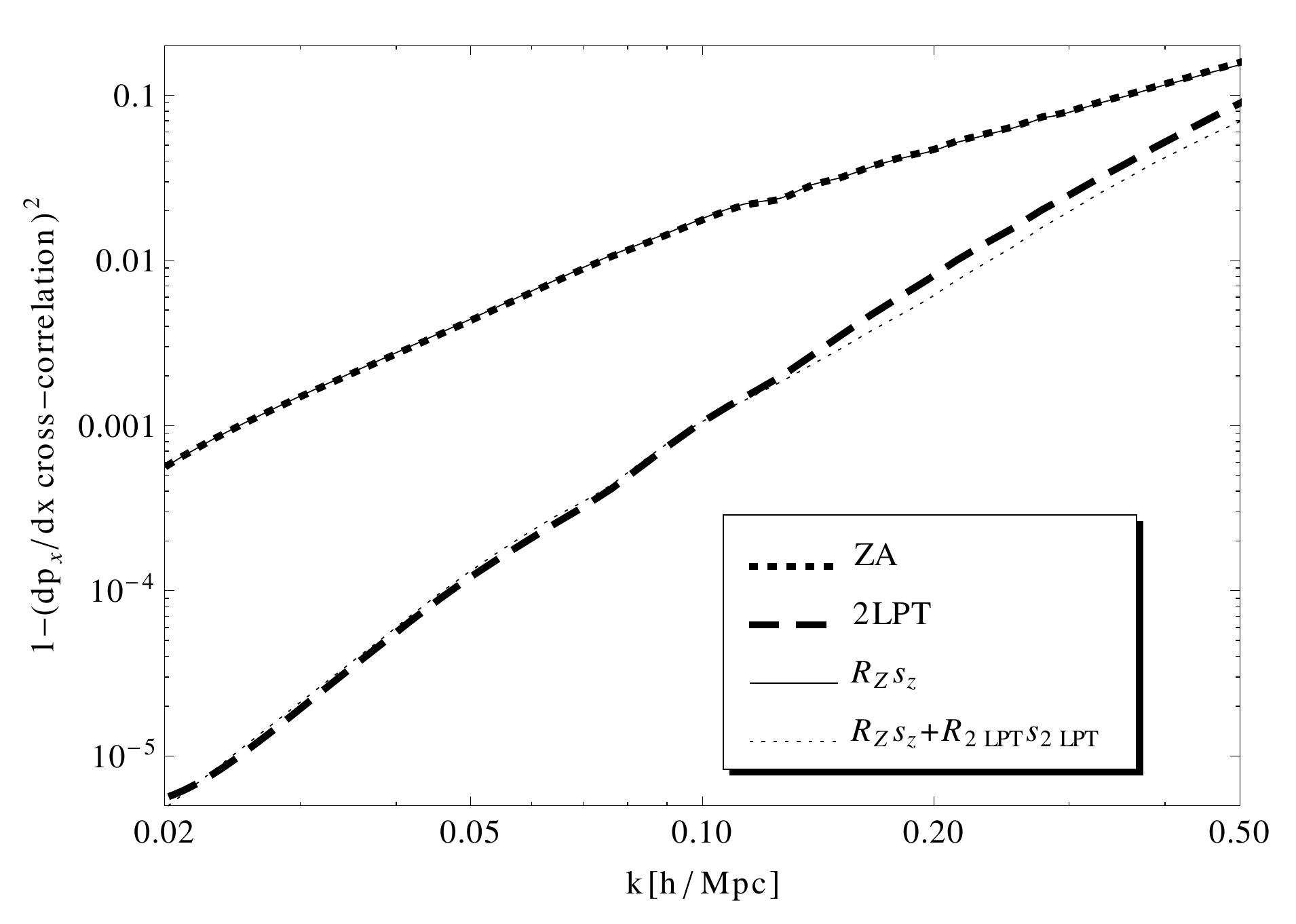}}
  \caption{The same as in Fig.~\ref{fig:displ} but for $k_{||}T^{(1)}_{||}$ which is proportional to the LoS derivative of the LoS momentum density: $dp_{||}/dx_{||}$.}
  \label{fig:dp}
\end{figure}

\subsection{LoS momentum and kinetic energy densities}

\begin{figure}
  \centering
  $z=0$\hspace{0.5\textwidth} $z=1$
  \\
  \subfloat{\includegraphics[width=0.5\textwidth]{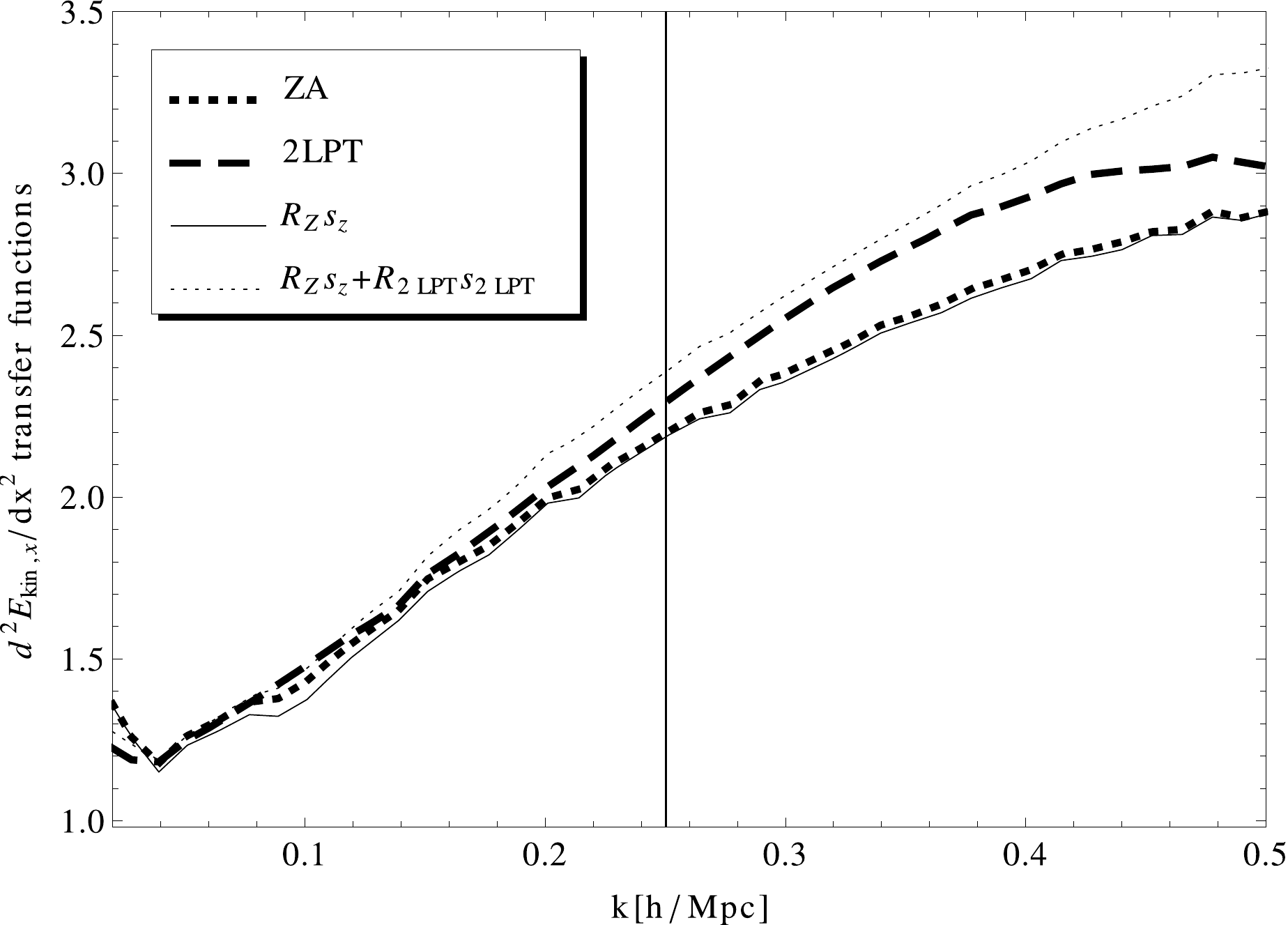}}
  \subfloat{\includegraphics[width=0.51\textwidth]{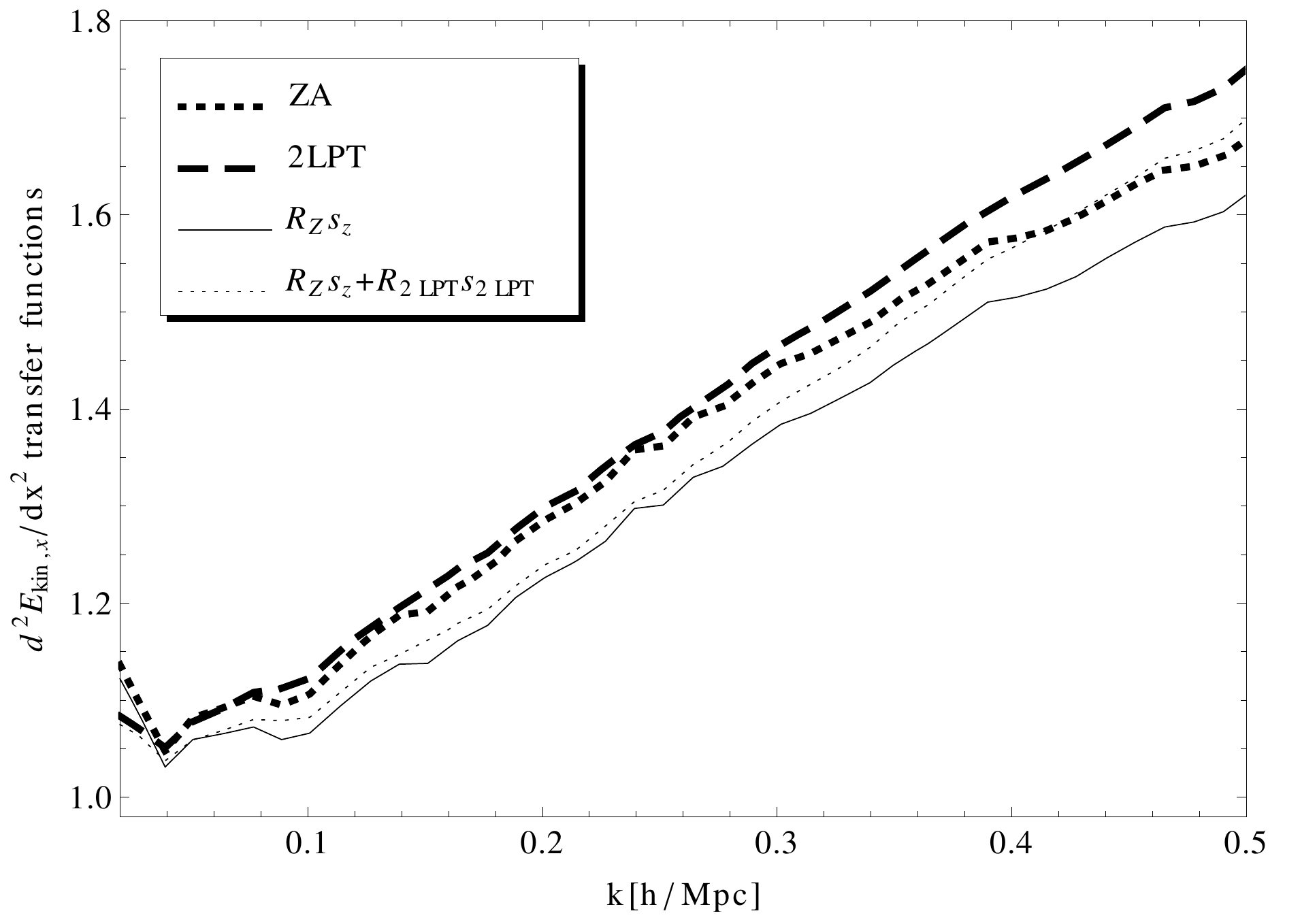}}
  \\
    \subfloat{\includegraphics[width=0.5\textwidth]{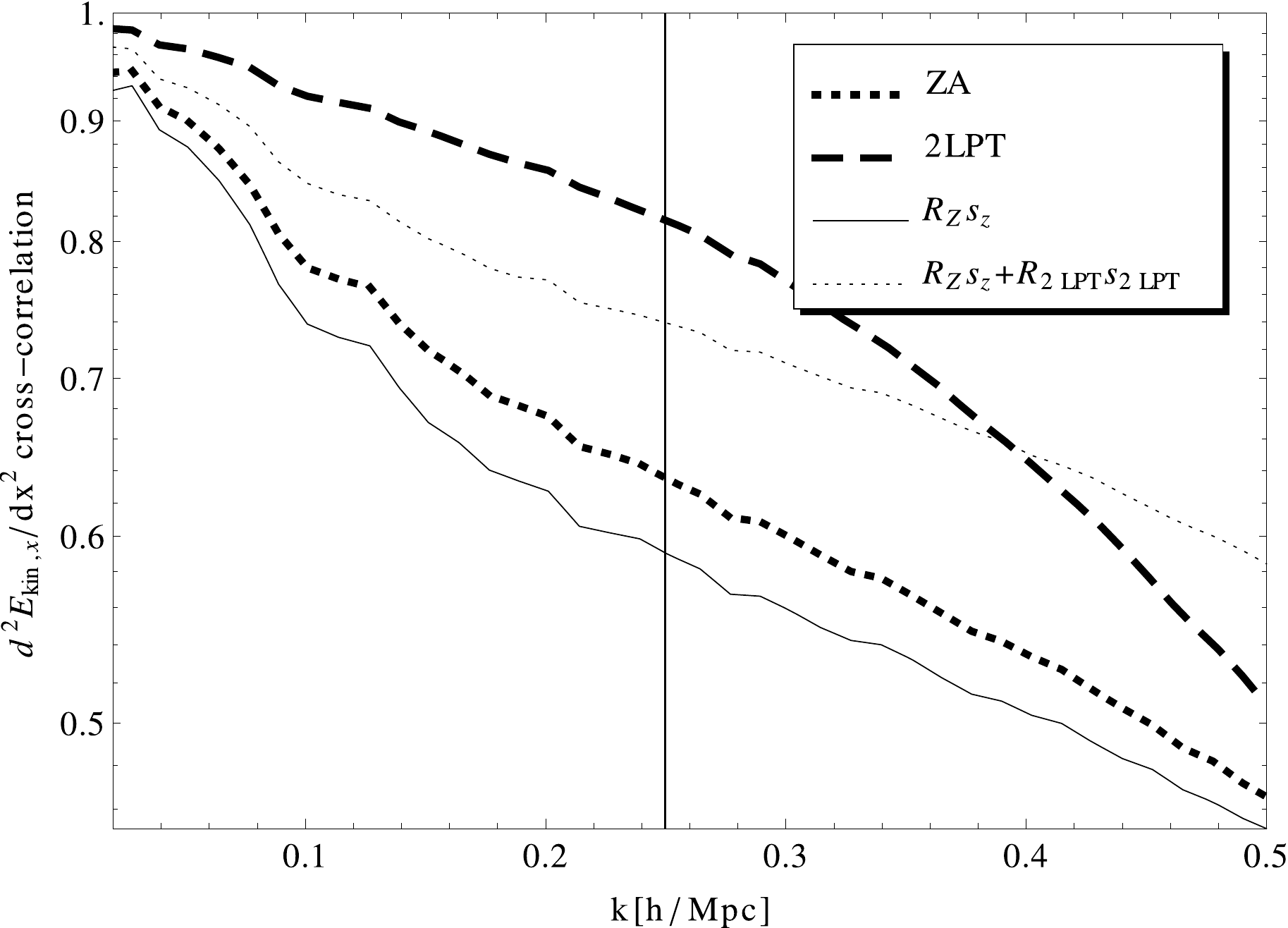}}
  \subfloat{\includegraphics[width=0.51\textwidth]{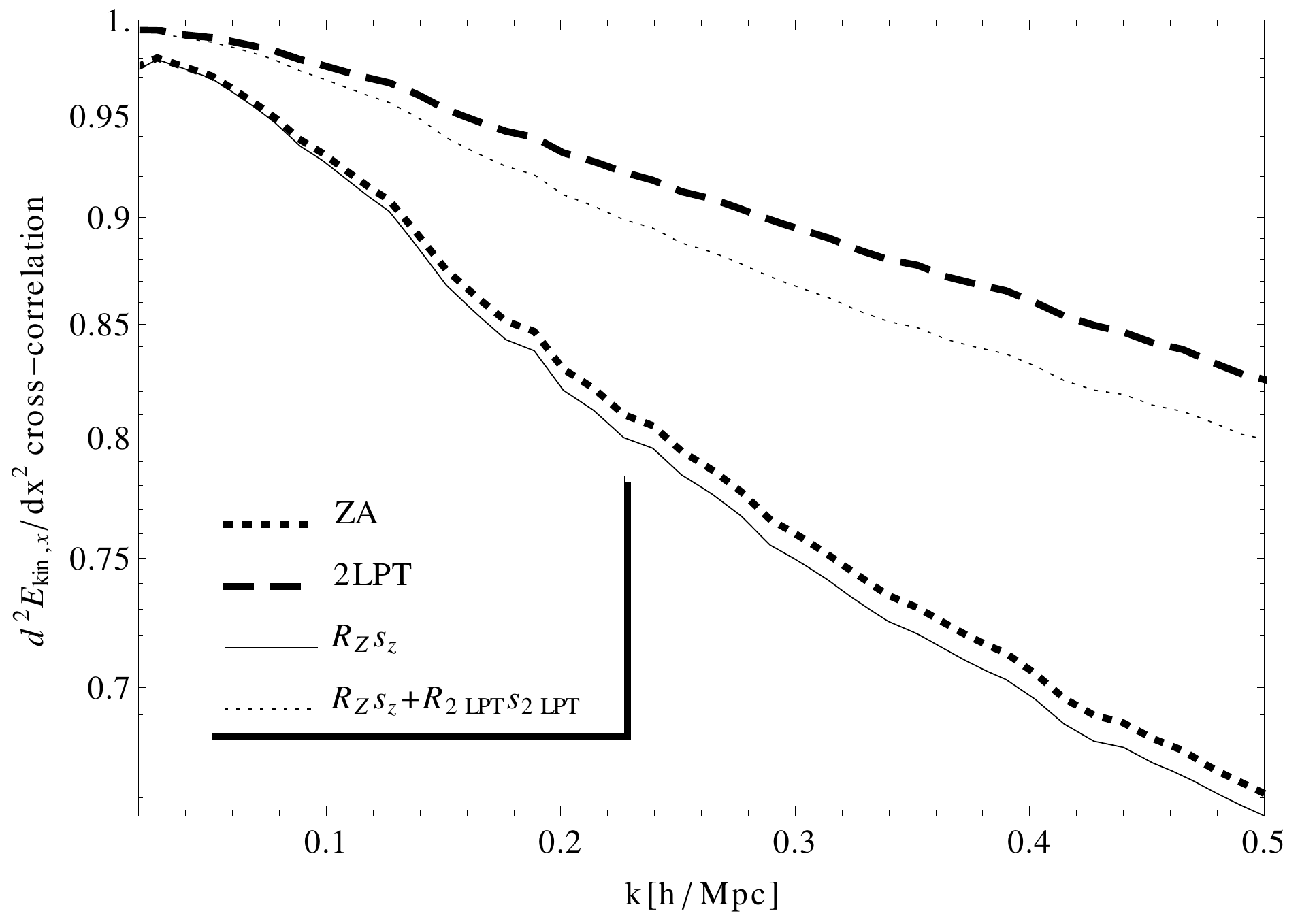}}
  \caption{The same as in Fig.~\ref{fig:displ} (dropping the bottom row) but for $k_{||}^2T^{(2)}_{||}$ which is proportional to the LoS second derivative of the LoS kinetic energy density: $d^2E_{k,||}/dx_{||}^2$}
  \label{fig:dE}
\end{figure}

Let us now focus again on the expansion of $\delta_s$ given by eq.~(\ref{redshiftDelta}). We already have a plot (Fig.~\ref{fig:dens}) of the transfer and cross-correlation functions for the zero order (in the velocity) $\delta_s$, which is simply $\delta$ in real space. The properties of the next two orders are plotted in Figures~\ref{fig:dp} and \ref{fig:dE}. For these plots, we make the flat sky approximation for simplicity.

As one can see by comparing Figures~\ref{fig:dens} and \ref{fig:dp}, the density and the momentum approximations share the same features.\footnote{The corresponding plots for the LoS momentum density approximations are very close to the ones for the LoS derivative of the LoS momentum density, Fig.~\ref{fig:dp}, which is why we do not include the former.} That is, apart from the fact that all momentum transfer and cross-correlation functions deviate from 1 at larger scales -- a direct consequence of the fact that the velocity transfer/cross-correlation function decays at larger scales than the corresponding displacement functions (see Section~\ref{sec:dv}).

From Fig.~\ref{fig:dE}, we can see that qualitatively new features appear for $\delta_s$ at second order (in the velocity), which corresponds to the second LoS derivative of the LoS kinetic energy,  $k_{||}^2T^{(2)}_{||}$. Most strikingly, \textit{none} of the $\rho$'s for this quantity saturate to 1 at large scales. Moreover, the $R$ and $R^v$ corrected approximation for $k_{||}^2T^{(2)}_{||}$ is well below the uncorrected 2LPT result (at least at large scales for $z=0$). The reason behind both of these results is that the kinetic energy even at large scales is sensitive to the high-$k$ velocity dispersion. This is because short-scale random motions (inside halos for example) do not cancel for the kinetic energy, as they do for the momentum where only CM motions matter at large scales. Since the uncorrected 2LPT result has a larger (albeit artificially larger) velocity dispersion than the result corrected with $R^v<1$, the kinetic energy coming from the short-scales is boosted, thus compensating for the missing short-scale power due to the missing $\v_{MC}$. 

To lowest order in Eulerian perturbation theory, one can check that the high-$k$ contribution to $k_{||}^2T^{(2)}_{||}$ goes as 
\be\label{highKE}
k_{||}^2T^{(2)}_{||}\sim k_{||}^2\int\limits_{k}^\infty d\tilde k P_v^2(\tilde k)\tilde k^2 \ \ \hbox{(high $k$ contribution only)}\ ,
\ee
where $P_v$ is the velocity power spectrum. Therefore,  for a given velocity approximation, $1-\rho_{k_{||}^2T^{(2)}_{||}}^2$ can be estimated by the fractional difference between the quantity in eq.~(\ref{highKE}) evaluated with the fully non-linear $P_v$, and the quantity in eq.~(\ref{highKE}) evaluated with the $P_v$ obtained from the said velocity approximation. We find that at small $k$ this fractional difference is indeed on the order of several percent for all approximations we consider, consistent with the bottom row of Fig.~\ref{fig:dE}.  

Since ${k_{||}^2T^{(2)}_{||}}$ contributes to $\delta_s$,  we can expect that $\rho_{\delta_s}$ will not be as well reconstructed as one might expect from the $\rho_\delta$ plots (Fig.~\ref{fig:dens}). We will see that that is indeed the case in the next Section.

\subsection{Density in redshift space}\label{sec:RSD}

\begin{figure}[t!]
  \centering
  $z=0$\hspace{0.5\textwidth} $z=1$
  \\
  \subfloat{\includegraphics[width=0.5\textwidth]{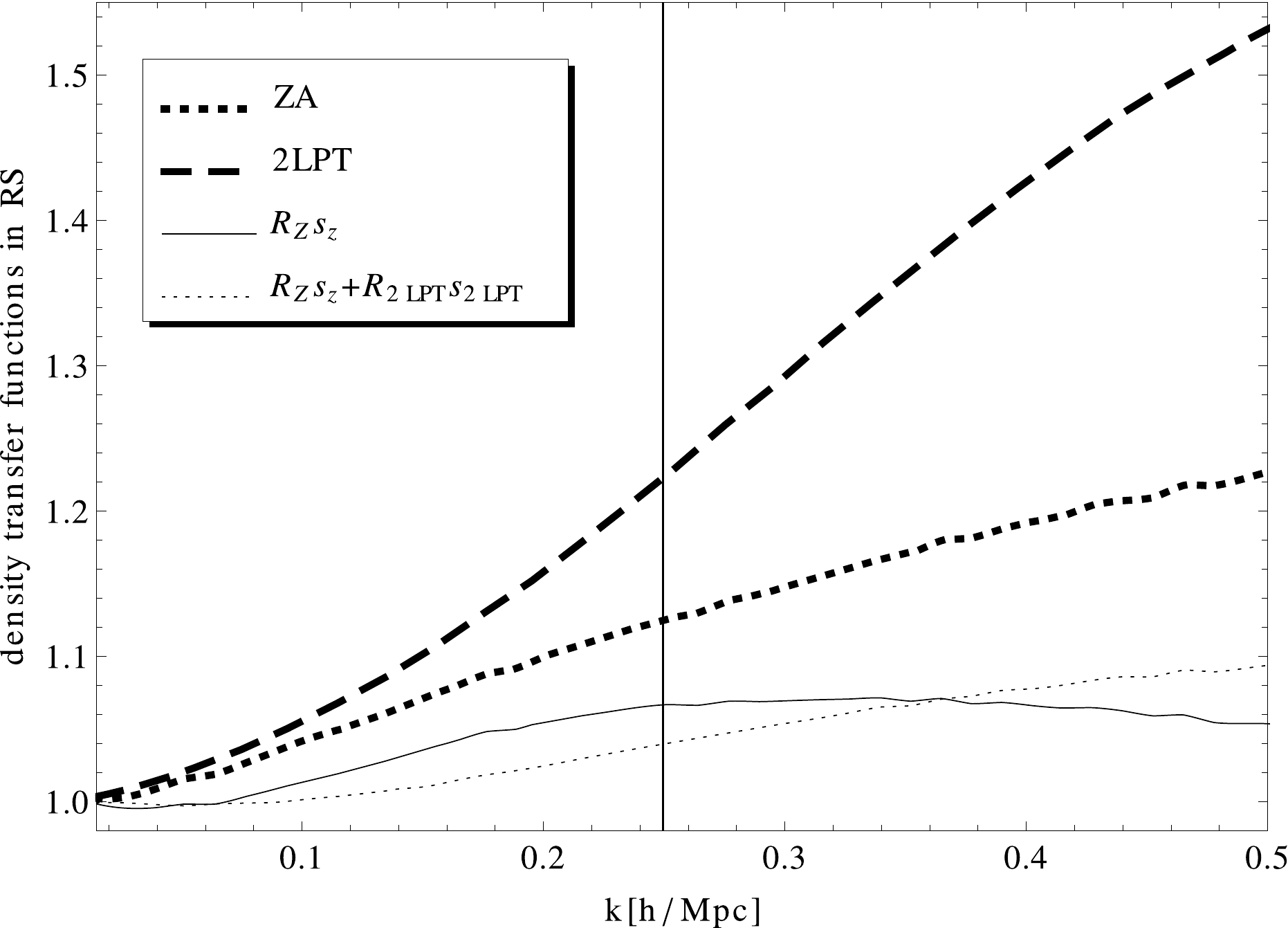}}                
  \subfloat{\includegraphics[width=0.5\textwidth]{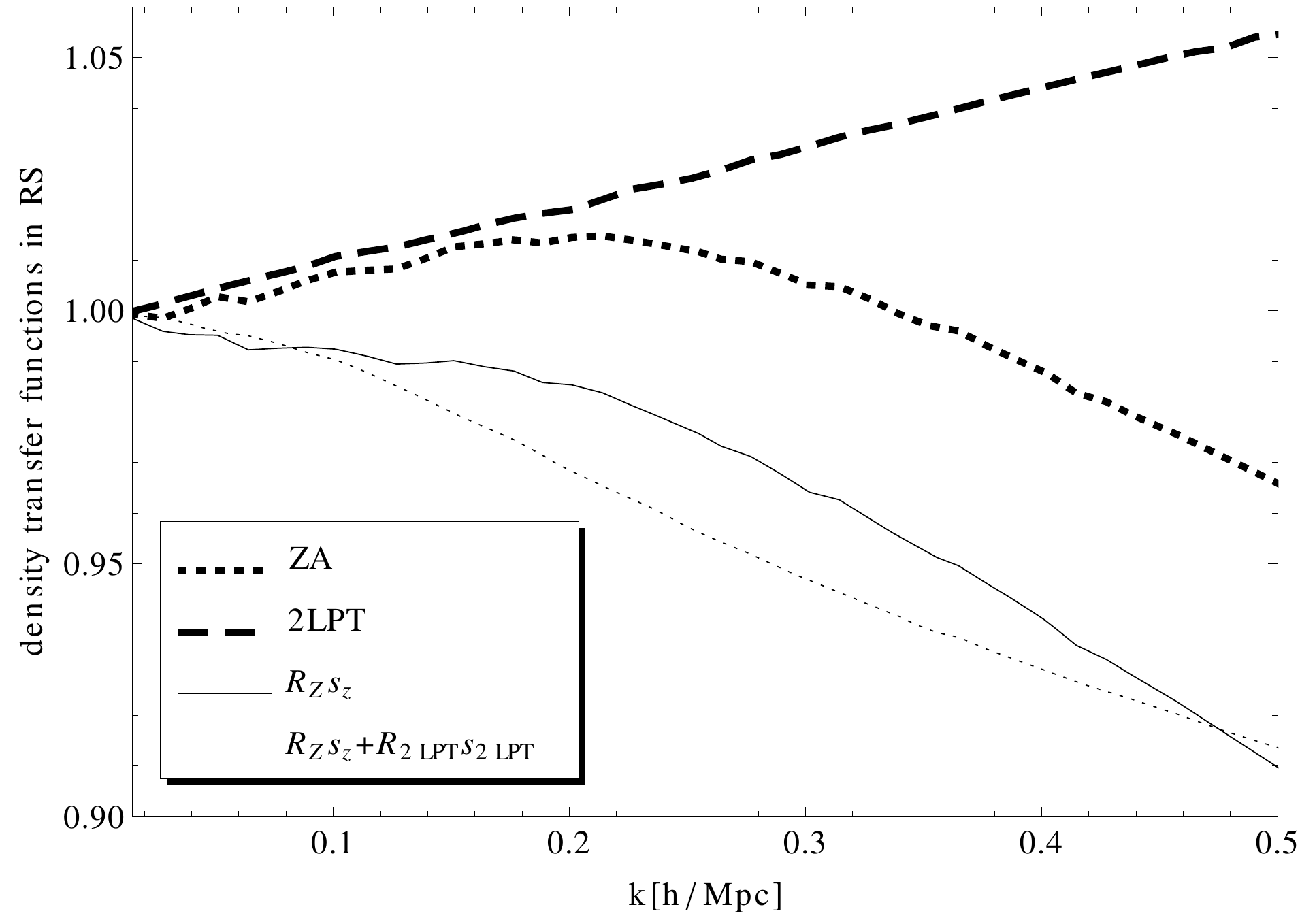}}
  \\
    \subfloat{\includegraphics[width=0.5\textwidth]{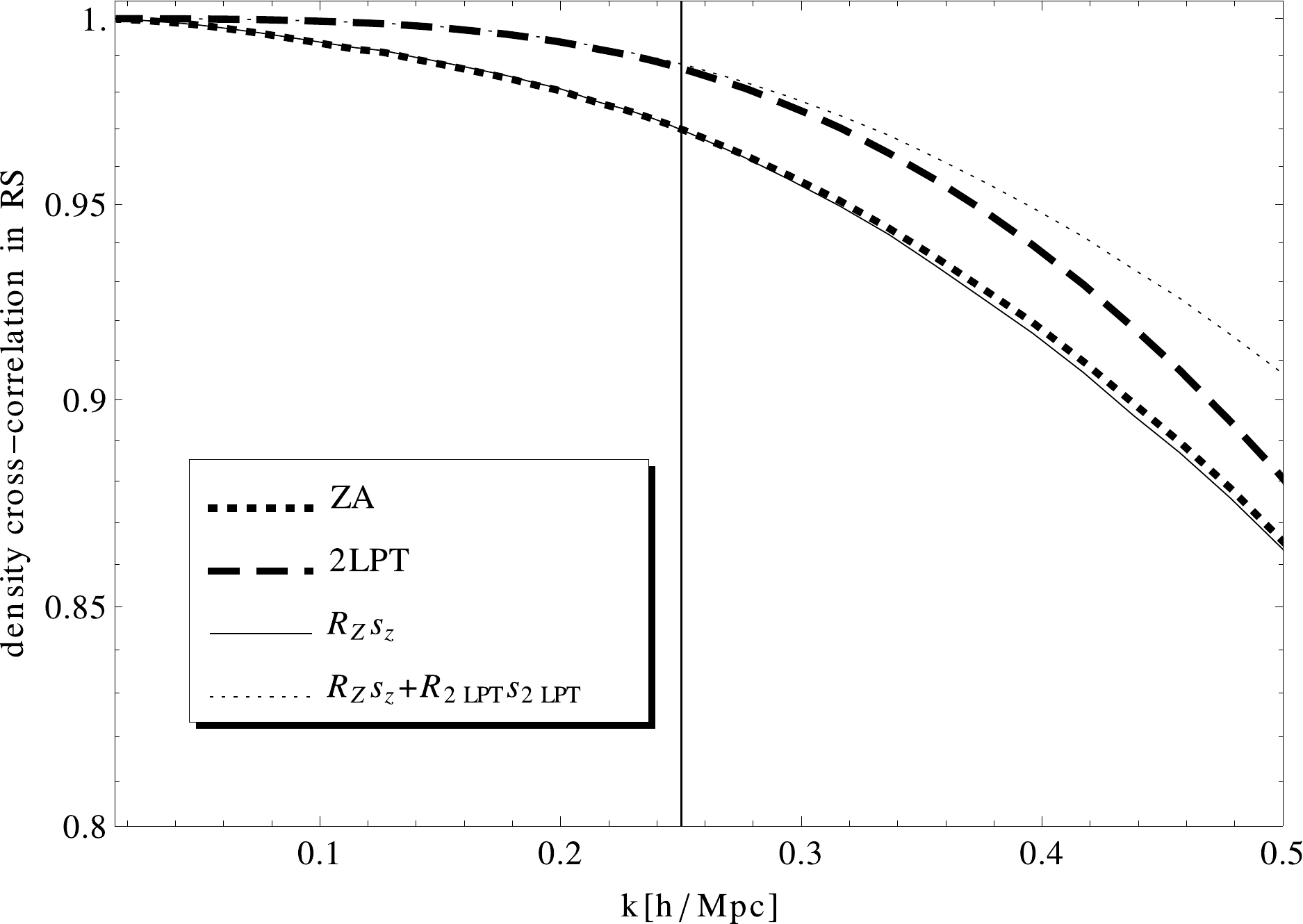}}                
  \subfloat{\includegraphics[width=0.5\textwidth]{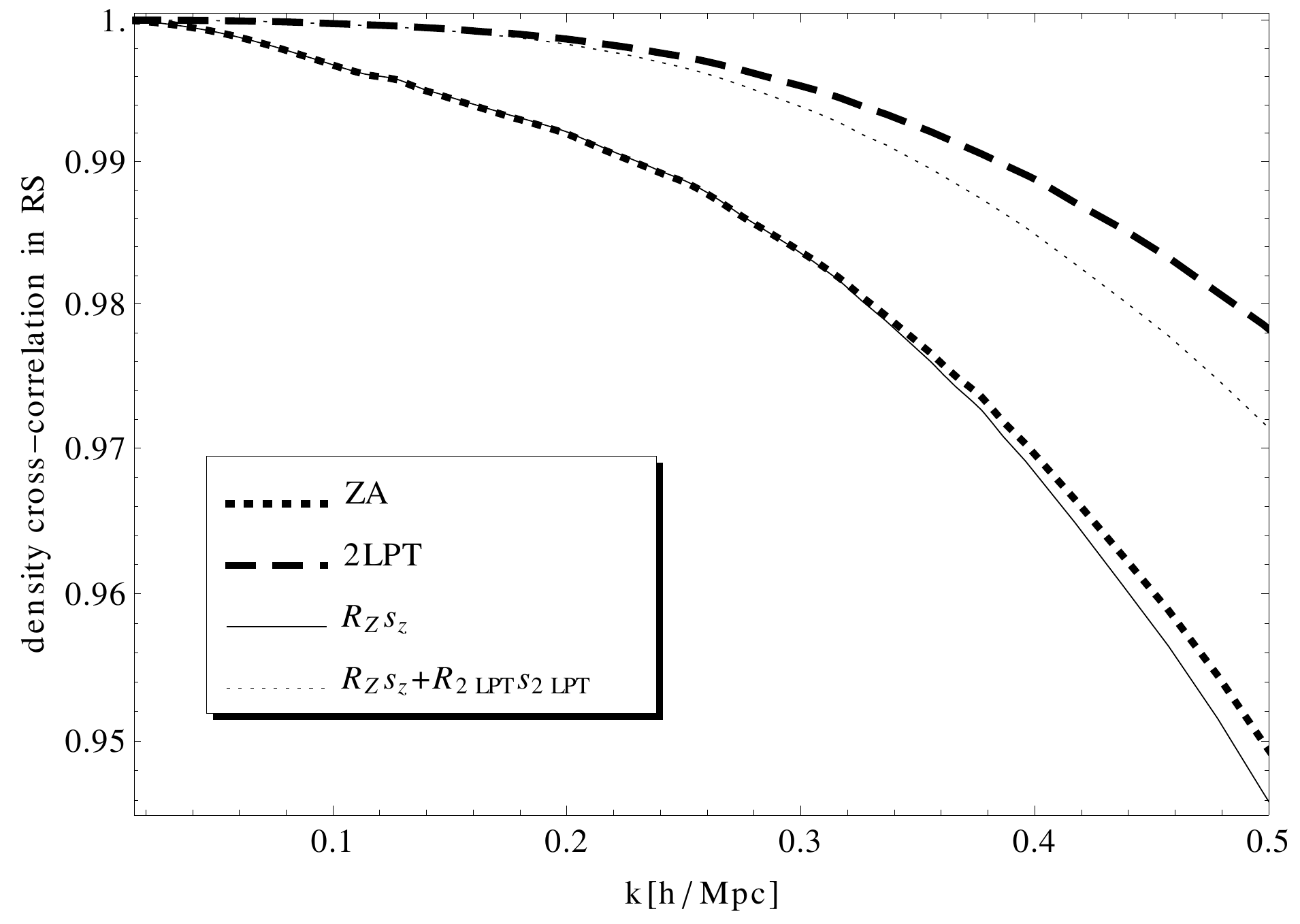}}
  \\
 \subfloat{\includegraphics[width=0.5\textwidth]{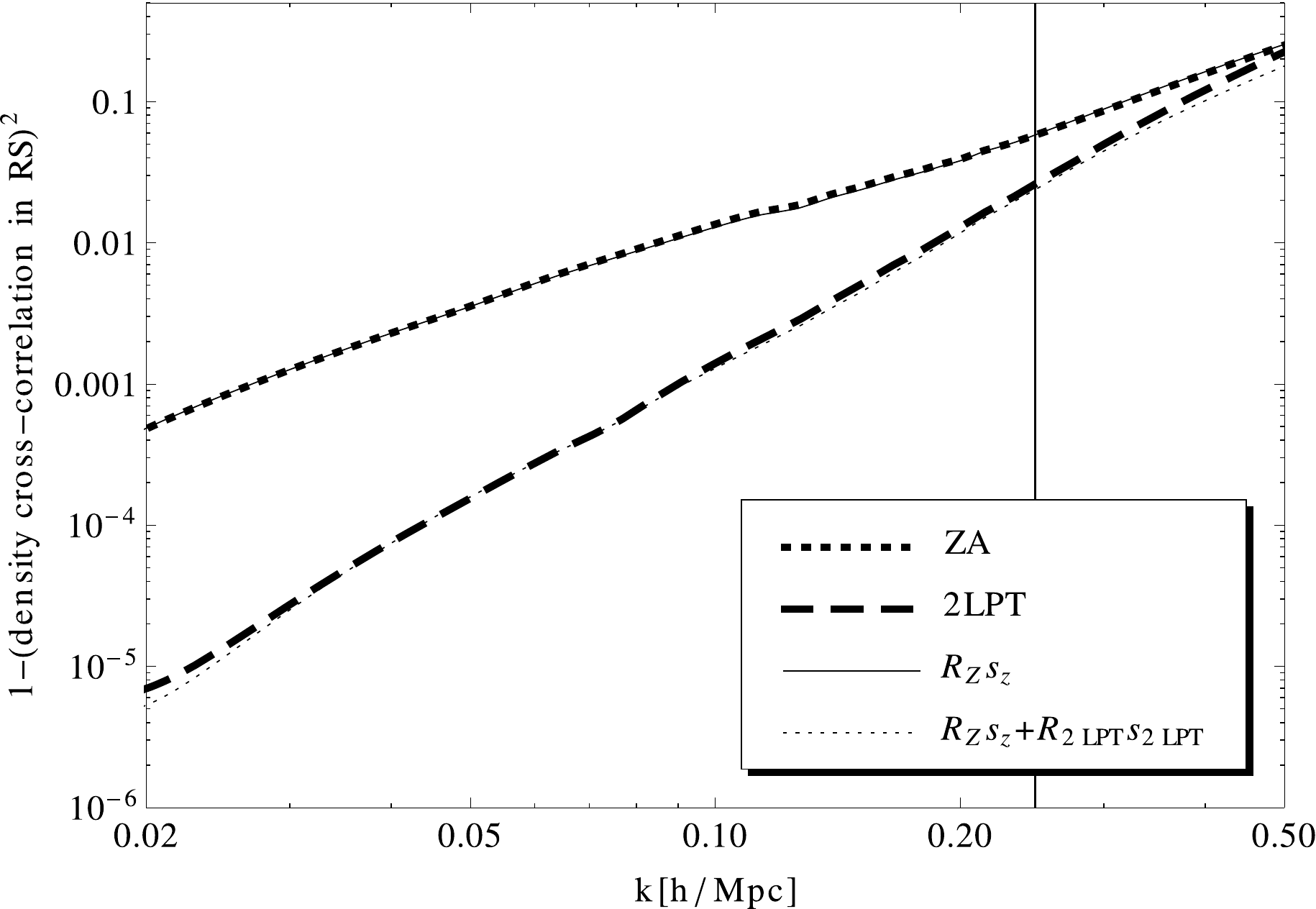}}                
  \subfloat{\includegraphics[width=0.52\textwidth]{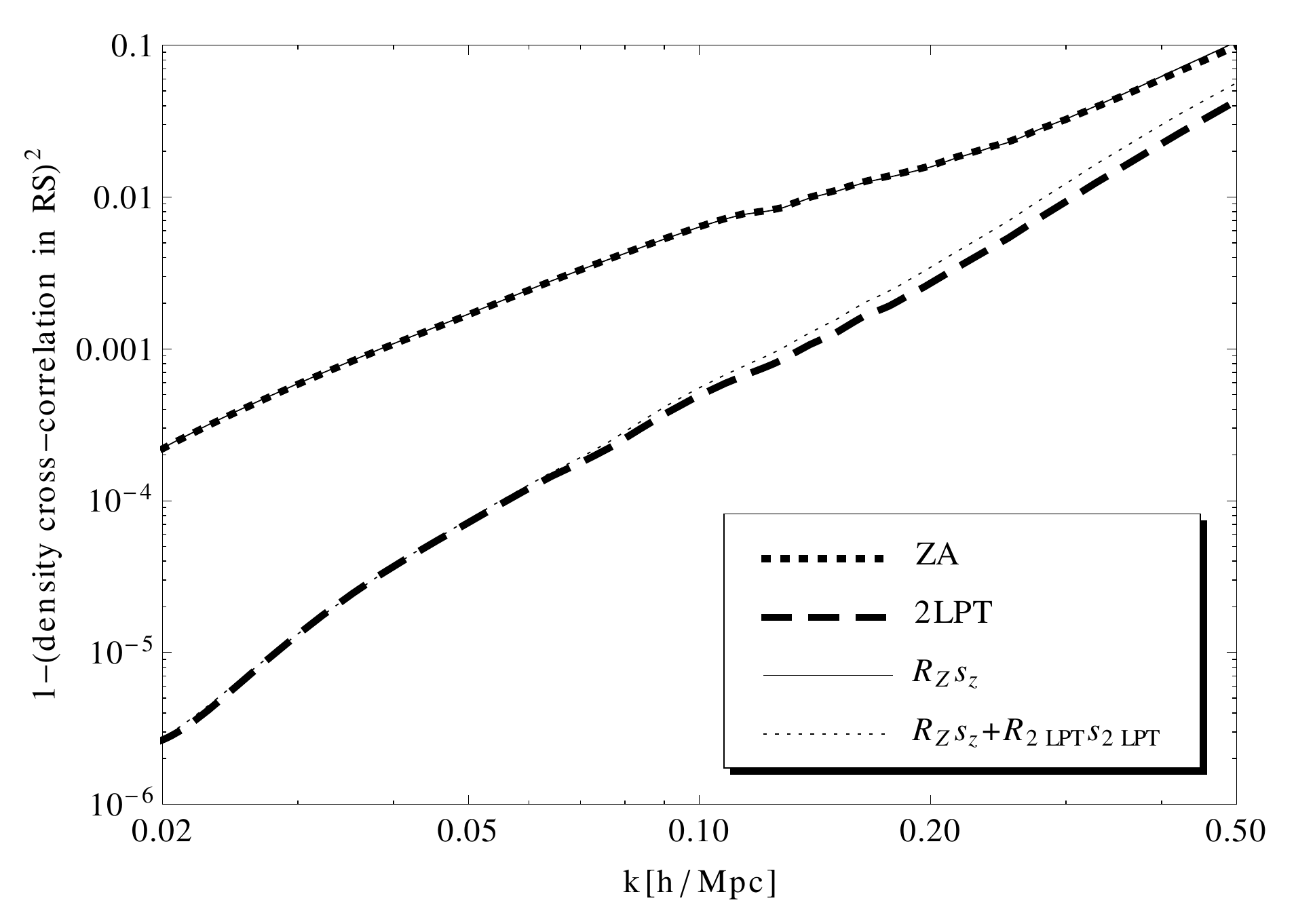}}    
  \caption{The same as in Fig.~\ref{fig:displ} but for the density monopole in redshift space.}
  \label{fig:densRSD}
\end{figure}

Now let us turn to Fig.~\ref{fig:densRSD} where we plot the same quantities as before but for the density monopole in redshift space for the four approximations for $\delta_s$ given in  eq.~(\ref{QETZ}, \ref{QE2}) (with Q replaced by $\delta_s$, which is given by eq.~(\ref{deltaRSD})). For these plots, we make the flat sky approximation for simplicity.

Comparing Fig.~\ref{fig:dens} and Fig.~\ref{fig:densRSD}, we find that including the displacement and velocity transfer functions does not improve $\rho_{\delta_s}$ significantly at $z=0$, while at $z=1$, the presence of the transfer functions makes $\rho_{\delta_s}$ even worse at high $k$. As we already showed in detail in the previous section, this effect is due entirely to the wrong reasons: Artificially setting $R^v=1$ captures by accident some of the effects of the mode-coupling velocities, which are missed by our best approximation (second line of eq.~(\ref{QE2})), for which $R^v<1$. We should stress that it is completely unclear whether this result will persist for halos or other tracers of the density field. 

Nevertheless, let us try to understand this result in more detail. We should remember that FoG effects move power from the NL to the MNL regime. By including the $R^v$'s, which decay at high $k$, we reduce the high-$k$ velocity power, which makes the FoG effect smaller. Thus, because of the non-linear mapping, eq.~(\ref{delta}), between displacements and velocities on the one side and $\delta_s$ on the other, we find that using $R^v<1$ at high $k$, affects the MNL $\delta_{s,\E}$. Those effects cannot be captured by the linear (isotropic) rescaling, $R_{\delta_s}$, and are instead dumped into $\delta_{s,MC}$. 

To test that this effect on $\rho_{\delta_s}$  is indeed due only to the $R^v$'s being smaller than one, we constructed a $\delta_{s,\E}$ by including only the $R^v$'s (i.e. setting $R_z=R_{2LPT}=1$) and then including only the displacement transfer functions (i.e. setting $R^v_z=R^v_{2LPT}=1$). We find that the former choice results in degradation of $\rho_{\delta_s}$, while the latter choice improves $\rho_{\delta_s}$ for both redshifts compared with the pure 2LPT result. Thus, we find that faking the high-$k$ power in $\v_{MC}$ by keeping $R^v=1$ improves  $\rho_{\delta_s}$. However, such a route is not consistent, since the mode-coupling and LPT velocities must have a vanishing 2-pt function. Instead, if one is determined to boost $\rho_{\delta_s}$ closer to 1, one may be better off including a Gaussian $\v_{MC}$, uncorrelated with $\sd_{z/2LPT}$. However, this would result in adding a velocity dispersion both in halos and in voids. Nevertheless, we tested that approach\footnote{We added $v_{MC}$ by adding both its solenoidal and irrotational components shown in Fig.~\ref{fig:MCs}. Doing the same for the displacements only further worsens the results.}, and indeed it results  in a reduced $\rho_{\delta_s}$, leading us to conclude that the higher order correlations between $\v_{MC}$ and $\sd_{z/2LPT}$ are important. Indeed looking at the bottom row of Fig.~\ref{fig:p}, we can see that the rms residual velocity is indeed correlated with the overdensity. However, further analysis along those lines will be warranted only if one calculates these transfer function for CDM halos (not the CDM matter density) or other biased tracers.

\section{Summary}\label{sec:summary}

In this paper, we consider an improvement to the formalism proposed in TZ for studying the mildly non-linear regime of structure formation. This improvement is obtained by constructing good approximations of the CDM particle trajectories. This is done by relating the displacement and velocity fields obtained in LPT (in Lagrangian space) to the non-linear ones by using time- and scale-dependent transfer functions. We work up to second order in LPT. Compared to the pure 2LPT results, after including the transfer functions we find an improvement in the approximated center of mass position of a given cell (in Eulerian space) which is about a factor of 2 better ($\lesssim 1\,$Mpc$/h$). 

These improvements in approximating the particle positions result in an improved approximation, $\delta_\E$, for the fully non-linear density, $\delta$. The cross-correlation coefficient between $\delta_\E$ and $\delta$ reaches a level of $\rho_\delta=0.95$ at $k\approx 0.37h/$Mpc for the 2LPT approximation (the best approximation used in TZ), while for our current best $\delta_\E$, which includes the displacement transfer functions to second order, the corresponding scale is $k\approx 0.45h/$Mpc (for $z=0$), almost twice smaller than the non-linear scale. For comparison, the corresponding scale for the cross-correlation between the density from linear theory and $\delta$ is $k\approx 0.075h/$Mpc, a factor of 6 larger, than our best density approximation.  Thus, our results allow for a speed-up of an order of magnitude or more in the scanning of the cosmological parameter space with N-body simulations for the scales relevant for the baryon acoustic oscillations. 

We also investigated the effect on using our trajectory approximations for calculating the density in redshift space. We find that compared to $\delta_s$ obtained in pure 2LPT, the resulting $\delta_s$ approximation has only slightly improved cross-correlation with the NL $\delta_s$ for $z=0$. That cross-correlation is even slightly worse than the 2LPT result for $z=1$. However, we argue that this slightly better cross-correlation for 2LPT is due to an inconsistency in the 2LPT approximation, and therefore one should still use our best approximation. Moreover, it is unclear whether this result will hold for halos or other biased tracers -- an analysis which we postpone to future work.

As an application of the results presented here, in a companion paper \cite{OptimalRec}, we utilize the models of the non-linear density field constructed in this paper, to develop a quasi-optimal reconstruction scheme of the BAO peak.


\acknowledgments ST would like to thank Daniel Eisenstein and David Spergel for useful discussions. We would like to thank Martin White for useful comments on the manuscript. The work of MZ is supported by NSF grants PHY-0855425, AST-0506556 \& AST-0907969, and by the David \& Lucile Packard and the John D. \& Catherine T. MacArthur Foundations.

\bibliography{mildly_NL_v1}

\providecommand{\href}[2]{#2}\begingroup\raggedright\begin{thebibliography}{10}

\bibitem{jain}
B.~{Jain} and E.~{Bertschinger}, {\it {Second-order power spectrum and
  nonlinear evolution at high redshift}},  {\em \apj} {\bf 431} (Aug., 1994)
  495--505, [\href{http://xxx.lanl.gov/abs/astro-ph/9311070}{{\tt
  astro-ph/9311070}}].

\bibitem{2005ApJ...633..560E}
D.~J. {Eisenstein}, I.~{Zehavi}, D.~W. {Hogg}, R.~{Scoccimarro}, M.~R.
  {Blanton}, R.~C. {Nichol}, R.~{Scranton}, H.~{Seo}, M.~{Tegmark}, Z.~{Zheng},
  S.~F. {Anderson}, J.~{Annis}, N.~{Bahcall}, J.~{Brinkmann}, S.~{Burles},
  F.~J. {Castander}, A.~{Connolly}, I.~{Csabai}, M.~{Doi}, M.~{Fukugita}, J.~A.
  {Frieman}, K.~{Glazebrook}, J.~E. {Gunn}, J.~S. {Hendry}, G.~{Hennessy},
  Z.~{Ivezi{\'c}}, S.~{Kent}, G.~R. {Knapp}, H.~{Lin}, Y.~{Loh}, R.~H.
  {Lupton}, B.~{Margon}, T.~A. {McKay}, A.~{Meiksin}, J.~A. {Munn}, A.~{Pope},
  M.~W. {Richmond}, D.~{Schlegel}, D.~P. {Schneider}, K.~{Shimasaku},
  C.~{Stoughton}, M.~A. {Strauss}, M.~{SubbaRao}, A.~S. {Szalay}, I.~{Szapudi},
  D.~L. {Tucker}, B.~{Yanny}, and D.~G. {York}, {\it {Detection of the Baryon
  Acoustic Peak in the Large-Scale Correlation Function of SDSS Luminous Red
  Galaxies}},  {\em \apj} {\bf 633} (Nov., 2005) 560--574,
  [\href{http://xxx.lanl.gov/abs/astro-ph/0501171}{{\tt astro-ph/0501171}}].

\bibitem{Angulo:2007fw}
R.~E. {Angulo}, C.~M. {Baugh}, C.~S. {Frenk}, and C.~G. {Lacey}, {\it {The
  detectability of baryonic acoustic oscillations in future galaxy surveys}},
  {\em \mnras} {\bf 383} (Jan., 2008) 755--776,
  [\href{http://xxx.lanl.gov/abs/astro-ph/0702543}{{\tt astro-ph/0702543}}].

\bibitem{2009PhRvD..79f3523P}
N.~{Padmanabhan}, M.~{White}, and J.~D. {Cohn}, {\it {Reconstructing baryon
  oscillations: A Lagrangian theory perspective}},  {\em \prd} {\bf 79} (Mar.,
  2009) 063523--+, [\href{http://xxx.lanl.gov/abs/0812.2905}{{\tt
  arXiv:0812.2905}}].

\bibitem{2010ApJ...720.1650S}
H.~{Seo}, J.~{Eckel}, D.~J. {Eisenstein}, K.~{Mehta}, M.~{Metchnik},
  N.~{Padmanabhan}, P.~{Pinto}, R.~{Takahashi}, M.~{White}, and X.~{Xu}, {\it
  {High-precision Predictions for the Acoustic Scale in the Nonlinear Regime}},
   {\em \apj} {\bf 720} (Sept., 2010) 1650--1667,
  [\href{http://xxx.lanl.gov/abs/0910.5005}{{\tt arXiv:0910.5005}}].

\bibitem{tassev}
S.~{Tassev} and M.~{Zaldarriaga}, {\it {The mildly non-linear regime of
  structure formation}},  {\em \jcap} {\bf 4} (Apr., 2012) 13,
  [\href{http://xxx.lanl.gov/abs/1109.4939}{{\tt arXiv:1109.4939}}].

\bibitem{zeldovich}
Y.~B. {Zel'dovich}, {\it {Gravitational instability: An approximate theory for
  large density perturbations.}},  {\em \aap} {\bf 5} (Mar., 1970) 84--89.

\bibitem{catelan}
P.~{Catelan}, {\it {Lagrangian dynamics in non-flat universes and non-linear
  gravitational evolution}},  {\em \mnras} {\bf 276} (Sept., 1995) 115--124,
  [\href{http://xxx.lanl.gov/abs/astro-ph/}{{\tt astro-ph/}}].

\bibitem{rpt}
M.~{Crocce} and R.~{Scoccimarro}, {\it {Renormalized cosmological perturbation
  theory}},  {\em \prd} {\bf 73} (Mar., 2006) 063519--+,
  [\href{http://xxx.lanl.gov/abs/astro-ph/0509418}{{\tt astro-ph/0509418}}].

\bibitem{mlpt}
T.~{Matsubara}, {\it {Resumming cosmological perturbations via the Lagrangian
  picture: One-loop results in real space and in redshift space}},  {\em \prd}
  {\bf 77} (Mar., 2008) 063530--+,
  [\href{http://xxx.lanl.gov/abs/0711.2521}{{\tt arXiv:0711.2521}}].

\bibitem{BBKS}
J.~M. {Bardeen}, J.~R. {Bond}, N.~{Kaiser}, and A.~S. {Szalay}, {\it {The
  statistics of peaks of Gaussian random fields}},  {\em \apj} {\bf 304} (May,
  1986) 15--61.

\bibitem{SeljakRSD}
U.~{Seljak} and P.~{McDonald}, {\it {Distribution function approach to redshift
  space distortions}},  {\em \jcap} {\bf 11} (Nov., 2011) 39,
  [\href{http://xxx.lanl.gov/abs/1109.1888}{{\tt arXiv:1109.1888}}].

\bibitem{HHpaper}
S.~V. {Tassev}, {\it {The Helmholtz Hierarchy: phase space statistics of cold
  dark matter}},  {\em \jcap} {\bf 10} (Oct., 2011) 22,
  [\href{http://xxx.lanl.gov/abs/1012.0282}{{\tt arXiv:1012.0282}}].

\bibitem{gadget}
V.~{Springel}, {\it {The cosmological simulation code GADGET-2}},  {\em \mnras}
  {\bf 364} (Dec., 2005) 1105--1134,
  [\href{http://xxx.lanl.gov/abs/astro-ph/0505010}{{\tt astro-ph/0505010}}].

\bibitem{2006MNRAS.373..369C}
M.~{Crocce}, S.~{Pueblas}, and R.~{Scoccimarro}, {\it {Transients from initial
  conditions in cosmological simulations}},  {\em \mnras} {\bf 373} (Nov.,
  2006) 369--381, [\href{http://xxx.lanl.gov/abs/astro-ph/0606505}{{\tt
  astro-ph/0606505}}].

\bibitem{OptimalRec}
S.~{Tassev} and M.~{Zaldarriaga}, {\it {Towards an optimal reconstruction of
  baryon oscillations}},  {\em \jcap} {\bf 10} (Oct., 2012) 6,
  [\href{http://xxx.lanl.gov/abs/1203.6066}{{\tt arXiv:1203.6066}}].

\end{thebibliography}\endgroup

\end{document}